\documentclass[fleqn]{2023SCGE}
\usepackage{cite}     
\usepackage{subfigure}
\usepackage{stfloats}
\usepackage[thicklines]{cancel}

\begin{document}

\ensubject{subject}

\ArticleType{Article}
\SpecialTopic{SPECIAL TOPIC: Interfacial instability and flow transition (in Chinese) }
\Year{2024}
\Month{July}
\Vol{66}
\No{1}
\DOI{??}
\ArtNo{000000}
\ReceiveDate{March 24, 2024}
\AcceptDate{July 2, 2024}

\title{Significance of secondary baroclinic hydrodynamic instability on mixing enhancement in shock bubble interaction}

\author[1]{Xu Han}{}%
\author[1]{Bin Yu}{kianyu@sjtu.edu.cn}
\author[1]{Hong Liu}{}

\AuthorMark{Xu Han}

\AuthorCitation{Xu Han, Bin Yu and Hong Liu}

\address[1]{School of Aeronautics and Astronautics, Shanghai Jiao Tong University, Shanghai 200240, PR China}


\abstract{Different strength of hydrodynamic instability can be induced by the variations in the initial diffusion of shock bubble interaction (SBI), while the influence of hydrodynamic instability on variable-density mixing in SBI remains unclear. The present study aims to investigate the hydrodynamic instability of SBI through high-resolution numerical simulations. To isolate each factor within this instability, a circulation control method is employed to ensure consistent Reynolds number $Re$ and P\'eclect number $Pe$. An examination of the morphology of the bubbles and vorticity dynamics reveals that the hydrodynamic instability can be characterized by positive circulation. Through vorticity budget analysis, the positive circulation is dominated by the baroclinic torque. Therefore, the identified hydrodynamic instability is labeled as secondary baroclinic hydrodynamic instability (SBHI). Based on the dimensional analysis of vorticity transport equation, a new dimensionless parameter, the secondary baroclinic vorticity (SBV) number $\Omega_{sbv}$, is proposed to characterize the strength of SBHI. Regarding mixing characteristics, it is observed that cases with stronger SBHI exhibit higher mixing rates. Indicated by the temporal-averaged mixing rate with different $\Omega_{sbv}$, a scaling behavior is revealed: the mixing rate $\overline{\left\langle T_{SDR}\right\rangle}$ is increased proportionally to nearly the square of $\Omega_{sbv}$: $\overline{\left\langle T_{SDR}\right\rangle} \sim \Omega_{sbv}^{2}$. It is widely recognized that unstable flow can also be induced by a high Reynolds number $Re$. The distinction and connection of SBHI and high $Re$ unstable flow are further studied. It is observed that the scaling behavior of the mixing rate in SBHI is distinct from the Reynolds number $Re$ scaling: the mixing can hardly be altered effectively in the limit of large $Re$. The mechanisms behind these two scaling behaviors are inherently different with respect to the stretch term $-2\chi s_i \lambda_i^2$ in the transport equation of the mixing rate, where $s_i$ represents the principal strain and $\lambda_i$ represents the alignment. The SBHI, as characterized by $\Omega_{sbv}$, can enhance mixing by increasing the principal strain $s_1$ and decreasing the alignment $\lambda_1$ to intensify the stretch term. In contrast, the unstable flow induced by high $Re$ does increase the principal strain $s_1$ to a finite extent. However, the absence of significant change in alignment $\lambda_1$ leads to a negligible increase in the stretch term, thus resulting in minimal alteration to the mixing rate.
}



\maketitle


\begin{multicols}{2}
\section{Introduction}\label{section1}
The evolution of the interface between fluids of different densities subjected to shock can be classified as the Richtmyer–Meshkov (RM) instability \cite{richtmyer1960taylor,meshkov1969instability,brouillette2002richtmyer}. This instability is characterized by wave patterns, vortex dynamics, and turbulent mixing, making it a intricate fluid phenomenon. A typical example of the RM instability is the interaction between a shock wave and a cylindrical or spherical gas bubble with a density contrast from the ambient gas, which is known as shock bubble interaction (SBI) \cite{ranjan2011shock}. The misalignment of the pressure gradient ($\nabla p$) and density gradient ($\nabla \rho$) leads to the generation of baroclinic vorticity, depositing on the surface and triggering the formation of a primary vortex within the bubble. Subsequently, secondary instability arises and breaks the interface into smaller structures. Throughout the evolution of the bubble morphology, mixing between the bubble and the ambient gas takes place simultaneously. The complex evolution of the interface in SBI make it to be a typical case for testing the high-resolution numerical method \cite{pan2018high,pan2018conservative}. The investigation of mixing also holds a central position in SBI research due to its significant implications in inertial confinement fusion (ICF) \cite{lindl1992progress}, supernovas \cite{klein1994hydrodynamic}, and supersonic combustion \cite{marble1990shock,yu2020two}.

In a simplified conceptualization, the mixing in SBI can be outlined as follows: generally, following shock interaction, a large-scale primary vortex is formed. Driven by this main vortex, the initially concentrated bubble is tangentially stretched along the interface with the ambient air, leading to an increase in material line length. Simultaneously, as the interface undergoes stretching, the scalar gradient deposited along it is amplified, thereby intensifying molecular diffusion until the bubble is completely mixed with the surrounding air. This simplified mixing process is consistent with the one proposed by Villermaux et al. \cite{villermaux2019mixing}. According to this simplified mixing conception, the material line stretch rate and the bubble area serve as two critical parameters characterizing mixing, and have been widely used in SBI mixing research. Early experiments conducted by Jacobs \cite{jacobs1992shock} utilized the bubble area to depict the extent of mixing and revealed a constant mixing rate by investigating the temporal derivative of the bubble area ratio. Yang \cite{yang1993applications} employed the material line stretch rate to characterize the mixing rate, and discovered an exponential stretching of the material line in the early stage of SBI mixing. Kumar \cite{kumar2005stretching} also conducted experiments on SBI of $\rm{SF_6}$ cylinders using the planar laser-induced fluorescence (PLIF) technique to explore the material line stretch rate, and a strong dependence of the stretch rate on the configuration and orientation of the cylinders was observed. Niederhaus \cite{niederhaus2008computational} studied the mixing process using the mixedness defined by the bubble volume fraction in three-dimensional shock spherical interactions through numerical simulation. Ou and Zhai et al. \cite{ou2019effects} investigated the effect of aspect ratio on mixing in shock elliptic cylinder interaction. The bubble area, mean volume fraction and the mixing rate were presented to illustrate the increase of aspect ratio can promote mixing. Yu\;\cite{liu2020mixing} discovered the amplification effect of secondary baroclinic vorticity on the stretch rate and proposed a model to predict the variable-density mixing time.

Moreover, mixing is a process characterized by the molecular diffusion of mass fraction $Y$. Hence, in addition to the material line stretch rate, mixing indicators can also be directly defined from the mass fraction. In the experiment on the single vortex mixing, Cetegen \cite{cetegen1993experiments} proposed the mixedness $f=4Y(1-Y)$ to quantify the extent of mixing, which reaches maximum value when the bubble and surrounding air are in a 1:1 state. From the derivation of the transport equation for mixedness $f$, Buch et al. \cite{buch1996experimental} put forward the scalar dissipation rate (SDR) $\chi = \frac{\partial Y}{\partial x_i}\frac{\partial Y}{\partial x_i}$ to describe the mixing rate in turbulent shear flow, with $\frac{\partial Y}{\partial x_i}$ representing the scalar gradient. Tomkins \cite{tomkins2008experimental} utilized PLIF experiments of shock interaction with $\rm{SF_6}$ cylinders, and observed that the SDR is primarily distributed on the bridge structure. Shankar et al.\;\cite{shankar2011two} studied the effect of the presence of a third species on the evolution of SBI by using SDR as the mixing rate. Li et al. \cite{li2019gaussian} put forward a Gauss model to predict the late time evolution of SBI based on the definition of the mixedness. In consideration of the difference between variable density mixing and passive scalar mixing, Yu et al. \cite{yu2020scaling} defined the density accelerated mixing rate. Through an investigation of the scaling behavior of this mixing rate with the Reynolds number $Re$ and Schmidt number $S\!c$ \cite{yu2022effects}, it was discovered that with a constant P\'elect number $Pe$, the mixing rate experiences minimal alteration with increasing Reynolds number once $Re$ reaches a large value: $\overline{\left\langle\chi^*\right\rangle}/Pe^{\alpha} \approx \beta Re^{0}$. This scaling behavior was also confirmed in homogeneous isotropic turbulence mixing \cite{buaria2021turbulence}. Furthermore, the underlying mechanism of the scaling behavior can be attributed to the influence of flow on the mixing rate. Based on the derivation of the transport equation of SDR, Buch et al. \cite{buch1996experimental} proposed the stretch term $-2\frac{\partial Y}{\partial x_i} S_{ij} \frac{\partial Y}{\partial x_j}$ as the primary source of the growth of SDR, which has been extensively examined in various types of mixing, including shock-turbulence interaction \cite{tian2017numerical,tian2019density,gao2020parametric}, isotropic turbulence mixing \cite{danish2016influence}, and under-expanded jets \cite{buttay2016analysis}. The characteristic of the stretch term is summarized in the recent review \cite{han2024mixing}.

\begin{figure}[H]
  \centering
  \subfigure[]{\includegraphics[clip=true,width=.45\textwidth]{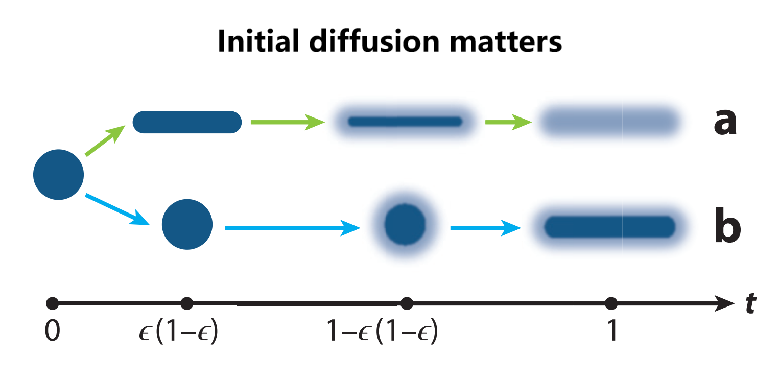}}\\
  \subfigure[]{\includegraphics[clip=true,width=.45\textwidth]{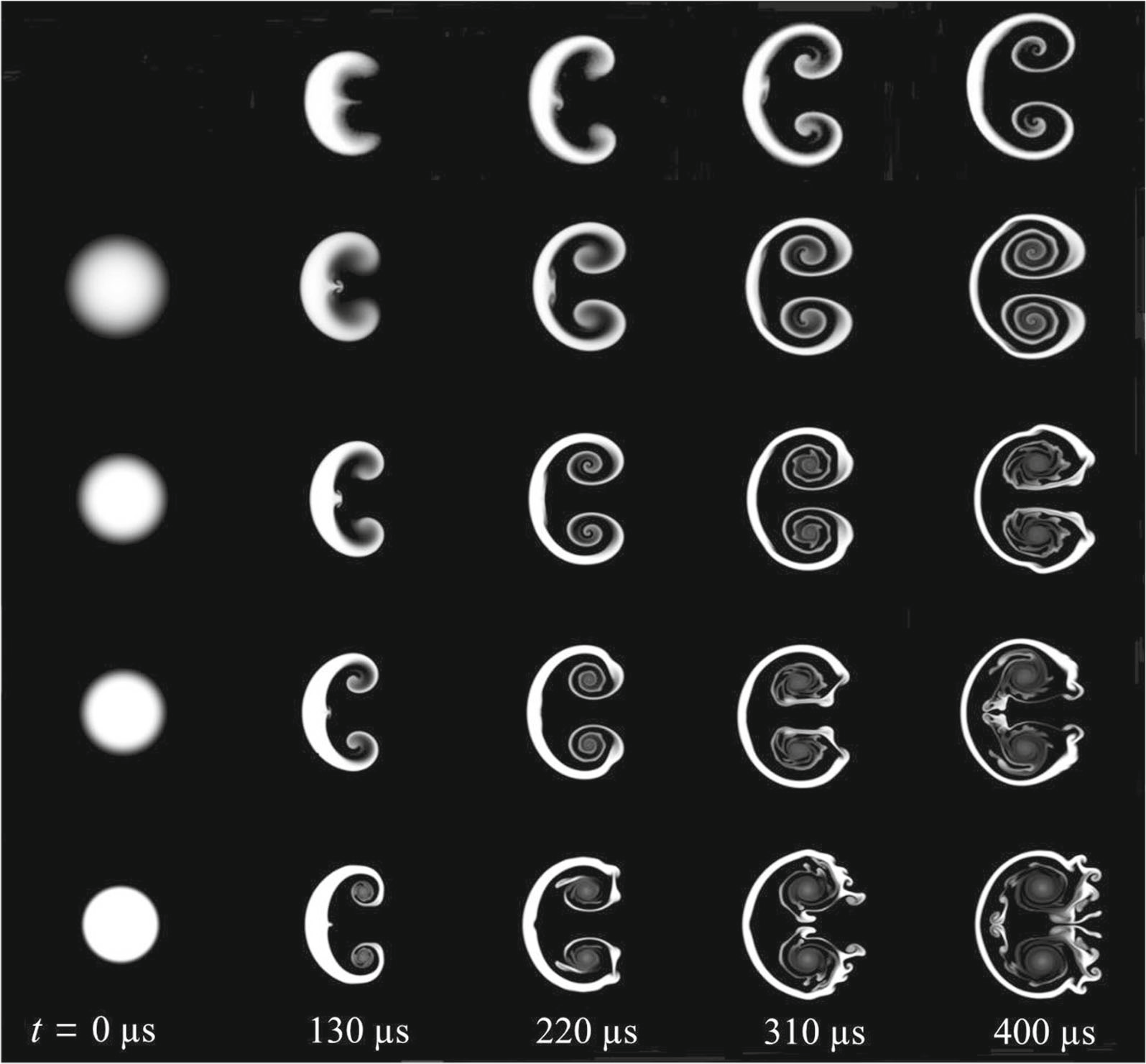}}
  \caption{The effect of initial diffusion on (a) mixing in simple shear flow \cite{villermaux2019mixing} , and (b) mixing in shock $\rm{SF}_6$ cylinder interaction \cite{li2019numerical}.}
  \label{the initial diffusion on mixing}
\end{figure}

With the interesting phenomenon that the initial diffusion can affect mixing even in simple shear flow described by Villermaux \cite{villermaux2019mixing} in Fig.\;\ref{the initial diffusion on mixing} (a), the initial diffusion also widely exists in SBI research. In previous SBI experiments, two common techniques have been used to create the gas cylinder: the membrane technique, such as soap films \cite{ou2019effects,ding2017interaction}, and the membraneless technique, such as jets \cite{jacobs1992shock,kumar2005stretching}. These techniques result in different states of initial diffusion, as indicated by the width of the diffusive layer. Differnt from the findings of Villermaux in Fig.\;\ref{the initial diffusion on mixing}\;(a), the variations in initial diffusion can lead to different strength of hydrodynamic instability, subsequently altering the evolution of the bubbles, as illustrated in Fig.\;\ref{the initial diffusion on mixing}\;(b) \cite{li2019numerical}. Therefore, Li \cite{li2019numerical,li2022effects} conducted a study on the impact of initial diffusion on mixing by examining the material line stretch rate and the bubble area. However, the total circulation was not controlled in these existing studies, resulting in variations in the Reynolds number $Re$ and P\'eclect number $Pe$ across the cases, which in turn made it difficult to distinguish each factor within this hydrodynamic instability. Therefore, the influence of the hydrodynamic instability on mixing remains unclear and can be summarized in the following three questions:

1) What is the source of the hydrodynamic instability induced by initial diffusion?

2) What is the effect of this instability on mixing?

3) What is the mechanism underlying the effect of this instability on mixing?

By investigating the aforementioned three questions, it was discovered that the hydrodynamic instability induced by the initial diffusion, which is characterized by a defined secondary baroclinc vorticity (SBV) number $\Omega_{sbv}$, can enhance mixing by amplifying the SDR stretch term $-2\frac{\partial Y}{\partial x_i}S_{ij}\frac{\partial Y}{\partial x_j}$. This paper is organized as follows: In Section \ref{sec:2}, we provide an overview of the numerical method and the setup of the cases designed to control circulation. Section \ref{sec:3} presents the numerical results and discussion, and the aforementioned three questions are elaborated in this section. Finally, our study is summarized, and potential future work is discussed in Section \ref{sec:4}.

\section{Numerical method and setup}\label{sec:2}
\subsection{Numerical method}
In this study, the governing equations used for simulating two-dimensional SBI are the compressible Navier-Stokes equations with multiple components, expressed as:
\begin{equation}
  \left\{
\begin{aligned}
& \frac{\partial\tilde{\rho}}{\partial \tilde t} + \frac{\partial (\tilde{\rho}\tilde{u_i})}{\partial \tilde{x_i}} = 0, \\
& \frac{\partial\tilde{\rho}\tilde{u_i}}{\partial \tilde t} + \frac{\partial (\tilde{\rho}\tilde{u_i}\tilde{u_j})}{\partial \tilde{x_j}} = -\frac{\partial\tilde{p}}{\partial\tilde{x_i}} + \frac{\partial \tilde{\sigma}_{ij}}{\partial \tilde{x_j}}, \\
&  \frac{\partial(\tilde{\rho}\tilde{E})}{\partial \tilde{t}} + \frac{\partial (\tilde{\rho}\tilde{u_i}\tilde{H})}{\partial \tilde{x_i}} = -\frac{\partial \tilde{q_i}}{\partial \tilde{x_i}} + \frac{\partial(\tilde{u_j}\tilde{\sigma}_{ij})}{\partial \tilde{x_i}},\\
&\frac{\partial(\tilde{\rho}\tilde{Y}_m)}{\partial \tilde{t}} + \frac{\partial(\tilde{\rho}\tilde{u_i}\tilde{Y}_m)}{\partial \tilde{x_i}} = \frac{\partial}{\partial \tilde{x_i}}\left(\tilde{\rho}\mathscr{D}\frac{\partial \tilde{Y}_m}{\partial \tilde{x_i}}\right),\\
& m = 1,2,\cdots,s-1.
\end{aligned} 
  \right.
  \label{NSequation}
\end{equation}
Here, $\tilde{(\cdot)}$ indicates dimensional parameters. The symbols $\tilde{\rho}, \tilde{u_i}, \tilde{p}, \tilde{E}, \tilde{H}$ denote the density, velocity, pressure, energy, and enthalpy, respectively. Furthermore, $\tilde{Y}_m$ refers to the mass fraction of species $m$, and $\tilde{t}$ and $\tilde{x_i}$ represent the corresponding time and space coordinates.

The viscous stress tensor and heat flux are defined as:
\begin{equation}
  \left\{
\begin{aligned}
& \tilde{\sigma}_{ij} = \mu\left(\frac{\partial \tilde{u_i}}{\partial \tilde{x_j}} + \frac{\partial \tilde{u_j}}{\partial \tilde{x_i}} -  \frac{2}{3}\delta_{ij}\frac{\partial \tilde{u_k}}{\partial \tilde{x_k}}\right), \\
& \tilde{q_i} = -\lambda \frac{\partial \tilde{T}}{\partial \tilde{x_i}}, \\
& \lambda = C_p \mu/Pr, 
\end{aligned} 
  \right.
  \label{Viscous tensor and heat flux}
\end{equation}
where $\mu$ and $\lambda$ denote the dynamic viscosity and heat conduction coefficient, respectively. The constant-pressure specific heat is represented by $C_p$, and the Prandtl number is set as a constant, i.e., $Pr = 0.72$. The kinetic viscosity is defined as $\nu = {\mu}/\,{\overline{\rho}}$, the average density $\overline{\rho} = (\rho_{1}' + \rho_{2}')/2$ corresponds to the average of the post-shock ambient air density $\rho_{1}'$ and post-shock helium bubble density  $\rho_{2}'$. These two parameters are obtained from one-dimensional shock dynamics \cite{ranjan2011shock}. In the multi-species transport equations as described in Eq.\;\ref{NSequation}, $\mathscr{D}$ represents the constant Fickian diffusivity. During the SBI simulation, the dynamic viscosity $\mu$ is set as a constant of $50 \times 10^{-6}\, \rm{Pa\cdot s}$, and the constant Fickian diffusivity 
$\mathscr{D}$ is set as $90 \times 10^{-6}\, \rm{m^2/s}$.

In this study, the compressible Navier-Stokes equations are numerically solved using an in-house high-resolution code, $ParNS3D$, which has undergone extensive validation \cite{wang2018scaling,liang2019hidden,liu2020optimal,liu2020mixing}. Temporal discretization is performed using the third-order Total Variation Diminishing (TVD) Runge-Kutta method. The fifth-order Weighted Essentially Non-Oscillatory (WENO) scheme is employed for the discretization of convection terms, while the second-order central difference method discretizes the viscous terms. The grid resolution study of the numerical results is provided in Appendix \ref{grid resolution}. Furthermore, the reliability of these results is verified through budget analysis, which will be presented in subsequent sections.

\subsection{Initial conditions and dimensionless parameters}
\begin{figure}[H]
  \centering
    \includegraphics[clip=true,width=.45\textwidth]{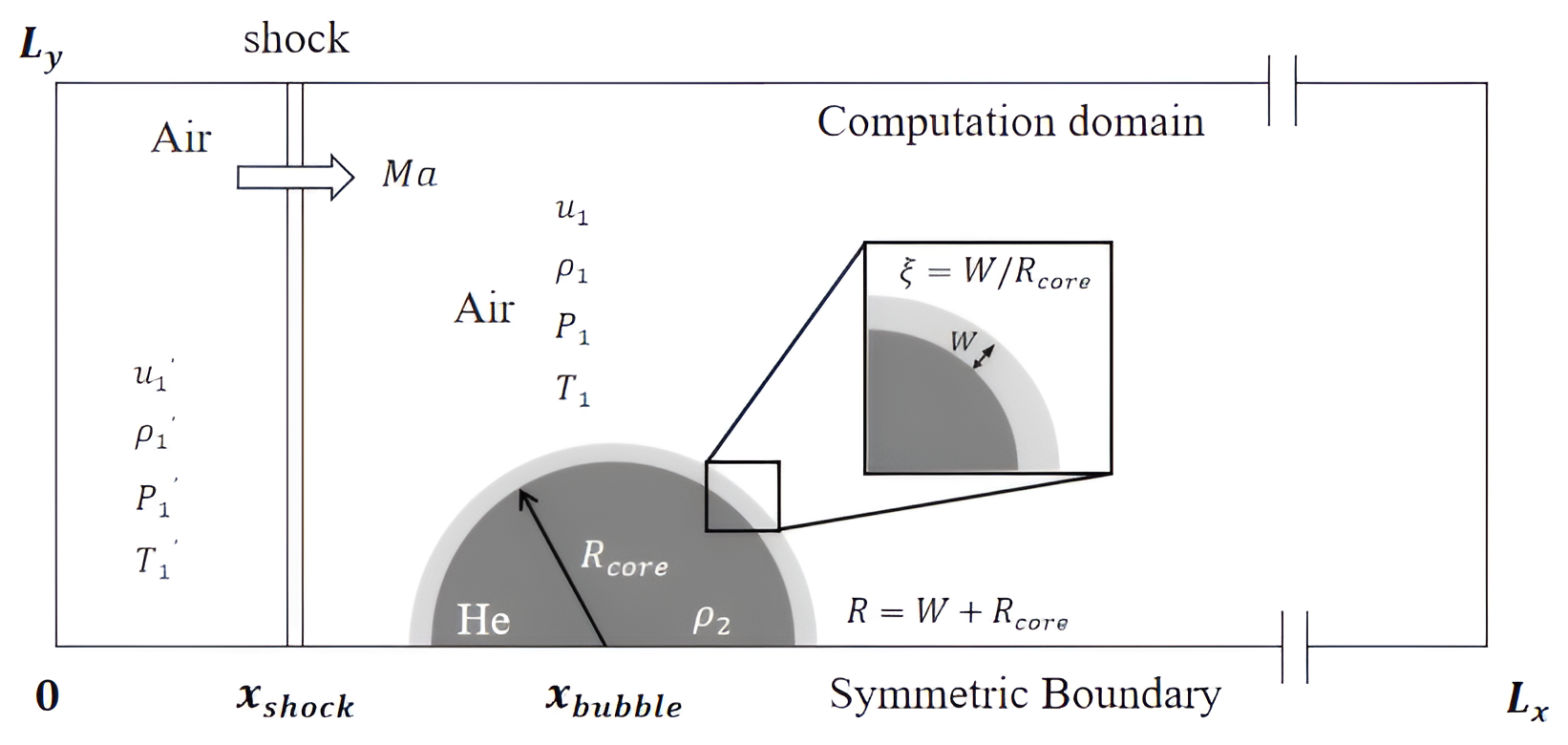}\\
    \caption{Schematic of the initial conditions of shock cylindrical bubble interaction.}
    \label{Initial conditions}
\end{figure}

The initial conditions for the numerical simulation are depicted in Fig.\;\ref{Initial conditions}. The computation domain encompasses a two-dimensional space in a Cartesian coordinate system, with $X \times Y = [0, L_x] \times [0, L_y]$, where the length is $L_x = 30 \,\rm{mm}$ and the height is $L_y = 5.5\,\rm{mm}$. A shock of strength $Ma = 2.4$ is positioned at $x_{shock} = 0.7\,\rm{mm}$. The pre-shock air has thermodynamic states of $1\,\rm{atm}$ pressure and $293\,\rm{K}$ temperature prior to being impacted by the shock. The post-shock air conditions ($u_{1}', \rho_{1}', P_{1}'$, and $T_{1}'$) are calculated from the Rankine-Hugoniot equation. The helium bubble, centered at the position $x_{bubble} = 4.5\,\rm{mm}$, consists of a core region with a radius of $R_{core}$ and a diffusive layer with a width of $W$. Then, the radius of the bubble is $R = R_{core} + W$. A new geometric parameter, $\xi = W/R_{core}$, is defined to represent the initial diffusion degree of the bubble. The diffusive layer is positioned at the edge of the core region, as illustrated in the inserted figure in Fig. \ref{Initial conditions}. Based on this configuration, the distribution of mass fraction of helium is set as: 
\begin{equation}
  Y_{He}(r) = \left\{
\begin{aligned}
& Y_{max}, & &r < R_{core}, \\
& Y_{max}e^{-\alpha[(r-R_{core})/W]^2}, & & R_{core} \leq r < R, \\
& 0. & &r \geq R, 
\end{aligned} 
  \right.
  \label{Mass fraction distribution}
\end{equation}
where $Y_{max} = 1.0$ and $\alpha = 5$, and $Y_{He}$ is simplified as $Y$ in the following content.

This distribution is consistent with that reported in Refs. \cite{wang2018scaling, liu2020mixing,yu2022effects}. The leftmost boundary is set as an inlet, while the upper and rightmost boundaries are conditioned with a fourth-order extra-interpolation to prevent pseudo-pressure reflection wave interference with the concerned flow structures. The symmetric boundary is applied at the bottom of the computational zone. Previous research has shown that the strength of hydrodynamic instability can be influenced by the initial diffusion condition of the bubble \cite{li2019numerical,li2022effects}. To manage this instability, the ratio $\xi = W/R_{core}$ is set at values of $0.1, 1.0, 5.0$, and $10.0$, respectively. 

Following the interaction of the shock wave with the bubble over a brief interval, a nearly constant total circulation is obtained from the spatial integration of the vorticity $\tilde{\omega} = \nabla \times \boldsymbol{u}$. The integration domain lies within the bubble, where the mass fraction of helium exceeds $0.05$ ($Y \geq 0.05$):
\begin{equation}
    \Gamma = \iint_{Y \geq 0.05} \tilde{\omega} \,d\tilde{V}.
    \label{circulation defination}
\end{equation}

Once the total circulation is determined, the non-dimensional Reynolds number can be defined as $Re = {\Gamma}/{\nu}$ \cite{glezer1988formation}. Given the constant average density, $\overline{\rho} = (\rho_{1}' + \rho_{2}')/2$, and constant dynamic viscosity, $\mu$, the kinetic viscosity $\nu = \mu/\,\overline{\rho}$ remains the same throughout the cases. Therefore, the careful selection of the bubble's geometric parameters, $R_{core}, W$ and $R$, is essential to control the circulation $\Gamma$ to manage the Reynolds number $Re$, as discussed in Appendix \ref{circulation control}. Finally, these parameters for the cases with the same Reynolds number $Re$, but different strength of hydrodynamic instability, are confirmed and listed in Table \ref{geometric parameters}. With the same circulation $\Gamma$, the equality of another dimensionless parameter related to mixing, the P\'eclet number $Pe = \Gamma/\mathscr{D}$ \cite{meunier2003vortices} is also guaranteed.

\begin{table}[H]
\caption{Geometric parameters for the two-dimensional SBI cases in this paper}
 \centering
  \begin{tabular}{c c c c}
    \hline
    $\xi$ & $R_{core}\,\rm{(mm)}$ & $W\,\rm{(mm)}$ & $R\,\rm{(mm)}$ \\ 
    \hline
    0.1 & 2.070 & 0.207 & 2.277 \\
    1.0 & 1.390 & 1.390 & 2.780 \\
    5.0 & 0.590 & 2.950 & 3.540 \\
    10.0 & 0.340 & 3.400 & 3.780 \\
    \hline
  \end{tabular}
  \label{geometric parameters}
\end{table}

Another aspect considered in this section is the selection of characteristic time and length to normalize the flow field for the cases outlined in Table \ref{geometric parameters}, each with different length scales. The details of this selection process are depicted in Appendix \ref{Dimensionless time}. The appropriate characteristic length $x^{}$ and time $t^{}$ considering the diffusive layer are defined as:
\begin{equation}
    \left\{
\begin{aligned}
& x^{*} = \sqrt{R R_{eff}}, \\
& t^{*} = \frac{R R_{eff}}{\Gamma}, \\
& R_{eff} = \Gamma/u_1^{'},
\end{aligned} 
  \right.
  \label{characteristic time and length}
\end{equation}
$R_{eff}$ indicates the effective radius. Utilizing the definition in Eq. \ref{characteristic time and length}, the characteristic velocity and vorticity are
\begin{equation}
    \left\{
\begin{aligned}
& u^{*} = \frac{x^{*}}{t^{*}} =  \frac{\Gamma}{\sqrt{RR_{eff}}}, \\
& \omega^{*} = \frac{u^{*}}{x^{*}} = \frac{\Gamma}{RR_{eff}}. \\
\end{aligned} 
  \right.
  \label{characteristic velocity and vorticity}
\end{equation}
Furthermore, the characteristic density $\rho^{*}$ can be given as the average density according to the definition of kinetic viscosity $\nu = {\mu}/\,{\overline{\rho}}$:
\begin{equation}
     \rho^{*} = ({\rho_1^{'}}+{\rho_2^{'}})/2.
\end{equation}

By using these definitions, the physical parameters $\tilde{x},\tilde{t},\tilde{u},\tilde{\omega},\tilde{\rho}$ are normalized as
\begin{equation}
    \left\{
\begin{aligned}
& x = \frac{\tilde{x}}{x^{*}}, \\
& t = \frac{\tilde{t} - t_0}{t^{*}}, \\
& u = \frac{\tilde{u}}{u^{*}}, \\
& \omega = \frac{\tilde{\omega}}{\omega^{*}}, \\
& \rho = \frac{\tilde{\rho}}{\rho^{*}},
\end{aligned} 
  \right.
  \label{normalized parameters}
\end{equation}
where $t_0$ is the moment when the shock wave passes through the right edge of the bubble. Once the physical parameters are non-dimensionalized, the numerical results with different length scales listed in Table \ref{geometric parameters} become comparable. Consequently, the influence of the hydrodynamic instability induced by initial diffusion on VD mixing in two-dimensional SBI can be investigated, as demonstrated in the next section.

\section{Results and discussions}\label{sec:3}
\subsection{Hydrodynamic characteristics of SBHI} \label{SBI on flow}
\subsubsection{Temporal morphology of the bubble}

\begin{figure*}[htbp]
    \centering
    \includegraphics[clip=true,width=.9\textwidth]{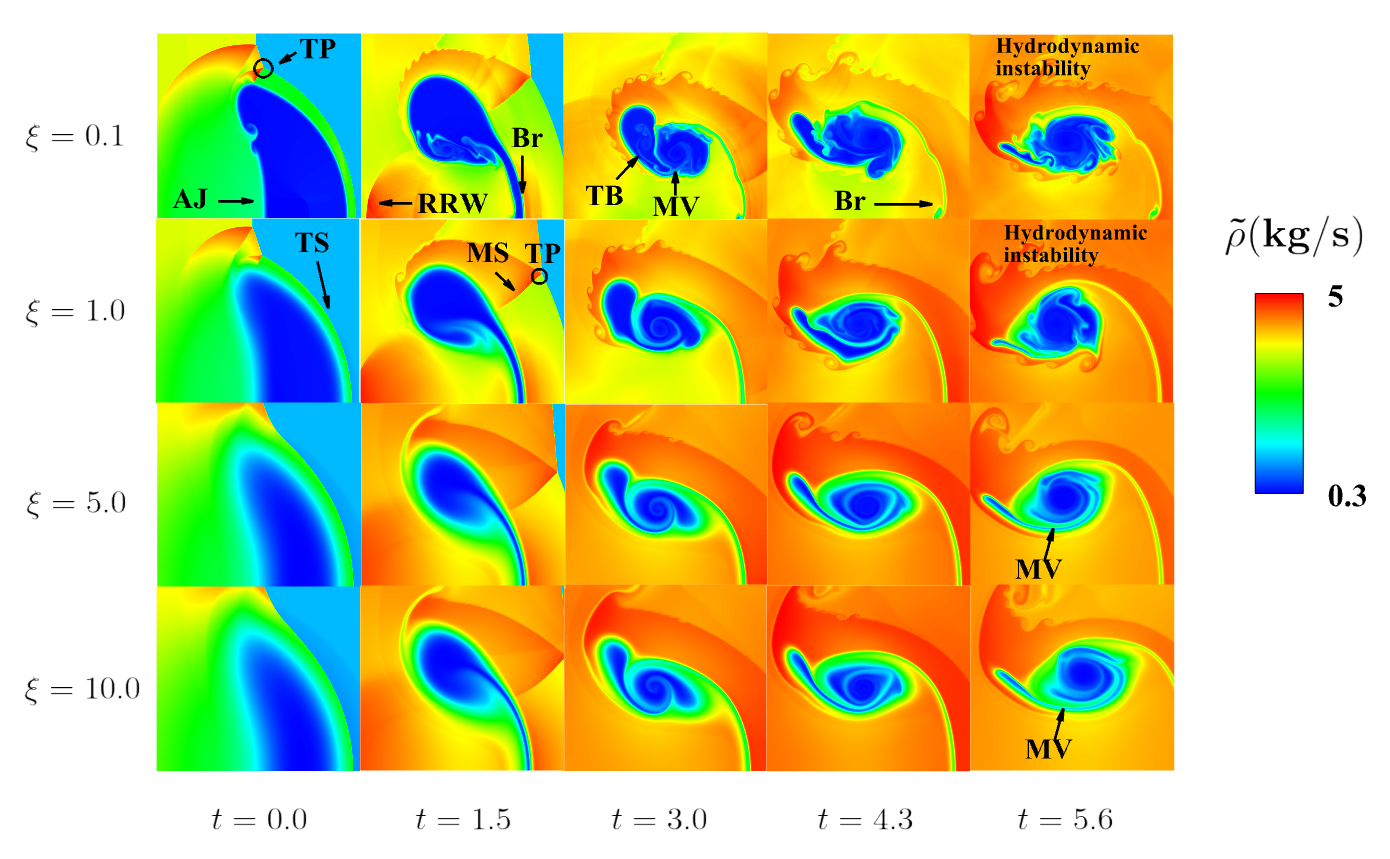}
    \caption{The temporal evolution of the density contour of the four cases with different diffusive layer ratio $\xi$.}
    \label{density contour}
\end{figure*}

\begin{figure*}[htbp]
  \centering
  \includegraphics[clip=true,width=.9\textwidth]{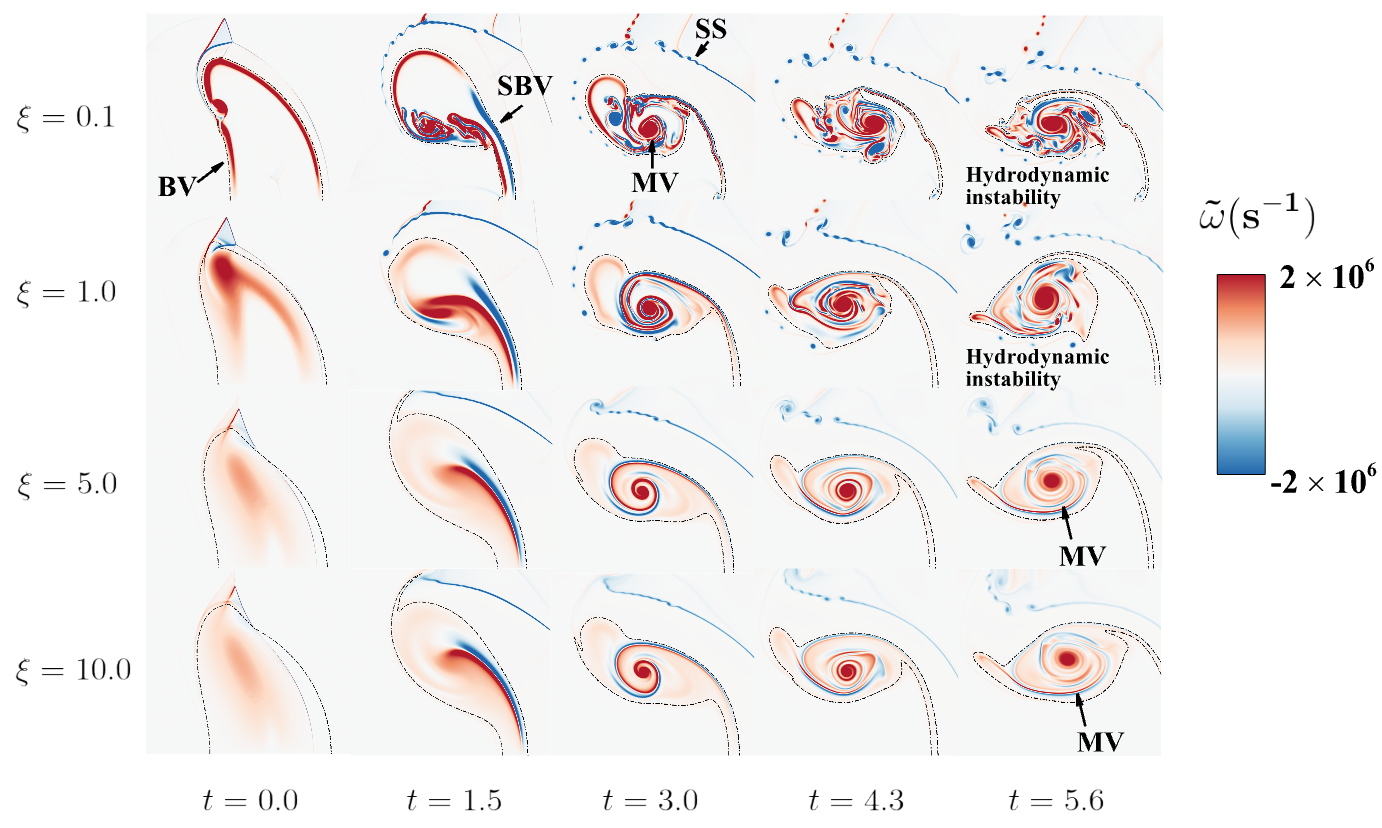}\\
  \caption{The temporal evolution of the vorticity contour of the four cases with different diffusive layer ratio $\xi$. The isoline of $Y=0.05$ is plotted as the black dashed-dot line. }
  \label{vorticity contour}
\end{figure*}

Using the dimensionless time $t$ in Eq.\;\ref{normalized parameters}, we qualitatively demonstrate the temporal evolution of the bubble's morphology. The density and vorticity contours for the cases with different diffusive layer ratios $\xi$ are presented in Fig.\;\ref{density contour} and Fig.\;\ref{vorticity contour}, respectively. It is evident that the bubble's morphology across different cases exhibits similar characteristics at the same dimensionless time, thus confirming the validity of the dimensionless time definition. 

Furthermore, these density and vorticity contours illustrate that SBI is a complex physical phenomenon strongly coupled with wave patterns, bubble deformation, and vortex dynamics. Regarding the wave patterns captured by the density contour in Fig.\;\ref{density contour}, the first relevant shock-related structure is the transmitted shock (TS), attributed to the faster propagation speed of the shock wave in the light bubble compared to the ambient air. The second structure is the reflected rarefaction wave (RRW), a result of the shock's interaction with the light bubble. Additionally, two canonical shock wave structures, Mach stem (MS) and triple points (TP), are also observed. The density contour in Fig.\;\ref{density contour} also reveals flow structures with respect to the bubble's deformation. The bubble is primarily stirred by a main vortex (MV), along with the presence of deformed trailing bubble (TB). The air jet (AJ) connects the bubble's downstream edge, forming the bridge structure (Br), as reported in studies on shock-heavy bubble interactions \cite{tomkins2008experimental}. The irregular interface, stemming from small-scale structures, is observed in the cases of $\xi = 0.1$ and $\xi = 1.0$, which is indicative of the presence of the hydrodynamic instability. Conversely, such instability is not evident in the cases of $\xi = 5.0$ and $\xi = 10.0$. The vorticity-related structures are depicted in the vorticity contour shown in Fig.\;\ref{vorticity contour}. Initially, at $t=0$, the baroclinic vorticity (BV) is located at the bubble edge, arising from the misalignment of the pressure gradient ($\nabla \tilde{p}$) and the density gradient ($\nabla \tilde{\rho}$). The width of the diffusive layer causes variation in the distribution of BV at $t = 0.0$. As the bubble evolves, secondary baroclinic vorticity (SBV) emerges, manifested as dominant-negative vorticity intensifying and rolling up with the positive vorticity. Subsequently, BV and SBV are aggregated together to form the main vortex (MV) structure. The vorticity deposited at the slipstream (SS) forms small-scale vortex, driven by the Kelvin–Helmholtz instability. The presence of small-scale vorticity structures in the cases of $\xi = 0.1$ and $\xi = 1.0$ also provides evidence of the occurrence of hydrodynamic instability, which is not evident in the cases of $\xi = 5.0$ and $\xi = 10.0$.

\subsubsection{Vorticity dynamics in SBHI}
\label{vorticity dynamics section}

After conducting a qualitative study of the morphology of bubble, we perform a qualitative analysis to explore the hydrodynamic characteristics of the instability induced by initial diffusion. Two fundamental parameters, the total circulation $\Gamma$, and the compression rate $\eta$, are presented firstly:
\begin{equation}
    \left\{
\begin{aligned}
& \Gamma = \iint_{Y \geq 0.05} \tilde{\omega} \,dV, \\
& \eta = \frac{\iint_{Y \geq 0.05} \,d\tilde{V}}{\tilde{V_0}} = \frac{\tilde{V}_{bubble}}{\tilde{V_0}}. \\
\end{aligned} 
  \right.
  \label{circulation and compressive rate}
\end{equation}

The circulation $\Gamma$ signifies the strength of the main vortex (MV) in Fig.\;\ref{density contour} and Fig.\;\ref{vorticity contour}, while the compression rate $\eta$ reflects the primary shock compression. The temporal evolution of $\Gamma$ and $\eta$ is depicted in Fig.\;\ref{circulation and compressive rate Fig}. Due to the careful case design detailed in Appendix \ref{circulation control}, after the dimensionless moment $t=0$, when the shock wave propagates through the right edge of the bubble, the circulation $\Gamma$ and compression rate $\eta$ stabilize at approximately $1.7\;\rm{m^2/s}$ and $0.4$, respectively.

\begin{figure}[H]
  \centering
    \includegraphics[clip=true,width=.45\textwidth]{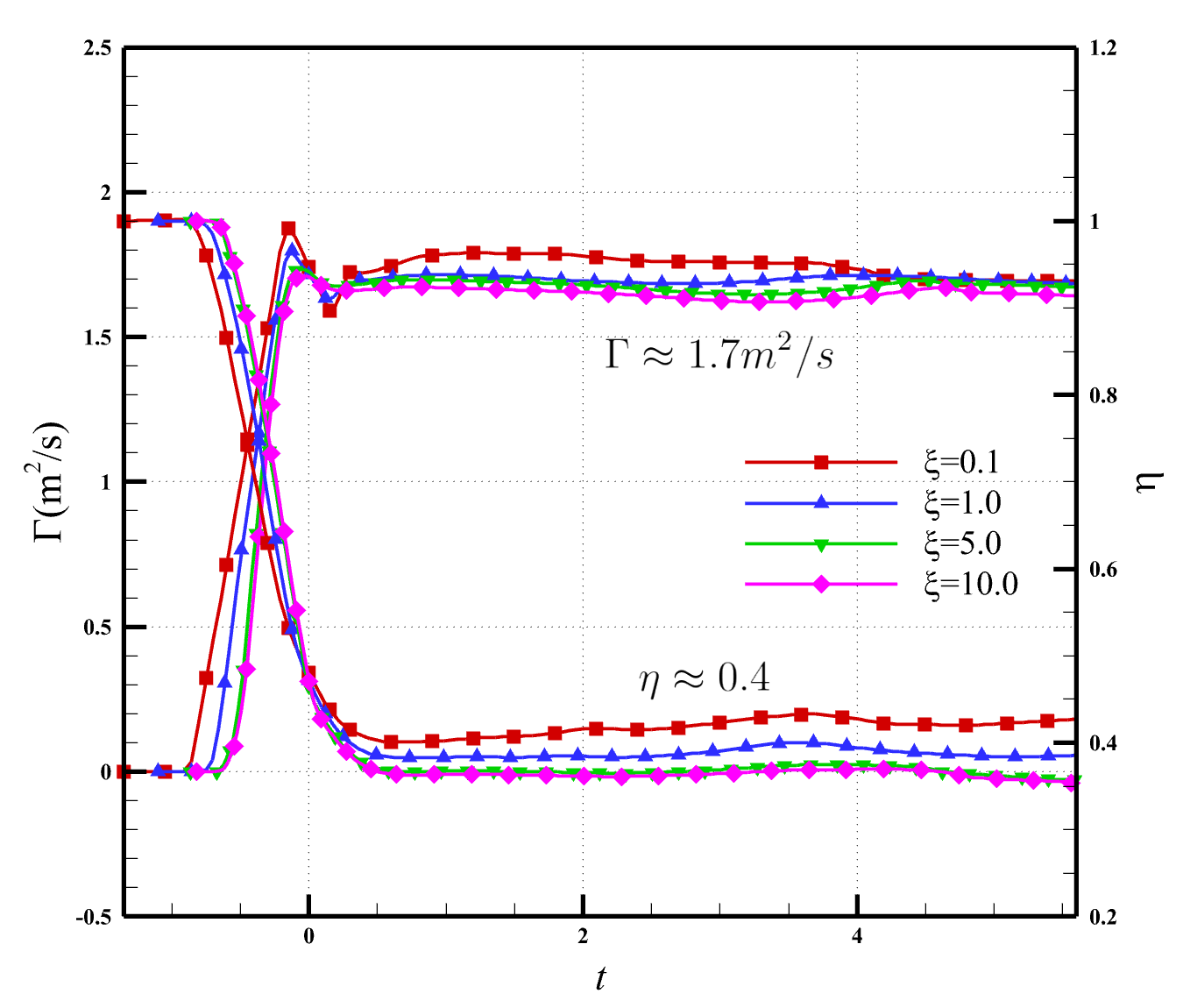}\\
    \caption{Comparison of the temporal variation of the circulation $\Gamma$ and the compression rate $\eta$ }
    \label{circulation and compressive rate Fig}
\end{figure}

Moreover, in Fig.\;\ref{vorticity contour}, we observe that the primary discrepancy within the vorticity field lies in the intensity of SBV. Thus the total circulation $\Gamma$ is decomposed to the positive part $\Gamma^+$ and negative part $\Gamma^-$:
\begin{equation}
    \left\{
    \begin{aligned}
& \Gamma = \Gamma^+ + \Gamma^-, \\
& \Gamma^+ = \left\langle \widetilde{\omega} \right\rangle|_{\tilde{\omega} > 0} =  \left(\iint_{Y \geq 0.05} \widetilde{\omega} \,d\tilde{V}\right)\Large|_{\tilde{\omega} > 0}, \\
& \Gamma^- = \left\langle \widetilde{\omega} \right\rangle|_{\tilde{\omega} < 0} =  \left(\iint_{Y \geq 0.05} \widetilde{\omega} \,d\tilde{V}\right)\Large|_{\tilde{\omega} < 0}. \\
\end{aligned} 
    \right.
    \label{positive and negative circulation equation}
\end{equation}

\begin{figure}[H]
  \centering
    \includegraphics[clip=true,width=.45\textwidth]{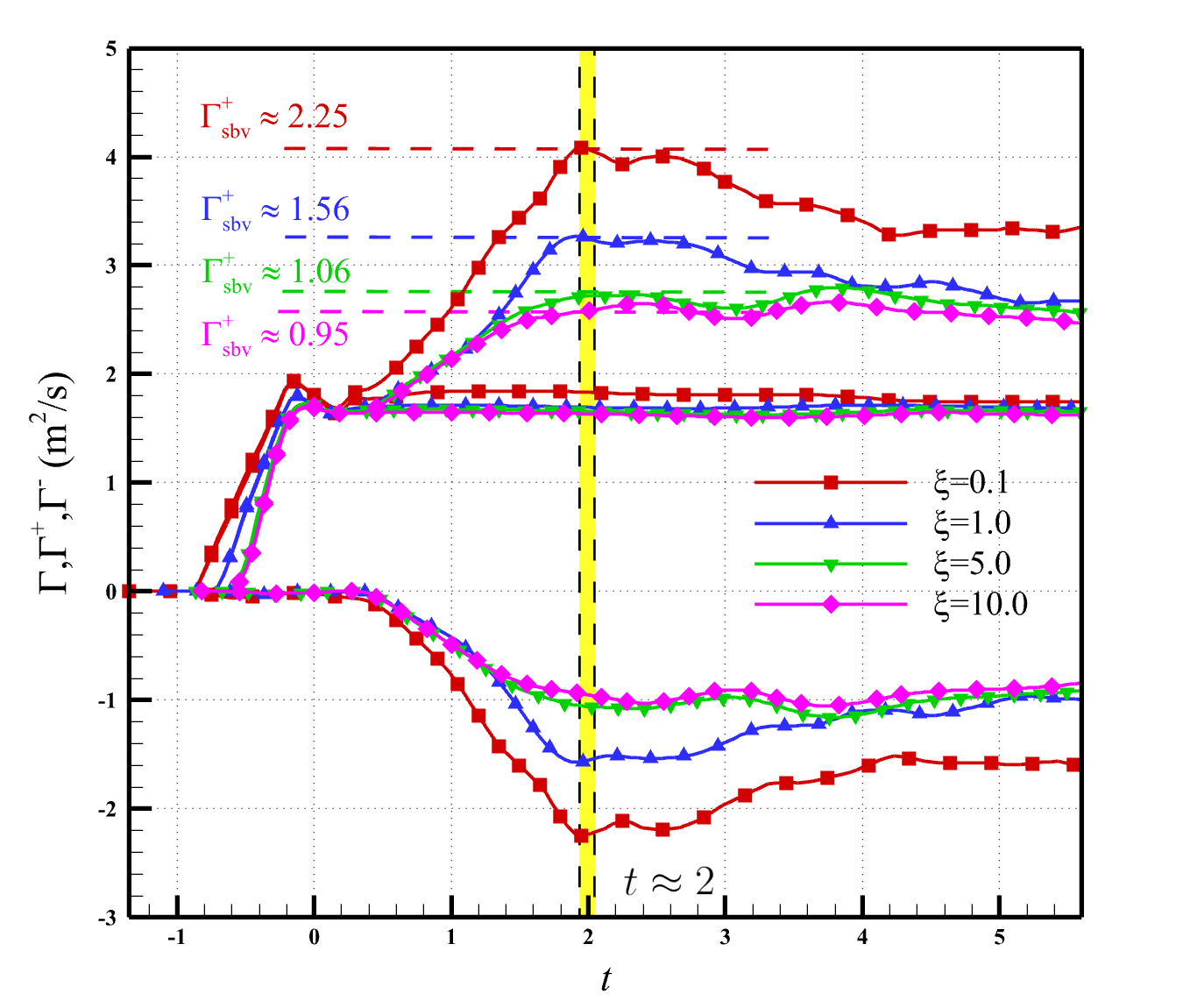}\\
    \caption{Temporal evolution of the circulation $\Gamma$, positive circulation $\Gamma^+$ and negative circulation $\Gamma^-$. $\Gamma_{sbv}^+ = \left(\Gamma^+\right)|_{peak} - \Gamma$ indicates the discrepancy between the peak value of positive circulation and the total circulation at $t=2$.}
    \label{positive and negative circulation}
\end{figure}

In Fig.\ \ref{positive and negative circulation}, following $t=0$, both the positive and negative circulation, denoted as $\Gamma^+$ and $\Gamma^-$, exhibit identical growth rates, which confirms the conservative nature of the total circulation $\Gamma$. It is also evident that at $t=2$, the positive circulation $\Gamma^+$ reaches its peak value, mirroring the behavior of the negative circulation $\Gamma^-$. Due to the growth rate and peak value of the positive circulation $\Gamma^+$ originate from the secondary baroclinic vorticity (SBV) shown in Fig.\;\ref{vorticity contour}, as will be validated by the budget analysis shortly, this hydrodynamic instability resulting from the initial diffusion is labeled as the secondary baroclinic hydrodynamic instability (SBHI), and can be measured by the discrepancy between the peak value of positive circulation and the total circulation $\Gamma_{sbv}^+ = \left(\Gamma^+\right)|_{peak} - \Gamma$.

We further analyze the positive circulation budget to identify the source of $\Gamma_{sbv}^+$. The vorticity transport equation is expressed as:
\begin{equation}
    \begin{aligned}
    \frac{D\widetilde{\omega_i}}{D\tilde{t}} &= -\left(\frac{\partial\widetilde{u_k}}{\partial\widetilde x_k}\right)\widetilde{\omega_i} + \widetilde{S_{ij}}\tilde{\omega_j} + \epsilon_{ijk}\frac{1}{\tilde{\rho}^2}\frac{\partial \tilde{\rho}}{\partial \tilde{x_j}}\frac{\partial \tilde{p}}{\partial \tilde{x_k}}\\
    &+ \nu \frac{\partial^2 \widetilde{\omega_i}}{\partial \widetilde{x_k}^2},
    \label{vorticity transport equation}
    \end{aligned}
\end{equation} The first term on the right side of Eq.\;\ref{vorticity transport equation}, $-\left(\frac{\partial\widetilde{u_k}}{\partial\widetilde x_k}\right)\widetilde{\omega_i}$, corresponds to the compression term resulting from the velocity divergence $\frac{\partial\widetilde{u_k}}{\partial\widetilde x_k}$ and is denoted as $\tilde{T}_{divU}$. The second term $\widetilde{S_{ij}}\tilde{\omega_j}$ represents the vortex stretch term, and this term vanishes in two-dimensional flow. The third term $\epsilon_{ijk}\frac{1}{\tilde{\rho}^2}\frac{\partial \tilde{\rho}}{\partial \tilde{x_j}}\frac{\partial \tilde{p}}{\partial \tilde{x_k}}$ constitutes the baroclinic torque originating from the misalignment of pressure gradient $\nabla \tilde{p}$ and density gradient $\nabla \tilde{\rho}$ and is indicated as $\tilde{B}$. The fourth term $\nu \frac{\partial^2 \widetilde{\omega_i}}{\partial \widetilde{x_k}^2}$ represents the viscous term and is labeled as $\widetilde{T_v}$. In two-dimensional SBI cases, the vorticity $\widetilde{\omega_i}$ only aligns with the Z coordinate, so it can be expressed as $\tilde{\omega}$. Then, Eq.\;\ref{vorticity transport equation} is simplified to:
\begin{equation}
    \frac{D\tilde{\omega}}{D\tilde{t}} = \tilde{T}_{divU} + \tilde{B} + \widetilde{T_{v}}.
    \label{vorticity transport equation 2}
\end{equation} According to the definition of positive circulation in Eq.\;\ref{positive and negative circulation equation}, the time derivative of $\Gamma^+$ is:
\begin{equation}
    \begin{aligned}
         \frac{D\Gamma^+}{D\tilde{t}} &=  \frac{D\left\langle \tilde{\omega} \right\rangle|_{\tilde{\omega}> 0}}{D\tilde{t}} = \left\langle \frac{D\tilde{\omega}}{D\tilde{t}} \right\rangle\Bigg|_{\tilde{\omega}> 0} + \left\langle \left(\frac{\partial\widetilde{u_k}}{\partial\widetilde x_k}\right)\widetilde{\omega} \right\rangle\Bigg|_{\tilde{\omega}> 0} \\
         &= \left\langle \frac{D\tilde{\omega}}{D\tilde{t}} \right\rangle\Bigg|_{\tilde{\omega}> 0} - \left\langle \tilde{T}_{divU} \right\rangle\Big|_{\tilde{\omega}> 0}.
    \end{aligned} 
    \label{derivative of positive circulation 1}
\end{equation} Substituting the vorticity transport equation Eq.\;\ref{vorticity transport equation 2}, Eq.\;\ref{derivative of positive circulation 1} transforms to: 
\begin{equation}
     \frac{D\Gamma^+}{D\tilde{t}}  = \left\langle \widetilde{B} \right\rangle\Big|_{\tilde{\omega}>0} + \left\langle \widetilde{T_v} \right\rangle\Big|_{\tilde{\omega}>0} = \left\langle \widetilde{B^+} \right\rangle + \left\langle \widetilde{T_v^+} \right\rangle
     \label{derivative of positive circulation 2}
\end{equation}
The left-hand side of Eq. \ref{derivative of positive circulation 2} can be normalized as following by employing the characteristic vorticity, time, and length in Eq.\;\ref{characteristic time and length} and Eq.\;\ref{characteristic velocity and vorticity}:
\begin{equation}
    \begin{aligned}
         \frac{D\Gamma^+}{D\tilde{t}} &= \frac{D\iint_{Y \geq 0.05} \tilde{\omega} \,d\tilde{V} }{D\tilde{t}}\Bigg|_{\tilde{\omega}> 0} = \frac{D\iint_{Y \geq 0.05} (\omega \omega^*) \,d({x^*}^2 V)}{D(t t^*)}\Bigg|_{\omega > 0} \\
         & = \frac{\Gamma^2}{RR_{eff}} \frac{D\left\langle \omega \right \rangle |_{\omega > 0}}{Dt} =  \frac{\Gamma^2}{RR_{eff}} \frac{D \Gamma_n^+}{Dt}. 
    \end{aligned}
\end{equation} $\Gamma_n^+ = \iint_{Y \geq 0.05} \omega \,dV = \left\langle \omega \right \rangle |_{\omega > 0}$ indicates the dimensionless positive circulation, which can also relate to the total circulation by: 
\begin{equation}
    \begin{aligned}
        \Gamma^+ &= \iint_{Y \geq 0.05} \frac{(\omega \omega^*) \,d({x^*}^2 V)}{D(t t^*)} \\
        & = \Gamma \iint_{Y \geq 0.05} \omega \,dV = \Gamma \left\langle \omega \right \rangle |_{\omega > 0} = \Gamma \Gamma_n^+.
    \end{aligned}
\end{equation}
Subsequently, Eq.\;\ref{derivative of positive circulation 2} is transformed to:
\begin{equation}
    \begin{aligned}
         \frac{D\Gamma_n^+}{D\tilde{t}} &= \frac{RR_{eff}}{\Gamma^2}\left( \left\langle \widetilde{B^+} \right\rangle + \left\langle \widetilde{T_v^+} \right\rangle\right) \\
         & = \left\langle B^+ \right\rangle + \left\langle T_v^+ \right\rangle.
    \end{aligned}
    \label{normalized positive circulation equation}
\end{equation}
In Eq.\;\ref{normalized positive circulation equation}, $\frac{D\Gamma_n^+}{D\tilde{t}}$ can be simplified as:
\begin{equation}
    \begin{aligned}
        \frac{D\Gamma_n^+}{Dt} &= \frac{D\left\langle \omega \right\rangle|_{\omega>0}}{Dt} = \left \langle \frac{D \omega}{Dt} \right \rangle \Bigg|_{\omega > 0} + \left \langle \omega\nabla\cdot \boldsymbol{u}\right \rangle \Big|_{\omega > 0} \\
        & = \left \langle \frac{\partial \omega}{\partial t} \right \rangle \Bigg|_{\omega > 0} + \left \langle \boldsymbol{u}\cdot \nabla \omega \right \rangle \Big|_{\omega > 0} + \left \langle \omega\nabla\cdot \boldsymbol{u}\right \rangle \Big|_{\omega > 0} \\
        & =  \left \langle \frac{\partial \omega}{\partial t} \right \rangle \Bigg|_{\omega > 0} + \left \langle \nabla \cdot (\omega \boldsymbol{u}) \right \rangle \Big|_{\omega > 0} = \frac{\partial \left \langle \omega \right \rangle |_{\omega > 0}}{\partial t}.
    \end{aligned}
\end{equation}
By utilizing the relationship that the spatial integration of the divergence of vorticity flux $\left \langle \nabla \cdot (\omega \boldsymbol{u}) \right \rangle = 0$ if the integration boundary is $\omega = 0$ \cite{yu2022effects}, we can replace the unavailable material derivative with the easily computed time derivative. Finally, the simplest form of Eq.\;\ref{normalized positive circulation equation} is:
\begin{equation}
    \frac{D\Gamma_n^+}{Dt} = \frac{\partial \left \langle \omega \right \rangle |_{\omega > 0}}{\partial t}  = \left\langle B^+ \right\rangle + \left\langle T_v^+ \right\rangle.
    \label{final form of circulation equation}
\end{equation}

\begin{figure*}[htbp]
  \centering
  \includegraphics[clip=true,width=.65\textwidth]{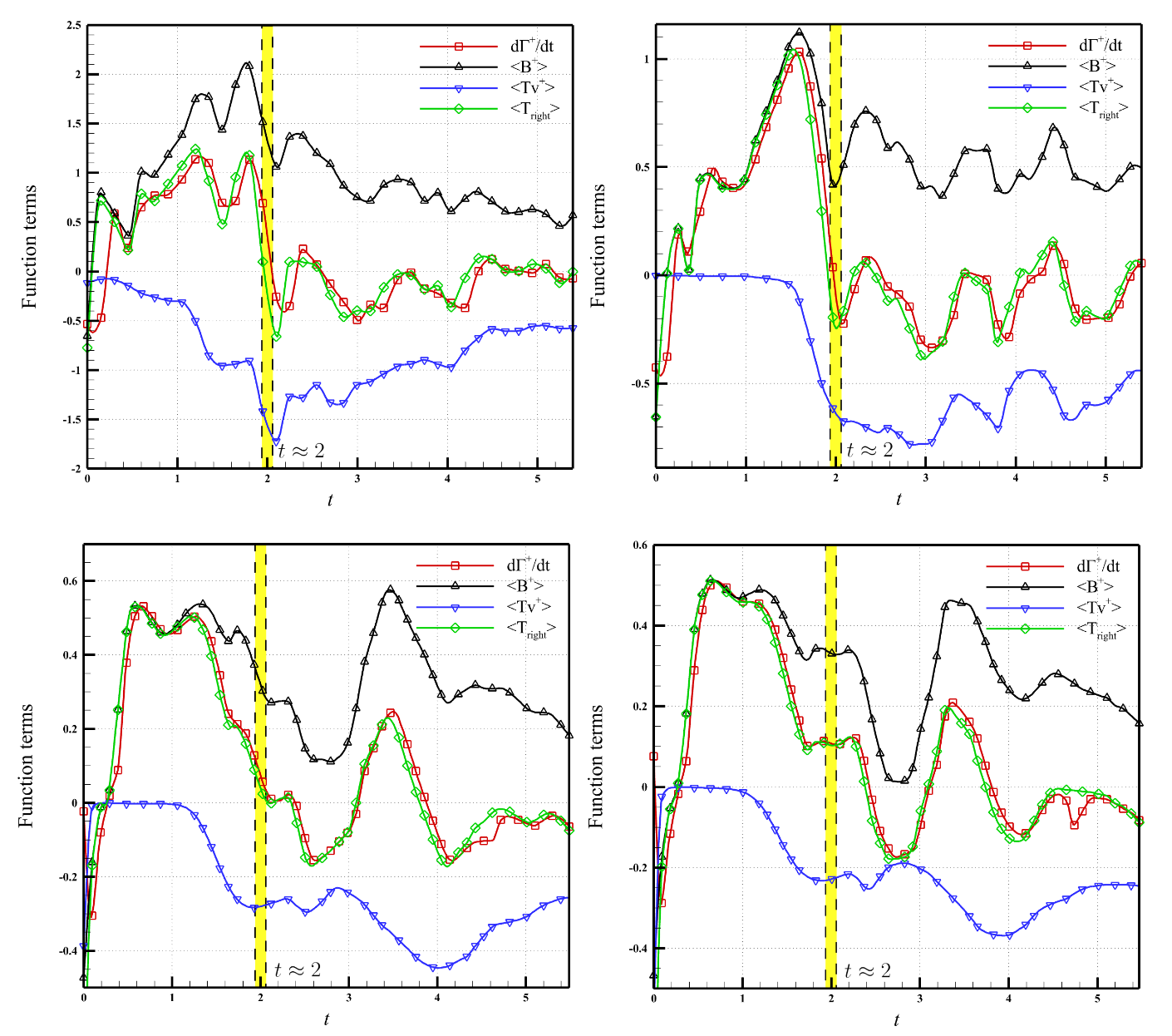}\\
  \caption{Time derivative of the normalized positive circulation $\Gamma_n^+$ and the corresponding source terms in Eq. \ref{final form of circulation equation}.}
  \label{time derivative of positive circulation}
\end{figure*}

\begin{figure}[H]
  \centering
    \includegraphics[clip=true,width=.45\textwidth]{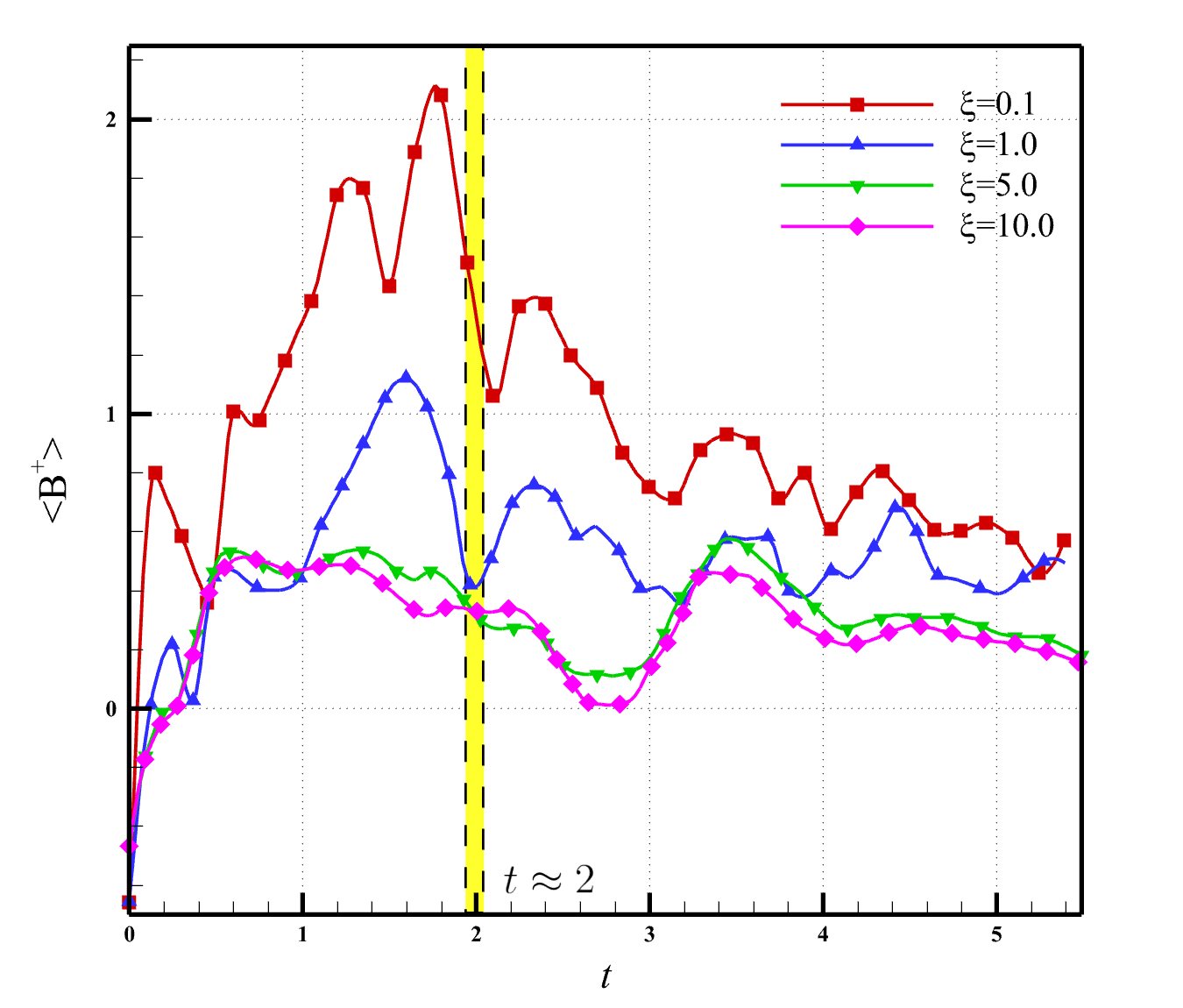}\\
    \caption{Comparison of the temporal evolution of baroclinic torque $\left\langle B^+\right\rangle$, which is the primary source term of positive circulation $\Gamma^+$.}
    \label{baroclinic torque}
\end{figure}

The time derivative of the normalized positive circulation $\Gamma_n^+$ and its corresponding source terms are depicted in Fig.\;\ref{time derivative of positive circulation}. It is evident that the green line, representing the sum of terms on the right-hand side of Eq.\;\ref{final form of circulation equation} $\left \langle T_{right} \right \rangle$, closely matches the red line depicting the time derivative of the dimensionless positive circulation, $\frac{D\Gamma_n^+}{Dt}$. This observation validates the reliability of numerical results in capturing vorticity dynamics within the grid, as demonstrated in Appendix \ref{grid resolution}.
It can also be observed before $t = 2$, the positive baroclinic torque $\left\langle B^+ \right\rangle$ dominates the growth of $\Gamma^+$; after $t=2$, the negative viscous term $\left\langle T_v^+ \right\rangle$ becomes the major factor hindering the growth of $\Gamma^+$. This observation explains why the moment when $\Gamma^+$ reaches its peak value occurs at $t=2$.
The temporal evolution of $\left\langle B^+ \right\rangle$ of different cases is compared in Fig.\;\ref{baroclinic torque}, and this comparison reveals the variation in $\Gamma_{sbv}^+$ arises from different baroclinic torque $\left\langle B^+ \right\rangle$. In summary, we can conclude that the hydrodynamic instability induced by the initial diffusion can be labeled as the SBHI here, as it originates from the baroclinic torque $\left\langle B^+ \right\rangle$, and we can measure the strength of SBHI by the discrepancy between the peak value of positive circulation and the total circulation $\Gamma_{sbv}^+$.

\subsection{Effect of SBHI on mixing} \label{SBI on mixing}
\subsubsection{Mixedness dynamics of SBHI}
\begin{figure*}[htbp]
  \centering
  \includegraphics[clip=true,width=.60\textwidth]{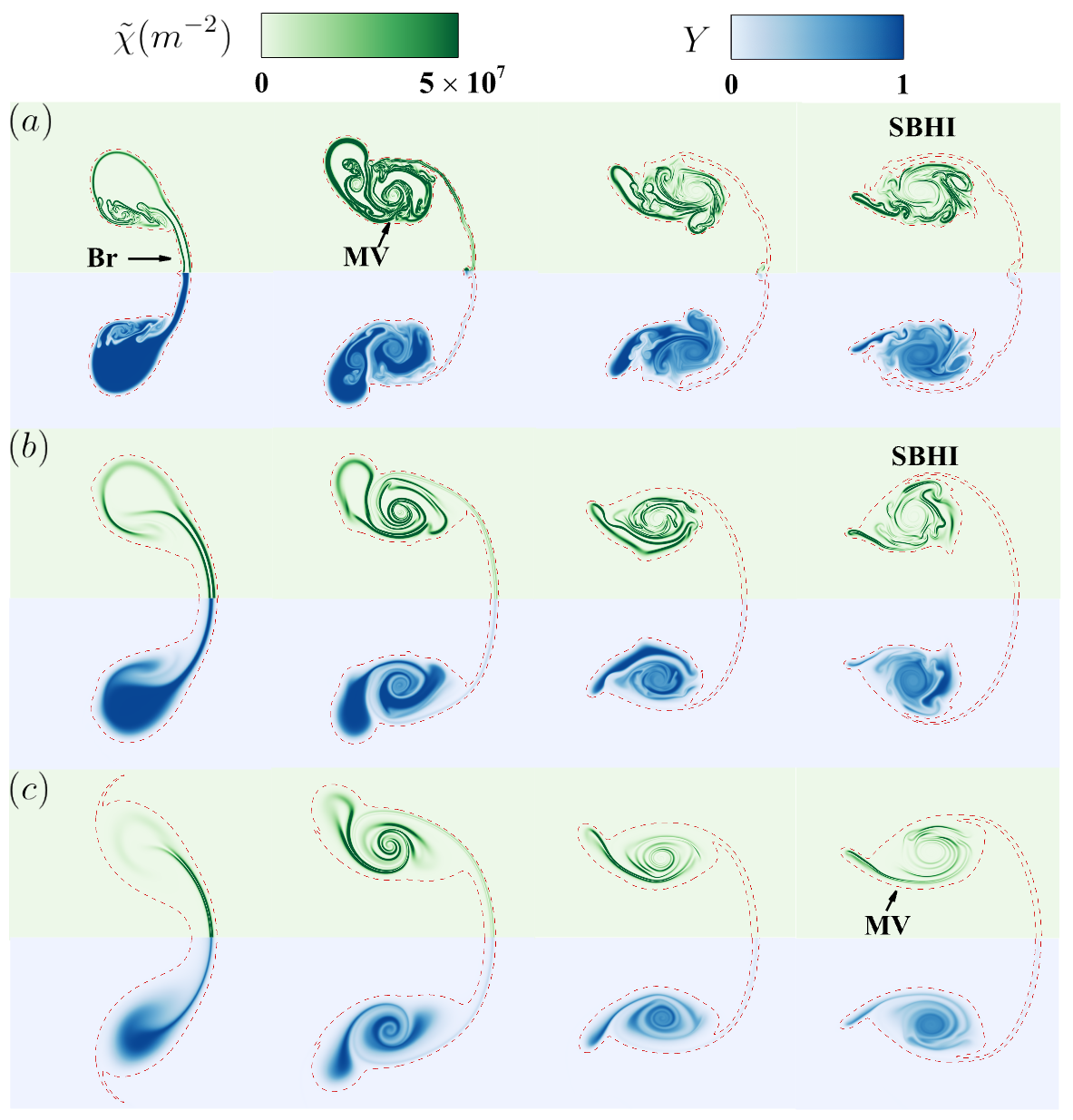}\\
  \caption{The temporal evolution of SDR $\tilde{\chi}$ contour (upper) and mass fraction contour $Y$ (lower) of three cases with different diffusive layer ratio (a) $\xi = 0.1$, (b) $\xi = 1.0$ and (c) $\xi = 5.0$. The isoline of $Y=0.05$ is plotted as red dashed-dot line.}
  \label{MixingField}
\end{figure*}

In section \ref{SBI on flow}, the hydrodynamic instability induced by the initial diffusion, which has been shown to originate from the baroclinic torque, is detonated as the SBHI. In this section, we focus on analyzing the effect of SBHI on the mixing characteristics. Two mixing parameters are employed to characterize the mixing process of SBI. Firstly, the mixedness $f$ is used to describe the local mixing extent of the mixture. Cetegon et al. \cite{cetegen1993experiments} defined mixedness $f$ by the equation:
\begin{equation}
    f = 4Y\left(1-Y\right).
    \label{mixedness}
\end{equation}
$Y$ represents the mass fraction of the relevant gas component, chosen as helium in this study. The mixedness $f$ reaches its maximum when the mixture is in a state where helium and the ambient air mix in a 1:1 proportion.

Secondly, the scalar dissipation rate (SDR) is considered. Buch et al. \cite{buch1996experimental} defined SDR $\tilde{\chi}$ as:
\begin{equation}
    \tilde{\chi} = \frac{\partial Y}{\partial \widetilde{x_i}}\frac{\partial Y}{\partial \widetilde{x_i}}.
    \label{scalar dissipation rate definition}
\end{equation}
$\frac{\partial Y}{\partial \widetilde{x_i}}$ indicates the spatial gradient of the mass fraction $Y$, referred to as the scalar gradient in the subsequent content. The SDR is the square of the modulus of the scalar gradient and represents the local mixing rate according to the mixedness transport equation:
\begin{equation}
    \frac{Df}{D\tilde{t}} = 8\mathscr{D}\tilde{\chi} + \mathscr{D}\frac{\partial^2 f}{\partial \widetilde{x_j}^2} + 4(1-2Y)\frac{\mathscr{D}}{\tilde{\rho}}\frac{\partial \tilde{\rho}}{\partial \widetilde{x_j}}\frac{\partial Y}{\partial \widetilde{x_j}}.
    \label{mixedness equation 1}
\end{equation}

The first term in Eq.\;\ref{mixedness equation 1}, $8\mathscr{D}\tilde{\chi}$, labeled as the SDR term $\tilde{T}_{SDR}$, explains why SDR $\tilde{\chi}$ represents the local mixing rate. The second term, $\mathscr{D}\frac{\partial^2 f}{\partial \widetilde{x_j}^2}$ is the diffusion term $\tilde{T}_{diff}$, which describes the transport of the mixedness $f$ by molecular diffusion. The third term, $4(1-2Y)\frac{\mathscr{D}}{\tilde{\rho}}\frac{\partial \tilde{\rho}}{\partial \widetilde{x_j}}\frac{\partial Y}{\partial \widetilde{x_j}}$, arises from the production of the density gradient $\frac{\partial \tilde{\rho}}{\partial \widetilde{x_j}}$ and the scalar gradient $\frac{\partial Y}{\partial \widetilde{x_j}}$, and is denoted as $\tilde{T}_{den}$.

The temporal evolution of SDR ($\tilde{\chi}$) contour and mass fraction ($Y$) contour is shown in Fig.\;\ref{MixingField}. Due to the similarities in the morphology of $\tilde{\chi}$ and $Y$ in the cases of $\xi = 5.0$ and $\xi = 10.0$, only the mixing field of $\xi = 5.0$ is illustrated. In the early stage of SBI mixing, a region with a high mixing rate is observed around the bridge (Br) structure, this observation is consistent with the previous experiments conducted by Tomkins \cite{tomkins2008experimental}. Subsequently, during the formation of the main vortex (MV), mixing is intensified by the stirring effect, leading to a high mixing rate inside the MV. As the mixing process approaches completion in the late stage, the SDR diminishes to a relatively low level. It is evident that the SBHI, characterized by the irregular interface originating from small-scale structures, amplifies the magnitude of SDR, thereby enhancing the mixing rate. This qualitative conclusion is supported by the quantitative analysis of mixing parameters in the following content.

To facilitate a quantitative comparison of mixing parameters, it is essential to transform the terms in Eq.\;\ref{mixedness equation 1} into the dimensionless form. Using the scaling method presented in Eq.\;\ref{normalized parameters}, each term in Eq.\;\ref{mixedness equation 1} can be normalized as:
\begin{equation}
    \left\{
    \begin{aligned}
& \frac{Df}{D\tilde{t}} = \frac{Df}{D(t^* t)} = \frac{\Gamma}{RR_{eff}}\frac{Df}{Dt}, \\
& 8\mathscr{D}\tilde{\chi} = 8\mathscr{D}\frac{\partial Y}{\partial \widetilde{x_j}}\frac{\partial Y}{\partial \widetilde{x_j}} = 8\mathscr{D}\frac{\partial Y}{\partial x^* x_j}\frac{\partial Y}{\partial x^* x_j} = 8\frac{\mathscr{D}}{R R_{eff}}\chi, \\
& \mathscr{D}\frac{\partial^2 f}{\partial \widetilde{x_j}^2} = \mathscr{D}\frac{\partial^2 f}{\partial ({x^*}^2 {x_j}^2)} = \frac{\mathscr{D}}{RR_{eff}}\frac{\partial^2 f}{\partial {x_j}^2}, \\
&  4(1-2Y)\frac{\mathscr{D}}{\tilde{\rho}}\frac{\partial \tilde{\rho}}{\partial \widetilde{x_j}}\frac{\partial Y}{\partial \widetilde{x_j}} = 4(1-2Y)\frac{\mathscr{D}}{RR_{eff}}\frac{\partial \rho}{\partial x_j}\frac{\partial Y}{\partial x_j},
\end{aligned} 
    \right.
    \label{mixedness equation 2}
\end{equation}
Substituting these normalized terms to Eq.\;\ref{mixedness equation 1}, the dimensionless mixedness transport equation is:
\begin{equation}
    \begin{aligned}
        \frac{Df}{Dt} &= 8Pe\chi + Pe\frac{\partial^2 f}{\partial {x_j}^2} + 4Pe(1-2Y)\frac{1}{\rho}\frac{\partial \rho}{\partial x_j}\frac{\partial Y}{\partial x_j}\\
        & = T_{SDR} + T_{diff} + T_{den}.
    \end{aligned}
    \label{mixedness equation 3}
\end{equation}
In this equation, $Pe = {\Gamma}/{\mathscr{D}}$ is the P\'eclet number, which remains constant due to the uniform circulation $\Gamma$ across the cases.

The spatial integration of mixedness $\left\langle f \right \rangle = \iint_{Y \geq 0.05} f \,dV$ is employed to reflect the global mixing extent, the corresponding transport equation is:
\begin{equation}
    \begin{aligned}
        \frac{D\left \langle f \right \rangle}{Dt} &= \left \langle \frac{Df}{Dt} \right \rangle + \left \langle f\left(\frac{\partial u_k}{\partial x_k}\right) \right \rangle\\
        &= \left \langle T_{SDR} \right \rangle + \left \langle T_{diff} \right \rangle  + \left \langle T_{den} \right \rangle + \left \langle T_{divU} \right \rangle \\
        & = \left \langle T_{SDR} \right \rangle + \left \langle T_{den} \right \rangle + \left \langle T_{divU} \right \rangle,
    \end{aligned}
    \label{mixedness equation 4}
\end{equation}
where $\left \langle T_{diff} \right \rangle = 0$ owing to the mixedness flux $\nabla f$ equals to 0 at the boundary the the integration region \cite{yu2020scaling}.

\begin{figure}[H]
  \centering
    {\includegraphics[clip=true,width=.38\textwidth]{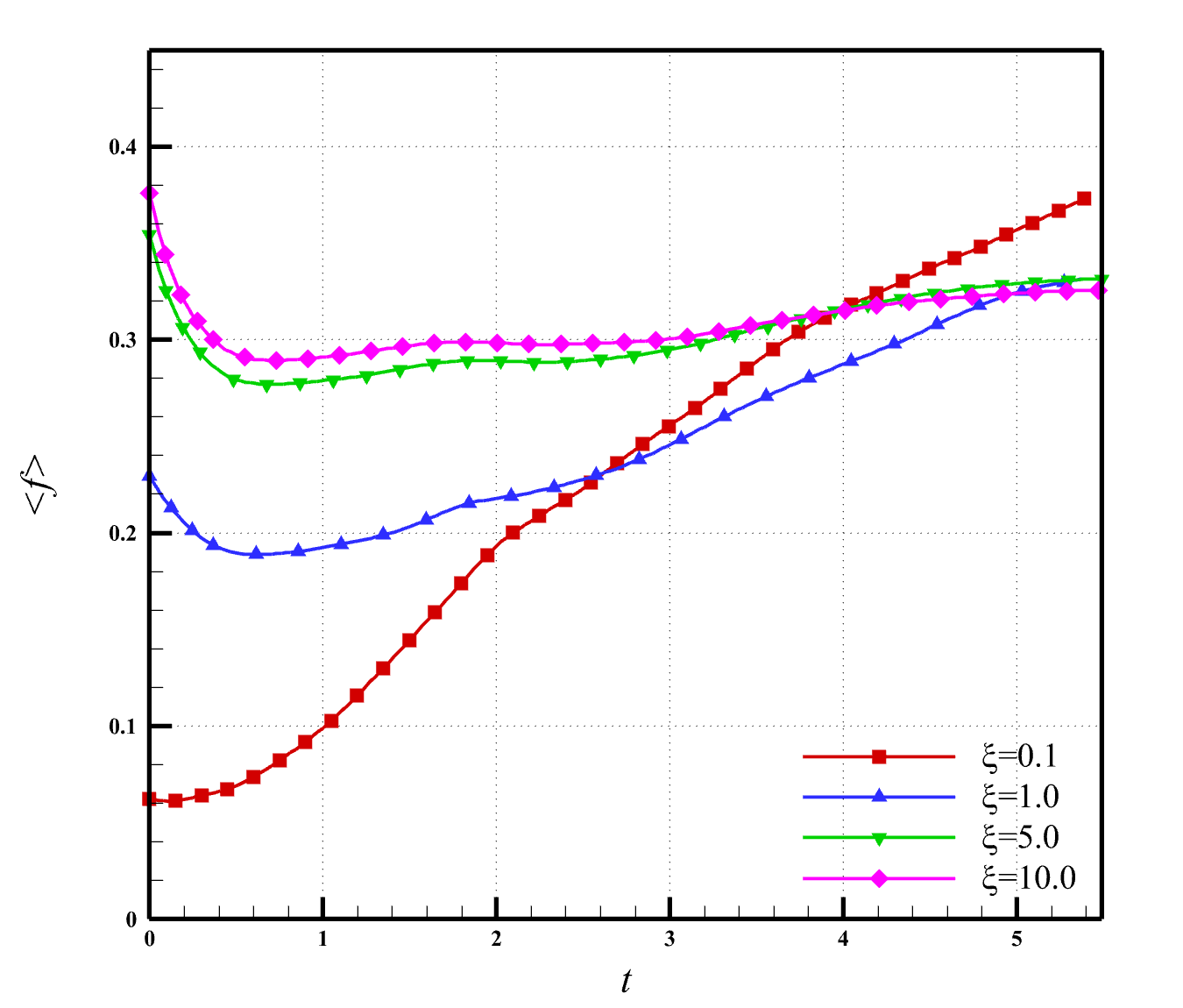}}
  \caption{Temporal evolution of the the spatial integration of mixedness $\left\langle f \right \rangle$ .}
  \label{mixedness evolution}
\end{figure}

Fig.\;\ref{mixedness evolution} displays the temporal evolution of $\left\langle f\right\rangle$. The diffusive layer ratio $\xi$ reflects the initial diffusion of the bubble, hence the initial global mixedness $\left\langle f\right\rangle$ at $t = 0$ is larger in cases with larger $\xi$. 
Subsequently, there is a decrease of $\left\langle f\right\rangle$ in the cases with larger $\xi$ during the early stage, and then an increase throughout the mixing process. Additionally, it can be observed that in cases with a smaller diffusive layer ratio $\xi$, representing stronger SBHI, the increase of $\left\langle f\right\rangle$ is more pronounced, thus confirming the quantitative conclusion obtained in Fig.\;\ref{MixingField}.

\begin{figure*}[htbp]
  \centering
  \includegraphics[clip=true,width=.72\textwidth]{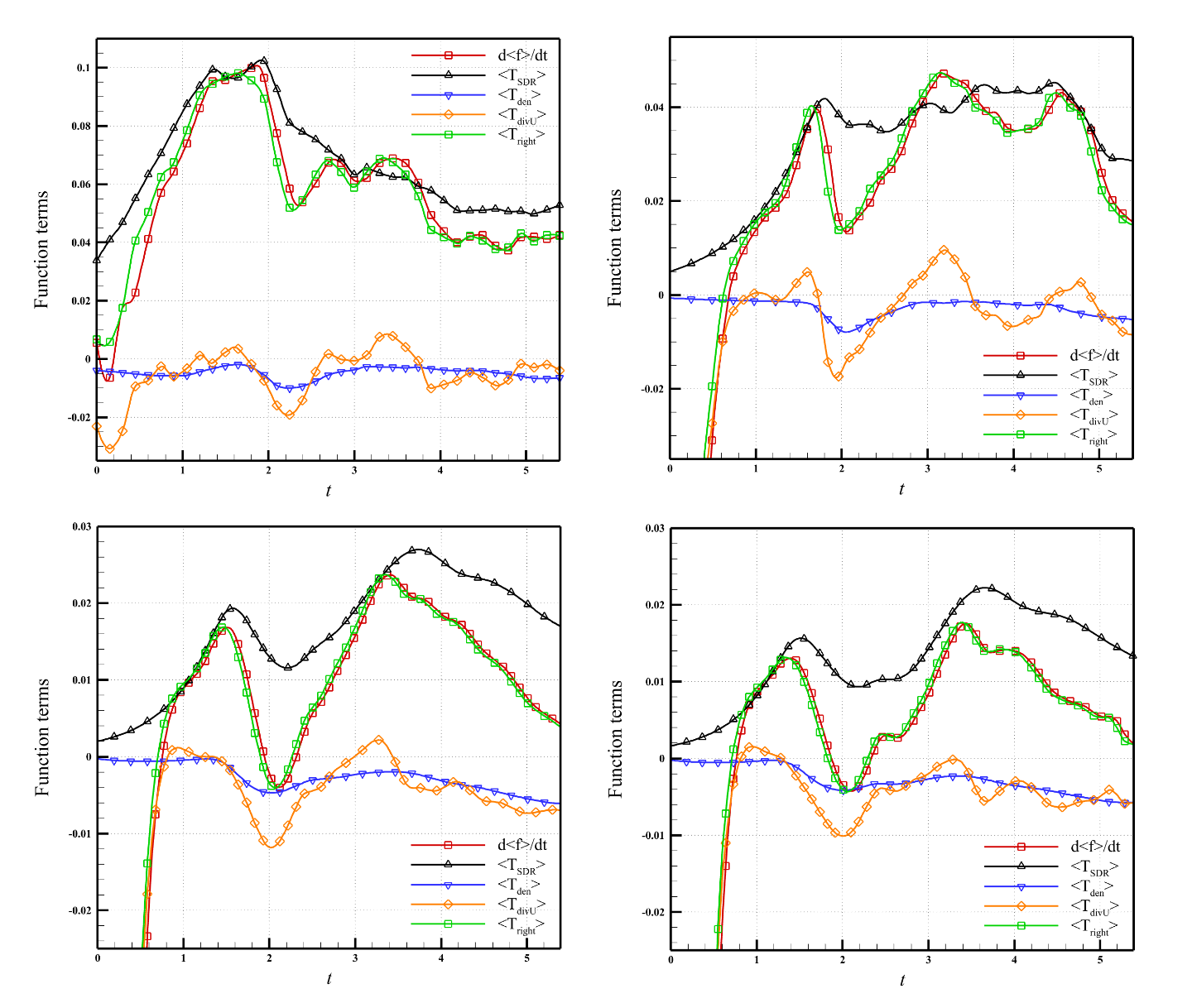}\\
  \caption{Time derivative of the spatial integration of mixedness $\left\langle f \right \rangle$ defined in Eq.\;\ref{mixedness} and the corresponding source terms in Eq.\;\ref{mixedness equation 4}.}
  \label{mixedness budget 1}
\end{figure*}

The budget analysis of Eq.\;\ref{mixedness equation 4} in Fig.\;\ref{mixedness budget 1} presents the sources contributing to the growth of global mixedness $\left\langle f \right \rangle$. The time derivative of global mixedness $\frac{D\left\langle f \right \rangle}{Dt}$, denoted by the red line, closely aligns with the green line representing $\left\langle T_{right} \right \rangle$, the sum of the terms on the right-hand side of Eq.\;\ref{mixedness equation 4}. This observation validates the trustworthiness of the numerical results in terms of the resolution of mixedness dynamics within the grid, as detailed in Appendix \ref{grid resolution}. Notably, the primary source of mixedness growth is the SDR term $\left\langle T_{SDR} \right \rangle$, depicted by the black line. This phenomenon explains the physical meaning of SDR $\chi$ as a measure of mixing rate. In contrast, the density term $\left\langle T_{den} \right \rangle$ illustrated by blue line approaches 0, while the velocity divergence term $\left\langle T_{divU} \right \rangle$ shown by orange line is predominantly negative throughout the mixing process, particularly during the early stage, contributing to the decrease in $\left\langle f \right \rangle$ in cases with large $\xi$. 

This budget analysis prompts a modification in the definition of global mixedness:

\begin{equation}
    \left\langle f^{*} \right \rangle = \left\langle f \right \rangle - \int_{0}^{t} \left\langle f\left( \frac{\partial u_k}{\partial x_k} \right) \right \rangle dt',
    \label{modified mixedness def}
\end{equation}
The transport equation of this parameter is:
\begin{equation}
    \begin{aligned}
        \frac{D\left \langle f^{*} \right \rangle}{Dt} =  \left \langle T_{SDR} \right \rangle + \left \langle T_{den} \right \rangle
    \end{aligned}
    \label{mixedness equation 5}
\end{equation}

\begin{figure}[H]
  \centering
    {\includegraphics[clip=true,width=.45\textwidth]{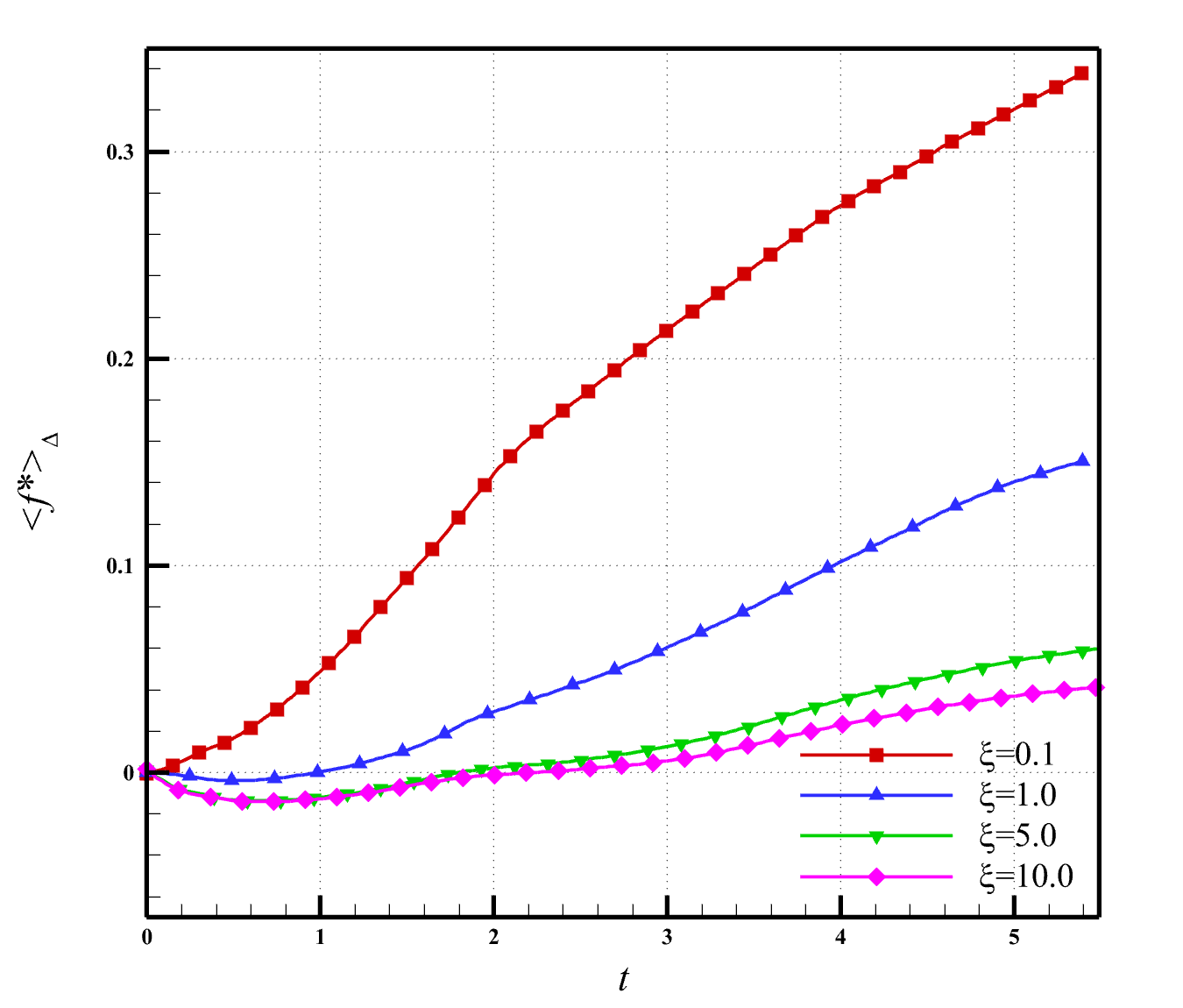}}
  \caption{Temporal evolution of the increment of spatial integration of modified mixedness $\left\langle f^{*} \right \rangle_{\Delta}$ .}
  \label{modified mixedness increment}
\end{figure}

\begin{figure*}[htbp]
  \centering
  \includegraphics[clip=true,width=.72\textwidth]{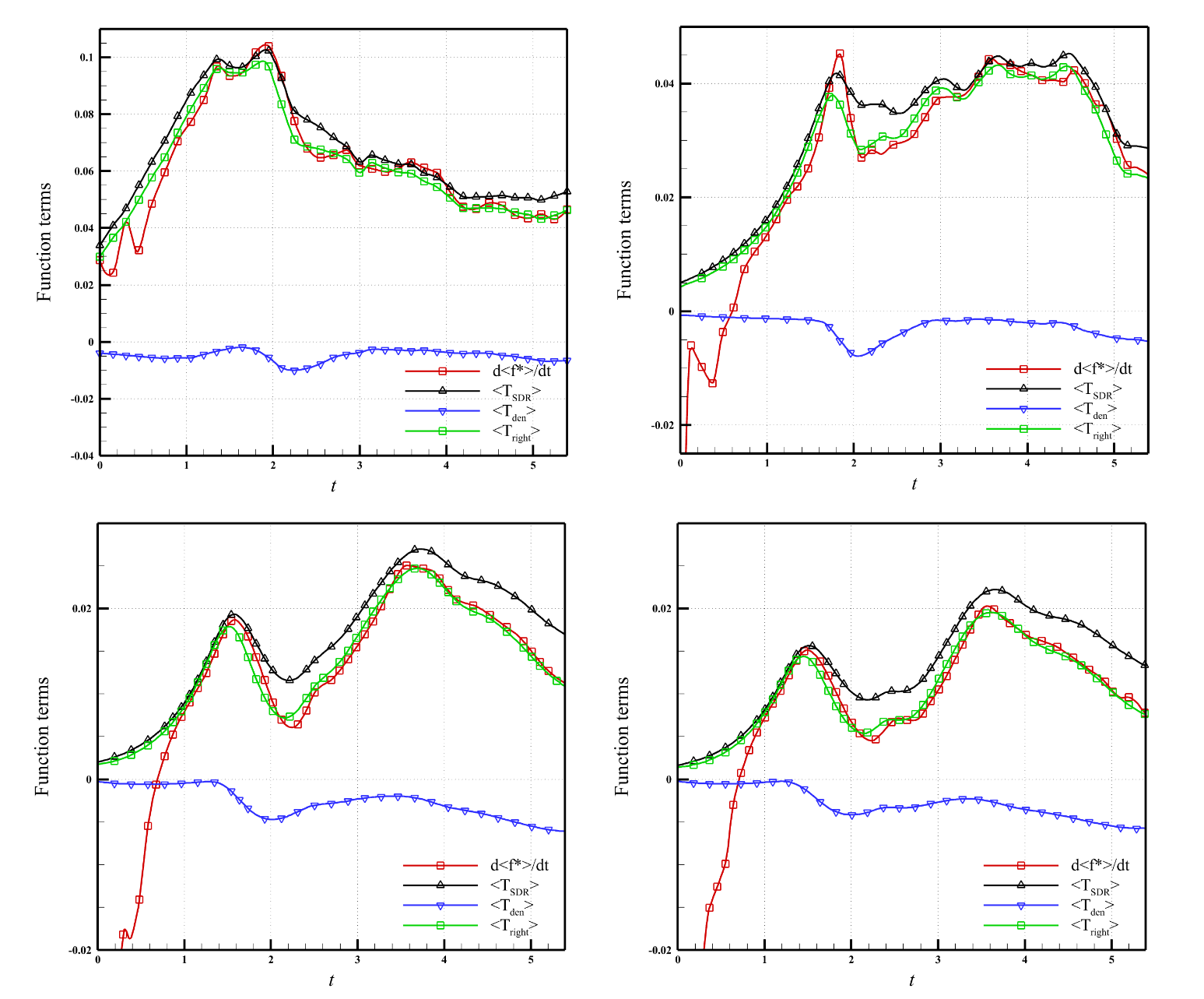}\\
  \caption{Time derivative of the spatial integration of modified mixedness $ \left\langle f^{*} \right \rangle$ defined in Eq.\;\ref{modified mixedness def} and the corresponding source terms in in Eq.\;\ref{mixedness equation 5}.}
  \label{mixedness budget 2}
\end{figure*}

\begin{figure}[H]
  \centering
  \includegraphics[clip=true,width=.45\textwidth]{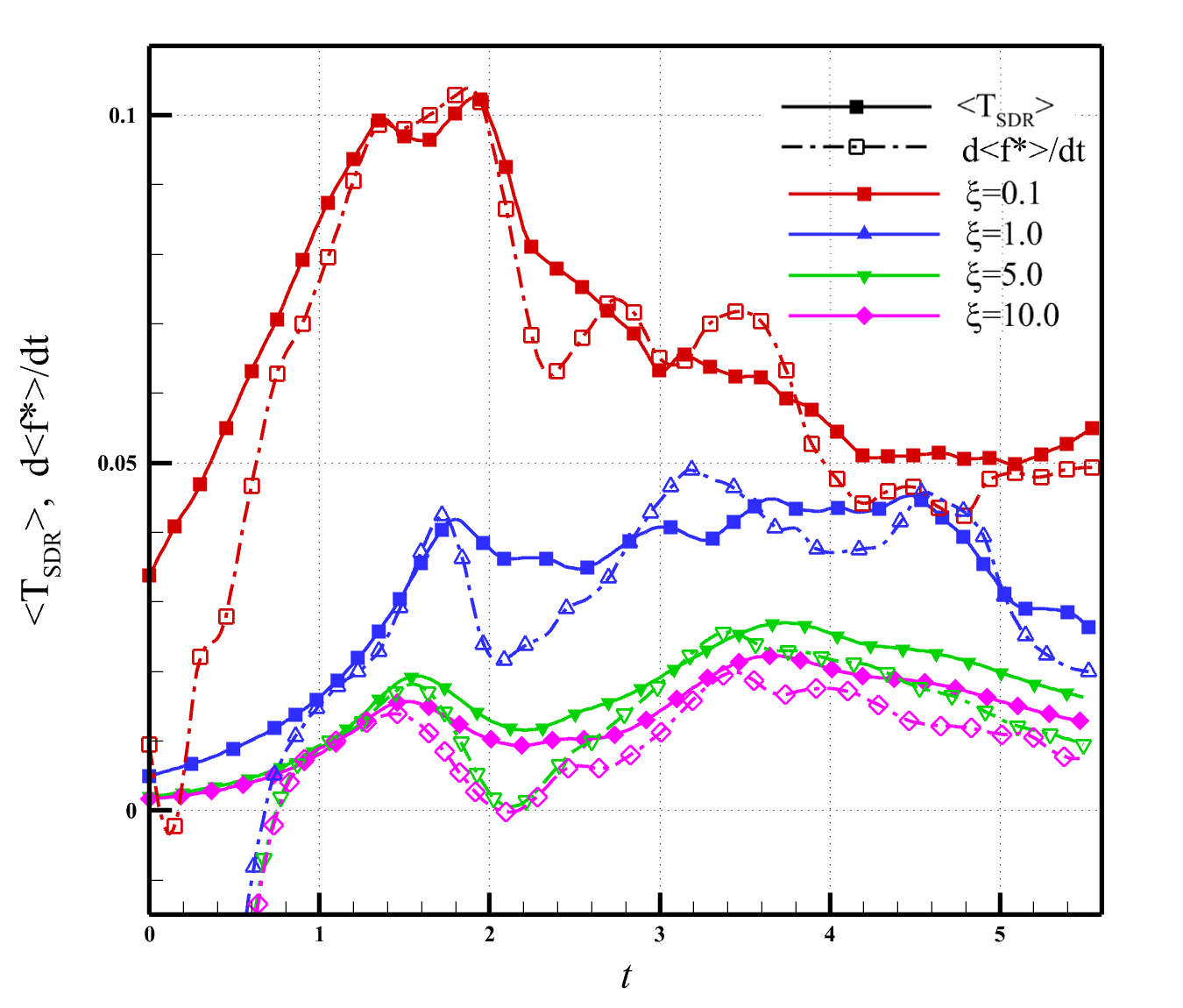}\\
  \caption{Comparison of the time evolution of the SDR term $\left\langle T_{SDR}\right\rangle$ and the time derivative of the modified mixedness ${d \left\langle f^*\right\rangle}/{dt}$.}
  \label{SDR term}
\end{figure}

The temporal evolution of the increment of the modified global mixedness, $\left\langle f^{*} \right \rangle_{\Delta} = \left\langle f^{*} \right \rangle - \left\langle f^{*} \right \rangle|_{t=0}$, is shown in Fig.\;\ref{modified mixedness increment}. Furthermore, the time derivative of the modified global mixedness $\left\langle f^{*} \right \rangle$ and its corresponding source term are illustrated in Fig.\;\ref{mixedness budget 2}. Additionally, the SDR term $\left\langle T_{SDR} \right \rangle$ and time derivative of modified mixedness are presented in Fig.\;\ref{SDR term}. From these figures, it can be deduced that the mixing rate, as indicated by the SDR term $\left\langle T_{SDR} \right \rangle$, is enhanced by SBHI, consequently leading to a larger increment of the modified global mixedness $\left\langle f^{*} \right \rangle_{\Delta}$.

\subsubsection{Scaling behavior of mixing rate on SBV number $\Omega_{sbv}$}

The observations in the aforementioned research indicate theSBHI, quantified by $\Gamma_{sbv}^+$, enhances the mixing rate as described by the SDR term $\left\langle T_{SDR}\right\rangle$ in Eq.\;\ref{mixedness equation 5}. This section introduces a new dimensionless number $\Omega_{sbv}$ to capture the strength of $\Gamma_{sbv}^+$, essentially measuring the intensity of SBHI. Additionally, the scaling behavior of the mixing rate in relation to the SBV number $\Omega_{sbv}$ is demonstrated.

As discussed in Section \ref{vorticity dynamics section}, the difference in $\Gamma_{sbv}^+$ originates from the variations of $\left \langle B^+ \right \rangle$ in Eq.\;\ref{derivative of positive circulation 2}. Hence, dimensional analysis is employed here to propose the SBV number $\Omega_{sbv}$.

The detailed form of Eq.\;\ref{derivative of positive circulation 2} is:
\begin{equation}
    \begin{aligned}
        \frac{D{\Gamma_n^+}}{Dt} & = \frac{D\left(\Gamma^+/\Gamma\right)}{Dt} = \left\langle B^+ \right\rangle + \left\langle {T_v}^+ \right\rangle \\
        & = \frac{RR_{eff}}{\Gamma^2}\left( \left\langle \epsilon_{ijk}\frac{1}{\tilde{\rho}^2}\frac{\partial \tilde{\rho}}{\partial \widetilde{x_j}}\frac{\partial \tilde{p}}{\partial \widetilde{x_k}} \right\rangle \Big|_{\tilde{\omega} > 0} + \left\langle \nu \frac{\partial^2 \tilde{\omega}}{\partial \widetilde{x_k}^2} \right\rangle \Big|_{\tilde{\omega} > 0} \right).
    \end{aligned}
    \label{derivative of positive circulation 3}
\end{equation}
Using the scaling method in Eq.\;\ref{normalized parameters}, the viscous term in this equation can be normalized:
\begin{equation}
    \begin{aligned}
        \left\langle {T_v}^+ \right\rangle &= \frac{RR_{eff}}{\Gamma^2} \iint_{\tilde{\omega} > 0} \nu \frac{\partial^2 \tilde{\omega_i}}{\partial \widetilde{x_k}^2} d\tilde{V} \\
        & = \frac{RR_{eff}}{\Gamma^2} \iint_{\omega > 0} \nu \frac{\partial^2 \omega \omega^*}{\partial {x_k}^2 {{x^*}^2}} {x^*}^2dV\\
        & = \iint_{\omega > 0} \frac{\nu}{\Gamma} \frac{\partial^2 \omega }{\partial {x_k}^2 }dV = \left\langle\frac{1}{Re}\frac{\partial^2 \omega}{\partial x_k^2}\right\rangle \Big|_{\omega>0}.
    \end{aligned}
    \label{viscous term}
\end{equation}
This dimensionless form indicates the Reynolds number $Re$ measures the viscous effect on the evolution of dimensionless positive circulation $\Gamma_n^+$.

Because the density $\tilde{\rho}$ and pressure $\tilde{p}$ distributions vary across the cases with different diffusive layer ratio $\xi$, the density gradient $\frac{\partial\tilde{\rho}}{\partial \widetilde{x_i}}$ and pressure gradient $\frac{\partial\tilde{p}}{\partial \widetilde{x_i}}$ need to be modeled to propose $\Omega_{sbv}$.

Regarding the density gradient, employing the canonical correlation between mass fraction and density in multispecies miscible flows \cite{tomkins2013evolution}:
\begin{equation}
\frac{1}{\tilde{\rho}} = \frac{1-Y}{\rho_1^{'}} + \frac{Y}{\rho_2^{'}},
\end{equation}
the relationship between the density gradient and the scalar gradient can be established as:
\begin{equation}
    \frac{\partial\tilde{\rho}}{\partial \widetilde{x_i}} = {\tilde{\rho}}^2\frac{\rho_2^{'} - \rho_1^{'}}{\rho_1^{'}\rho_2^{'}}\frac{\partial Y}{\partial \widetilde{x_i}}.
\end{equation}

Substituting this expression of density gradient into Eq.\;\ref{derivative of positive circulation 3}, the dimensionless baroclinic torque transforms to:
\begin{equation}
    \begin{aligned}
         \left\langle {B}^+ \right\rangle & = \frac{RR_{eff}}{\Gamma^2}\left\langle \epsilon_{ijk}\frac{1}{\tilde{\rho}^2}\frac{\partial \tilde{\rho}}{\partial \widetilde{x_j}}\frac{\partial \tilde{p}}{\partial \widetilde{x_k}} \right\rangle \Big|_{\tilde{\omega} > 0} \\
         & =  \frac{RR_{eff}}{\Gamma^2} \iint_{\tilde{\omega} > 0} \epsilon_{ijk}\frac{1}{\tilde{\rho}^2}\frac{\partial \tilde{\rho}}{\partial \widetilde{x_j}}\frac{\partial \tilde{p}}{\partial \widetilde{x_k}} d\tilde{V} \\
         & = \frac{RR_{eff}}{\Gamma^2} \iint_{\tilde{\omega} > 0} \epsilon_{ijk}\frac{\rho_2^{'} - \rho_1^{'}}{\rho_1^{'}\rho_2^{'}}\frac{\partial Y}{\partial \widetilde{x_j}}\frac{\partial \tilde{p}}{\partial \widetilde{x_k}} d\tilde{V}.
    \end{aligned}
    \label{baroclinic torque 1}
\end{equation}

The scalar gradient $\frac{\partial Y}{\partial \widetilde{x_j}}$ is modeled first. Similar to the definition of flame width between the mass fraction from $Y_a$ to $Y_b$ in the research of premixed-flame by Hamlinton et al. \cite{hamlington2011interactions}:
\begin{equation}
    \delta(Y_a,Y_b) = \int_{Y_a}^{Y_b} \left(\frac{\partial Y}{\partial n}\right)^{-1}\Big|_Y dY = \int_{Y_a}^{Y_b} {\tilde{\chi}}^{-1/2}(Y) dY,
\end{equation}
the definition of average width of bubble's mass fraction is:
\begin{equation}
    \overline{\delta} = \delta(0,1) = \int_{0}^{1} {\tilde{\chi}}^{-1/2}(Y) dY = \iint  {\tilde{\chi}}^{-1/2} d\tilde{V},
\end{equation}
by which the scalar gradient is scaled as:
\begin{equation}
    \begin{aligned}
        \frac{\partial Y}{\partial \widetilde{x_i}} &= \frac{1}{\overline{\delta}} \frac{\partial Y}{\partial x_i}.
    \end{aligned}
\end{equation}

\begin{figure}[H]
  \centering
  \includegraphics[clip=true,width=.45\textwidth]{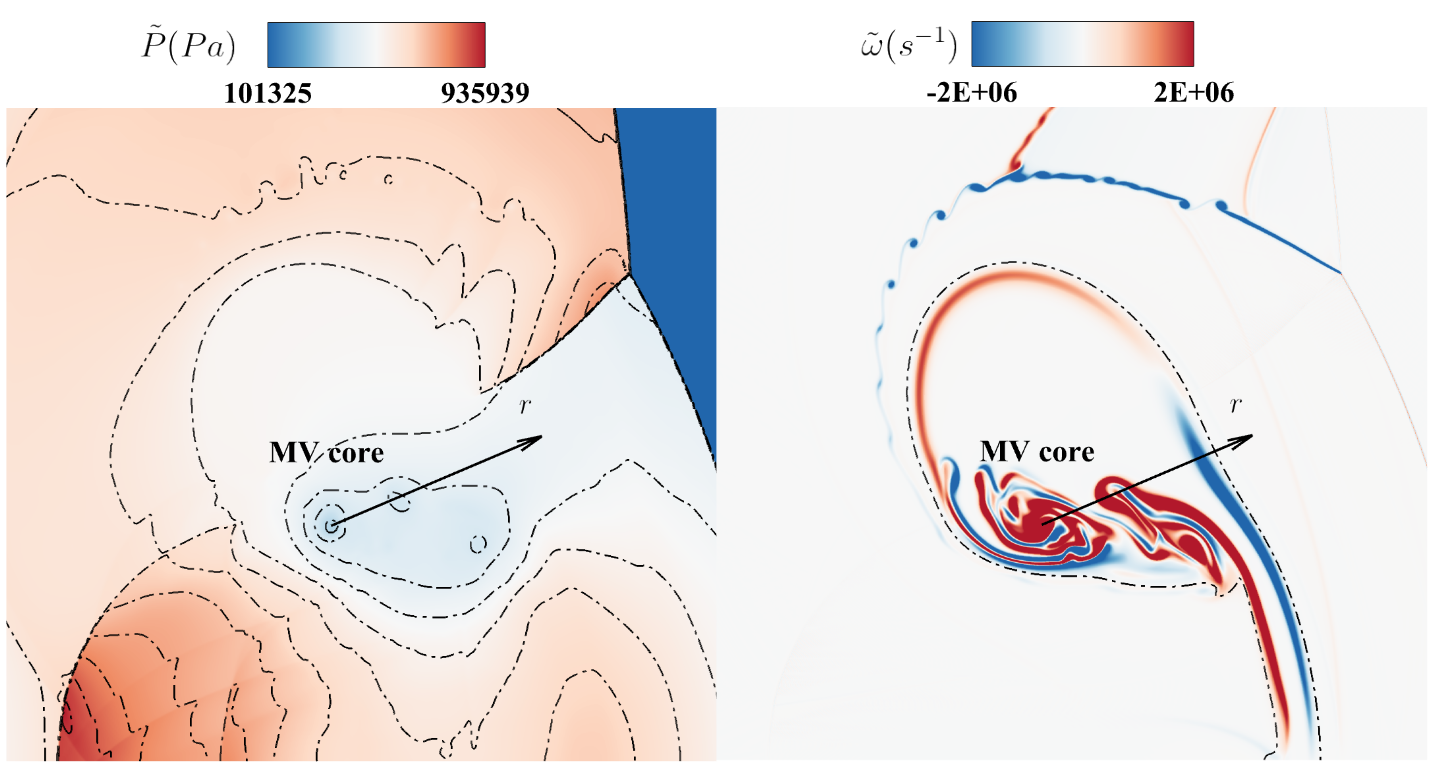}\\
  \caption{(left) The pressure contour and (right) vorticity contour in the frame of main vortex core of the SBI case with diffusive interface ratio $\xi = 0.1$ at time $t=1.3$.}
  \label{pressure model}
\end{figure}

The normalization of the pressure gradient is based on the assumption of a single vortex. Illustrated in Fig.\;\ref{pressure model}, the pressure contour suggests that the pressure field of SBI resembles that of a single vortex. By establishing a polar system with the main vortex (MV) core as the coordinate origin, the pressure gradient can be estimated using the Lamb-Oseen vortex model \cite{liu2020mixing,peng2021mechanism}:
\begin{equation}
    \frac{\partial \tilde{p}}{\partial \tilde{r}} = \frac{\tilde{\rho}\Gamma^2}{4\pi^2 {\tilde{r}}^3},
\end{equation}
therefore, the pressure gradient $\frac{\partial \tilde{p}}{\partial \widetilde{x_k}}$ is scales as:
\begin{equation}
    \frac{\partial \tilde{p}}{\partial \widetilde{x_i}} =  \frac{\rho^{*}\Gamma^2}{4\pi^2 {(R_{eff}R)}^{\frac{3}{2}}}\frac{\partial p}{\partial x_i} = \frac{\rho_1^{'} + \rho_2^{'}}{2}\frac{\Gamma^2}{4\pi^2 {(R_{eff}R)}^{\frac{3}{2}}}\frac{\partial p}{\partial x_i}.
\end{equation}

As the baroclinic torque $B^+$ is deposited on the bubble's area with positive vorticity, the characteristic length in Eq.\;\ref{characteristic time and length} is not appropriate to normalize $d\tilde{V}$ in Eq.\;\ref{baroclinic torque 1}, the area $A = \iint_{\tilde{\omega}>0}d\tilde{V}$ is defined to normalize $d\tilde{V}$, represented as: $d\tilde{V} = A dV$. With these normalized parameters, Eq.\;\ref{baroclinic torque 1} adopts the following form:
\begin{equation}
    \begin{aligned}
         &\left\langle B^+ \right\rangle 
          = \frac{RR_{eff}}{\Gamma^2} \iint_{\tilde{\omega} > 0} \epsilon_{ijk}\frac{\rho_2^{'} - \rho_1^{'}}{\rho_1^{'}\rho_2^{'}}\frac{\partial Y}{\partial \widetilde{x_j}}\frac{\partial \tilde{p}}{\partial \widetilde{x_k}} d\tilde{V} \\
         & = \frac{RR_{eff}}{\Gamma^2} \iint_{\tilde{\omega} > 0} \epsilon_{ijk}\frac{\rho_2^{'} - \rho_1^{'}}{\rho_1^{'}\rho_2^{'}}\frac{\partial Y}{\partial (\overline{\delta}x_j)}\frac{\frac{\rho_1^{'} + \rho_2^{'}}{2}\Gamma^2}{4\pi^2 {(R_{eff}R)}^{\frac{3}{2}}}\frac{\partial p}{\partial x_k}AdV \\
         & = \left(\frac{\rho_2^{'} - \rho_1^{'}}{\rho_1^{'}\rho_2^{'}}\frac{\rho_1^{'} + \rho_2^{'}}{2}\right)\frac{A}{4\pi^2\overline{\delta}\sqrt{RR_{eff}}}\iint_{\omega>0}\epsilon_{ijk}\frac{\partial Y}{\partial x_j}\frac{\partial p}{\partial x_k}dV \\
         & = \frac{(\sigma - 1)(\sigma + 1)}{2\sigma}\frac{A}{4\pi^2\overline{\delta}\sqrt{RR_{eff}}}\left\langle\frac{\partial Y}{\partial x_j}\frac{\partial p}{\partial x_k}\right\rangle \Big|_{\omega > 0}.
    \end{aligned}
    \label{dimensionless baroclinic torque}
\end{equation}
In this equation, $\sigma = {\rho_2^{'}}/{\rho_1^{'}} \approx 0.083$ is the density ratio to the post-shock helium and ambient air. 

Substituting the dimensionless baroclinic torque in Eq.\;\ref{dimensionless baroclinic torque} and viscous term in Eq.\;\ref{viscous term}, Eq.\;\ref{derivative of positive circulation 3} transforms to: 
\begin{equation}
    \begin{aligned}
         \frac{D\left(\Gamma^+/\Gamma\right)}{Dt} &= \frac{(\sigma - 1)(\sigma + 1)}{2\sigma}\frac{A}{4\pi^2\overline{\delta}\sqrt{RR_{eff}}}\left\langle\frac{\partial Y}{\partial x_j}\frac{\partial p}{\partial x_k}\right\rangle \Big|_{\omega > 0} \\
        &+  \left\langle\frac{1}{Re}\frac{\partial^2 \omega}{\partial x_k^2}\right\rangle \Big|_{\omega>0}.
    \end{aligned}
    \label{derivative of positive circulation 4}
\end{equation}

Referring to the budget analysis of the growth of positive circulation in Fig.\;\ref{time derivative of positive circulation}, $\Gamma_{sbv}^+$ is primarily influenced by the baroclinic torque $\left\langle B^+ \right\rangle$. Furthermore, as the Reynolds number $Re$ remains constant across the cases, the viscous effect on the variance of $\Gamma_{sbv}^+$ is neglected. Hence, $\frac{\Gamma_{sbv}^+}{\Gamma}$ is estimated as:
\begin{equation}
    \begin{aligned}
        \frac{\Gamma_{sbv}^+}{\Gamma} &\sim \left|\int_{0}^{2} \frac{(\sigma - 1)(\sigma + 1)}{2\sigma}\frac{A}{4\pi^2\overline{\delta}\sqrt{RR_{eff}}}dt\right| \\
        &= \frac{(1 - \sigma)(1 + \sigma)}{2\sigma}\frac{1}{4\pi^2\sqrt{RR_{eff}}}\int_{0}^{2}\frac{A}{\overline{\delta}}dt.
    \end{aligned}
\end{equation}

\begin{figure}[H]
  \centering
  \subfigure[]{\includegraphics[clip=true,width=.45\textwidth]{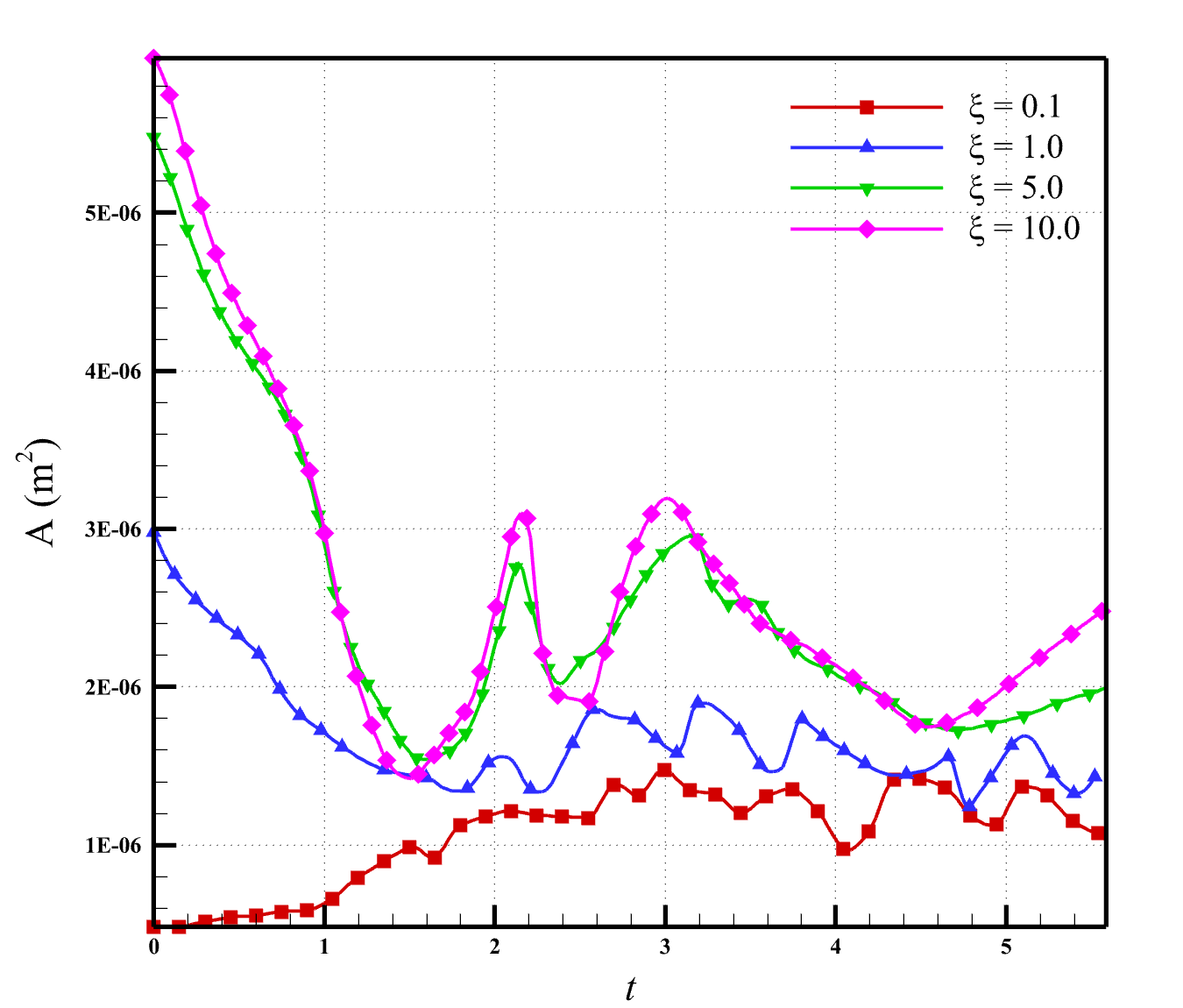}}\\
  \subfigure[]{\includegraphics[clip=true,width=.45\textwidth]{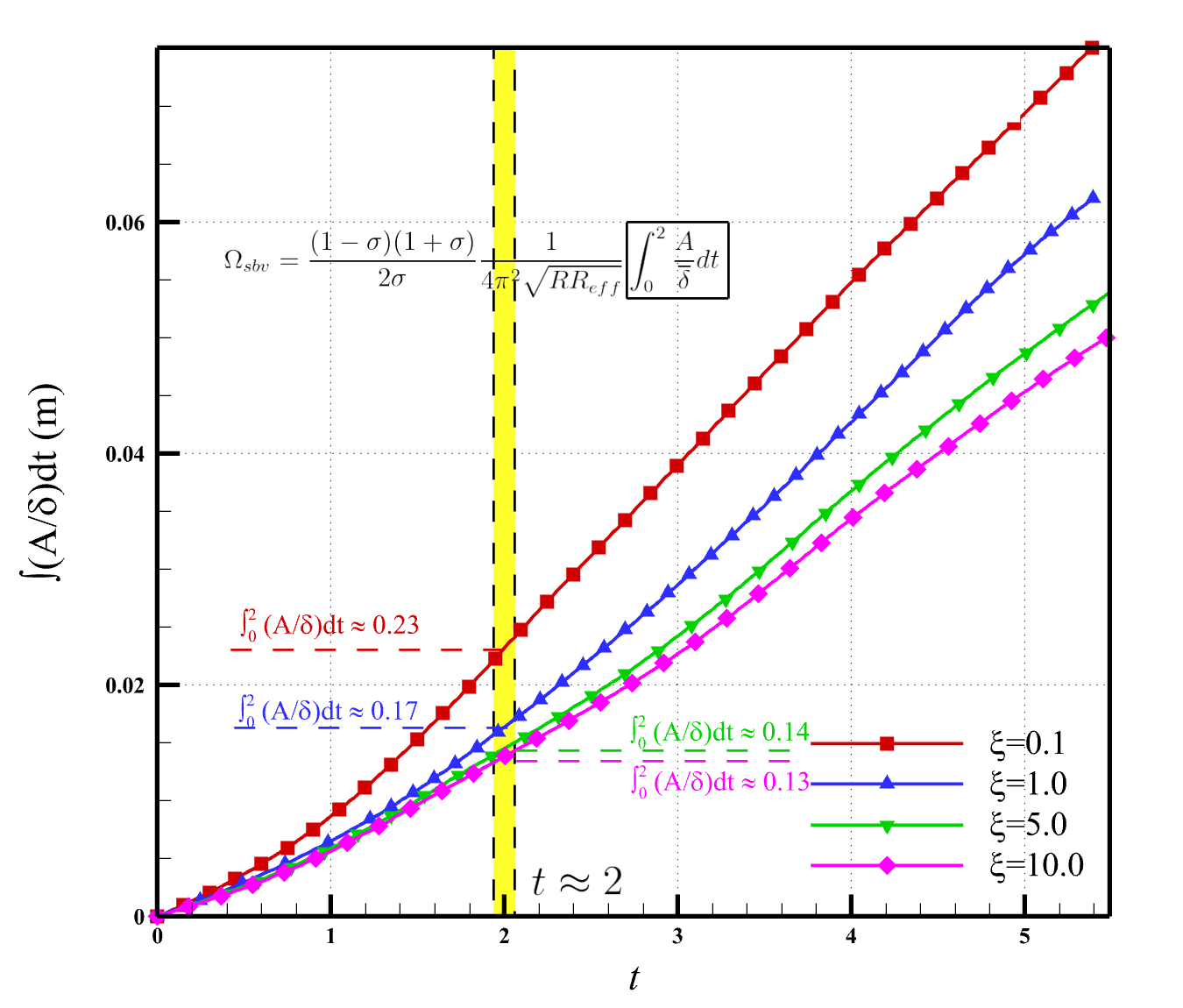}}
  \caption{Comparison of the time evolution of (a) the area $A$ over which the positive vorticity $\omega^+$ is distributed, and (b) the time integrated result of area over the mean width of the bubble's mass fraction $\int ({A}/{\overline{\delta}})dt$.}
  \label{model evolution curve}
\end{figure}

A new dimensionless SBV number $\Omega_{sbv}$ is put forward:
\begin{equation}
    \Omega_{sbv} = \frac{(1 - \sigma)(1 + \sigma)}{2\sigma}\frac{1}{4\pi^2\sqrt{RR_{eff}}}\int_{0}^{2}\frac{A}{\overline{\delta}}dt,
    \label{M_sbv definition}
\end{equation}
where the area $A$ and integration $\int_{0}^{2}\frac{A}{\overline{\delta}}dt$ are determined based on their evolution curve plotted in Fig.\;\ref{model evolution curve}. This definition of $\Omega_{sbv}$ has similarity to the SBV model proposed in our previous research \cite{liu2020mixing}: $\Omega_{sbv}=\frac{|At^+|D}{\pi \delta'}\left(\frac{1}{1-\Delta U t_b D} -1\right)$. However, the previous model lacks the capacity to measure $\frac{\Gamma_{sbv}^+}{\Gamma}$ since it only models the baroclinic torque distributed in the Br structure shown in Fig.\;\ref{vorticity contour}.

\begin{figure}[H]
  \centering
  \subfigure[]{\includegraphics[clip=true,width=.45\textwidth]{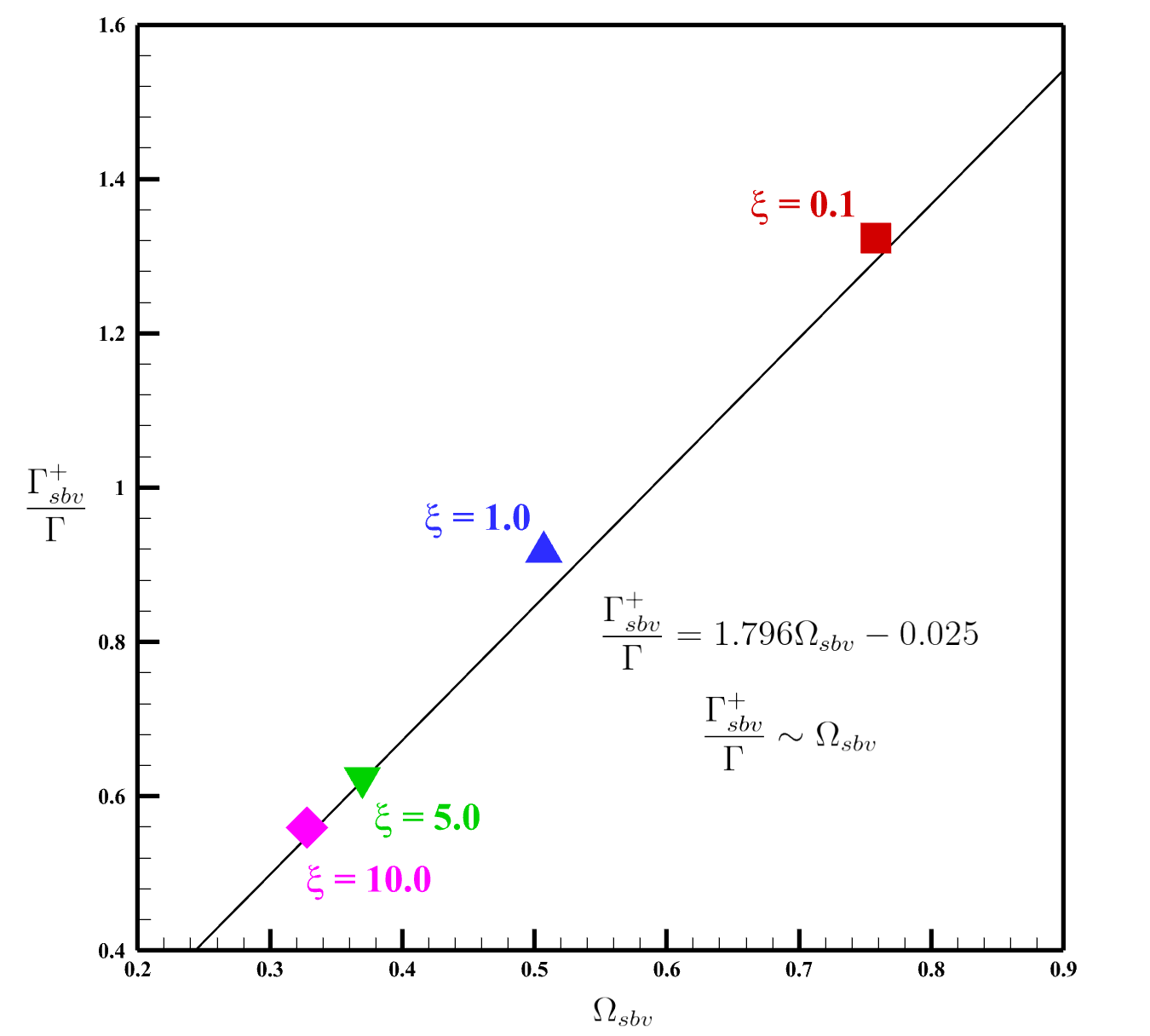}}\\
  \subfigure[]{\includegraphics[clip=true,width=.45\textwidth]{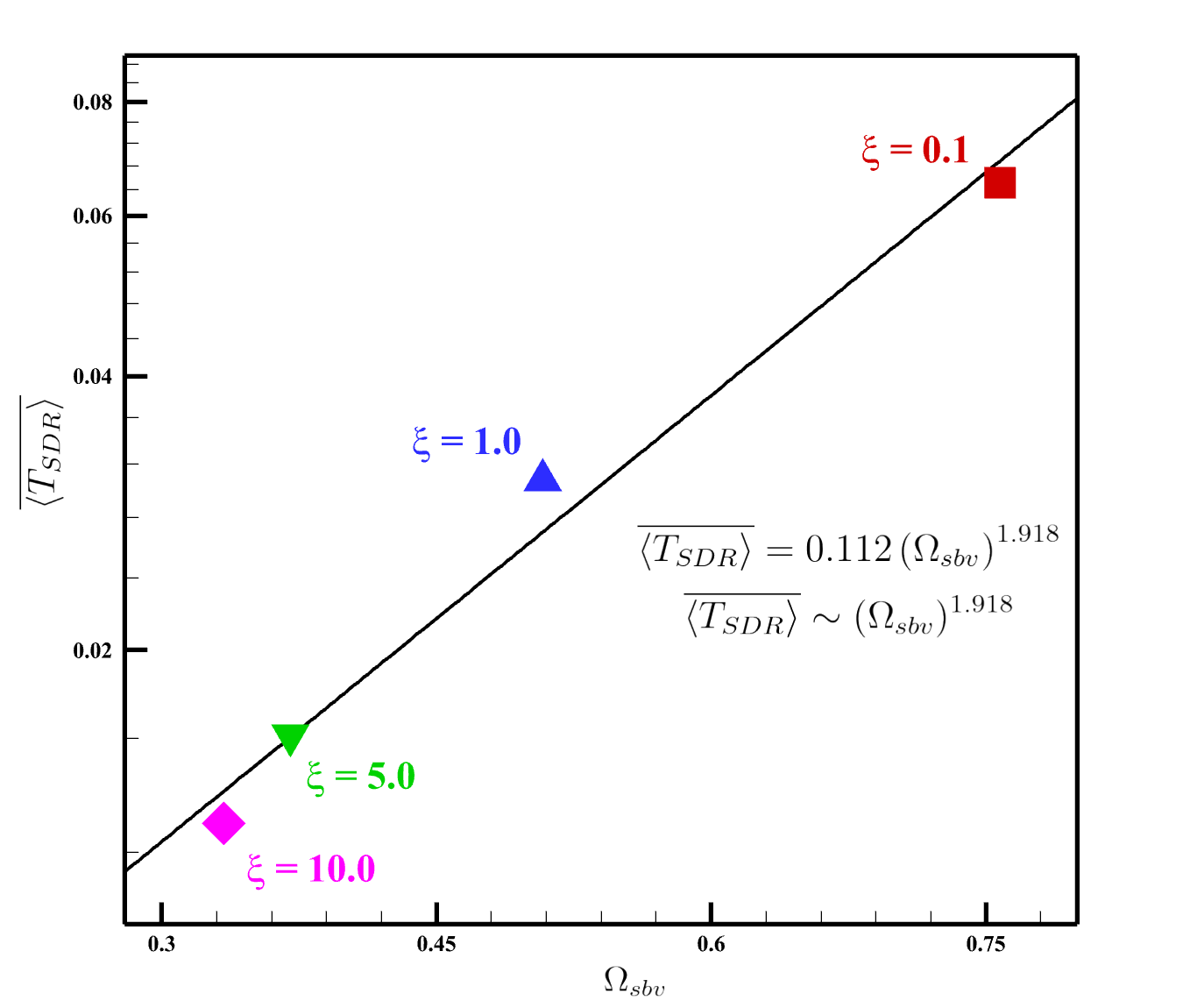}}
  \caption{(a) The dimensionless number $\Omega_{sbv}$ versus the ratio of the positive circulation $\Gamma^+$ and the total circulation $\Gamma$. (b) The scaling of the time averaged SDR term $\overline{\left\langle T_{SDR} \right\rangle}$ on the dimensionless number $\Omega_{sbv}$.}
  \label{M_sbv validation figure and scaling}
\end{figure}

The validation is required to assess the reasonability of the definition of $\Omega_{sbv}$ in Eq.\;\ref{M_sbv definition}. The validation process involved the use of Table \ref{M_sbv validation table} and Fig.\;\ref{M_sbv validation figure and scaling} (a), which present the values of $\Omega_{sbv}$ and the corresponding ${\Gamma_{sbv}^+}/{\Gamma}$ in different cases. It's observed that the ratio between ${\Gamma_{sbv}^+}/{\Gamma}$ and $\Omega_{sbv}$ remains almost constant, indicating a linear correlation between these two variables. Consequently, the SBV number $\Omega_{sbv}$ demonstrates effectiveness in quantifying ${\Gamma_{sbv}^+}/{\Gamma}$, and therefore has the capacity to measure the strength of SBHI.

Once the SBV number $\Omega_{sbv}$ is proposed, a scaling analysis of the mixing rate's dependency on $\Omega_{sbv}$ is presented. As shown in  Eq.\;\ref{mixedness equation 5}, the global mixing rate is represented by the SDR term $\left\langle T_{SDR} \right\rangle$. The time-averaged mixing rate is expressed as:
\begin{equation}
\overline{\left\langle T_{SDR} \right\rangle} = \frac{1}{T}\int_{0}^{T} \left\langle T_{SDR} \right\rangle dt,
\end{equation}
where $T = 5.6$ is the total SBI evolution time. The scaling behavior of $\overline{\left\langle T_{SDR} \right\rangle}$ with respect to the SBV number $\Omega_{sbv}$ is illustrated in Table \ref{M_sbv validation table} and Fig.\;\ref{M_sbv validation figure and scaling} (b). It is observed that the time-averaged mixing rate $\overline{\left\langle T_{SDR} \right\rangle}$ is positively correlated with $\Omega_{sbv}$. This positive correlation demonstrates the enhancement effect of SBHI on the mixing rate, which can be summarized as:

\begin{equation}
    \left\{
    \begin{aligned}
        &\frac{\Gamma_{sbv}}{\Gamma} \sim \Omega_{sbv},\\
        &\overline{\left\langle T_{SDR}\right\rangle} \sim \Omega_{sbv}^{2}.
    \end{aligned}
    \right.
\end{equation}

\begin{table}[H]
\caption{The SBV number $\Omega_{sbv}$, dimensionless positive circulation at $t=2$ $\Gamma_{sbv}^+/\Gamma$, ${\rm ratio_1} {\rm\; is\; defined\; as\;} \left(\Gamma_{sbv}^+/\Gamma\right)/\Omega_{sbv}$, the time-averaged mixing rate $\overline{\left\langle T_{SDR}\right\rangle}$, and ${\rm ratio_2} {\rm\; is\; defined\; as\;} \overline{\left\langle T_{SDR}\right\rangle}/\Omega_{sbv}^{1.918}$.}
 \centering
  \begin{tabular}{c c c c c c}
    \hline
    $\xi$ & $\Omega_{sbv}$ & $\Gamma_{sbv}^+/\Gamma$ & $\rm{ratio}_1$ & $\overline{\left\langle T_{SDR}\right\rangle}$ & $\rm ratio_2$ \\ 
    \hline
    0.1 & 0.758 & 1.32 & 1.741 & 0.0652 & 0.111 \\
    1.0 & 0.507 & 0.92 & 1.815 & 0.0308 & 0.113 \\
    5.0 & 0.370 & 0.62 & 1.676 & 0.0161 & 0.109\\
    10.0 & 0.328 & 0.56 & 1.707 & 0.0129 & 0.109 \\
    \hline
  \end{tabular}
  \label{M_sbv validation table}
\end{table}

\subsection{The underlying mechanisms of the scaling behavior} \label{scaling behavior and mechanism}
\subsubsection{The scaling of mixing rate on Reynolds number $Re$} \label{scaling on Re}
In the previous section, turbulent-like flow resulting from SBHI is observed to have the capacity to enhance mixing. However, previous research \cite{yu2020scaling,yu2022effects,buaria2021turbulence} have suggested that the unstable flow originating from high Reynolds number $Re$ does not significantly contribute to mixing when the P\'eclect number $Pe$ is held constant. In this study, additional simulations are conducted to investigate the scaling behavior of the mixing rate with respect to the Reynolds number $Re$. The geometric parameters outlined in Table \ref{geometric parameters} remains unchanged to ensure that the additional cases have the same SBV number $\Omega_{sbv}$ as the original set. However, the dynamic viscosity $\mu$ is varied in order to obtain different Reynolds numbers $Re = \frac{\Gamma}{\nu} = \frac{\rho^{*}\Gamma}{\mu}$. The parameters for the original and additional cases are set as in Table \ref{parameters in original and new cases}.  

\begin{figure}[H]
  \centering
  \subfigure[]{\includegraphics[clip=true,width=.43\textwidth]{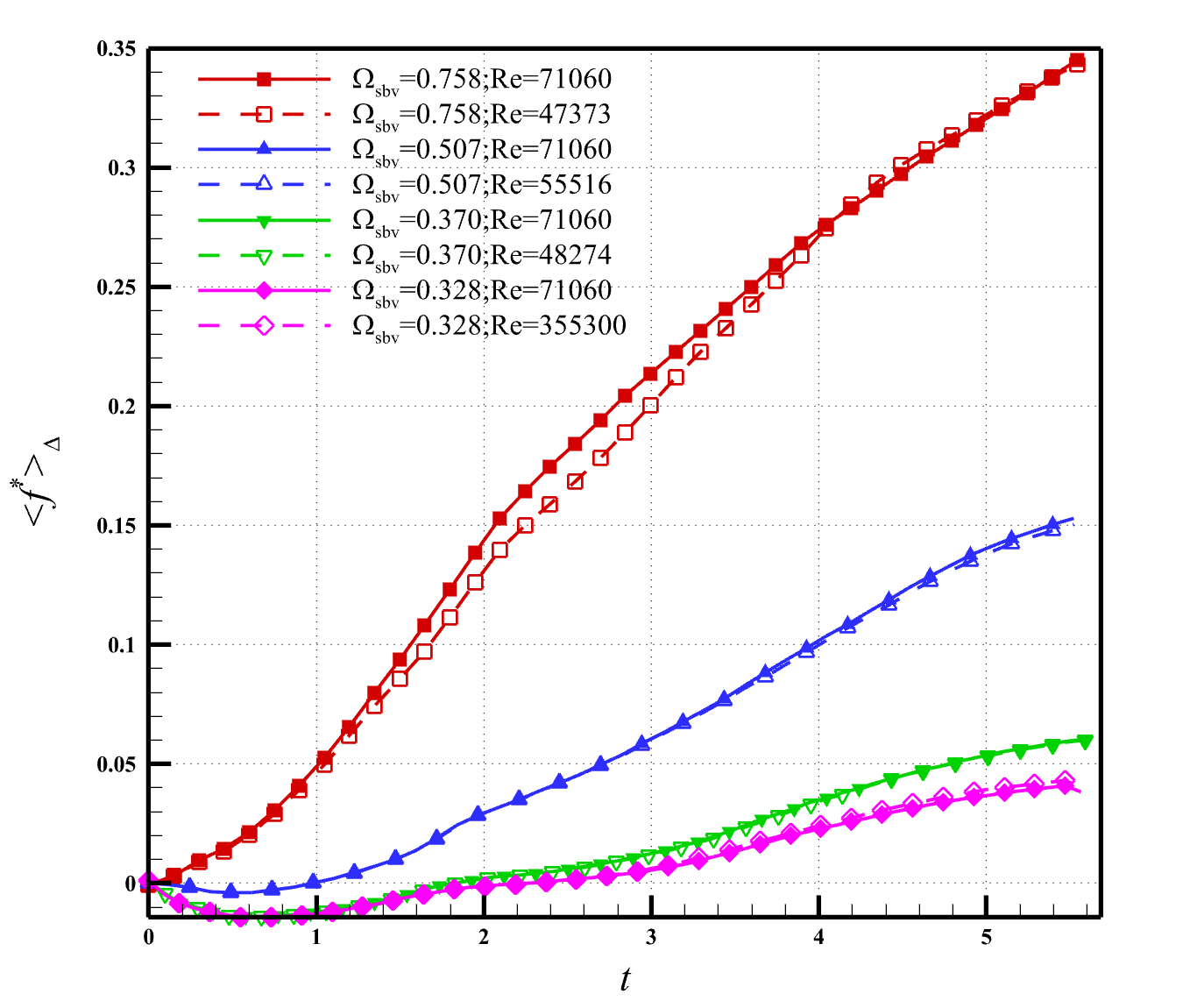}}\\
  \subfigure[]{\includegraphics[clip=true,width=.43\textwidth]{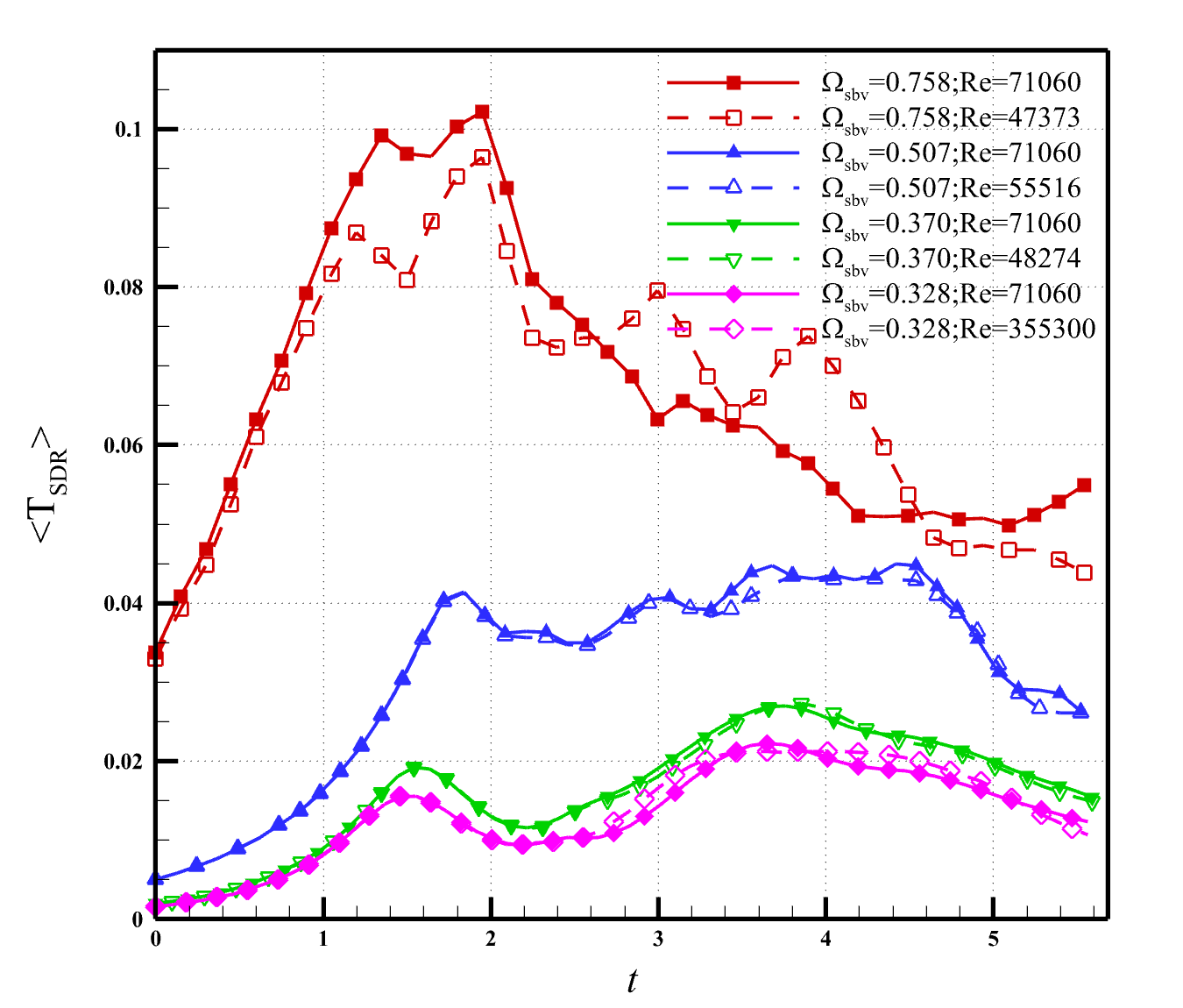}}
  \caption{Comparison of the temporal evolution of (a) the increment of modified mixedness $ \left\langle f^* \right \rangle_{\Delta}$ and (b) SDR term $\left\langle T_{SDR} \right\rangle$ of SBI cases with different $\Omega_{sbv}$ and Reynolds number $Re$.}
  \label{Re mixedness and SDR}
\end{figure}

\begin{table}[H]
\setlength{\tabcolsep}{6pt}
\caption{The parameters for the original and additional cases}
 \centering
  \begin{tabular}{c c c c c}
    \hline
    $\xi$ & $R_{core}\,(\rm{mm})$ & $\mu\,( \rm{Pa\cdot s^{-1}})$ & $\Omega_{sbv}$ & $Re$  \\ 
    \hline
    0.1 & 2.070 & $5.0\times 10^{-5}$ & 0.758 & 71060 \\
    0.1 & 2.070 & $7.5\times 10^{-5}$ & 0.758 & 47373 \\ 
    1.0 & 1.390 & $5.0\times 10^{-5}$ & 0.507 & 71060 \\
    1.0 & 1.390 & $6.4\times 10^{-5}$ & 0.507 & 55516 \\
    5.0 & 0.590 & $5.0\times 10^{-5}$ & 0.370 & 71060 \\
    5.0 & 0.590 & $7.4\times 10^{-5}$ & 0.370 & 48274 \\
    10.0 & 0.340 & $5.0\times 10^{-5}$ & 0.328 & 71060 \\
    10.0 & 0.340 & $1.0\times 10^{-5}$ & 0.328 & 355300 \\
    \hline
  \end{tabular}
  \label{parameters in original and new cases}
\end{table}

\begin{figure*}[htbp]
  \centering
  \includegraphics[clip=true,width=.8\textwidth]{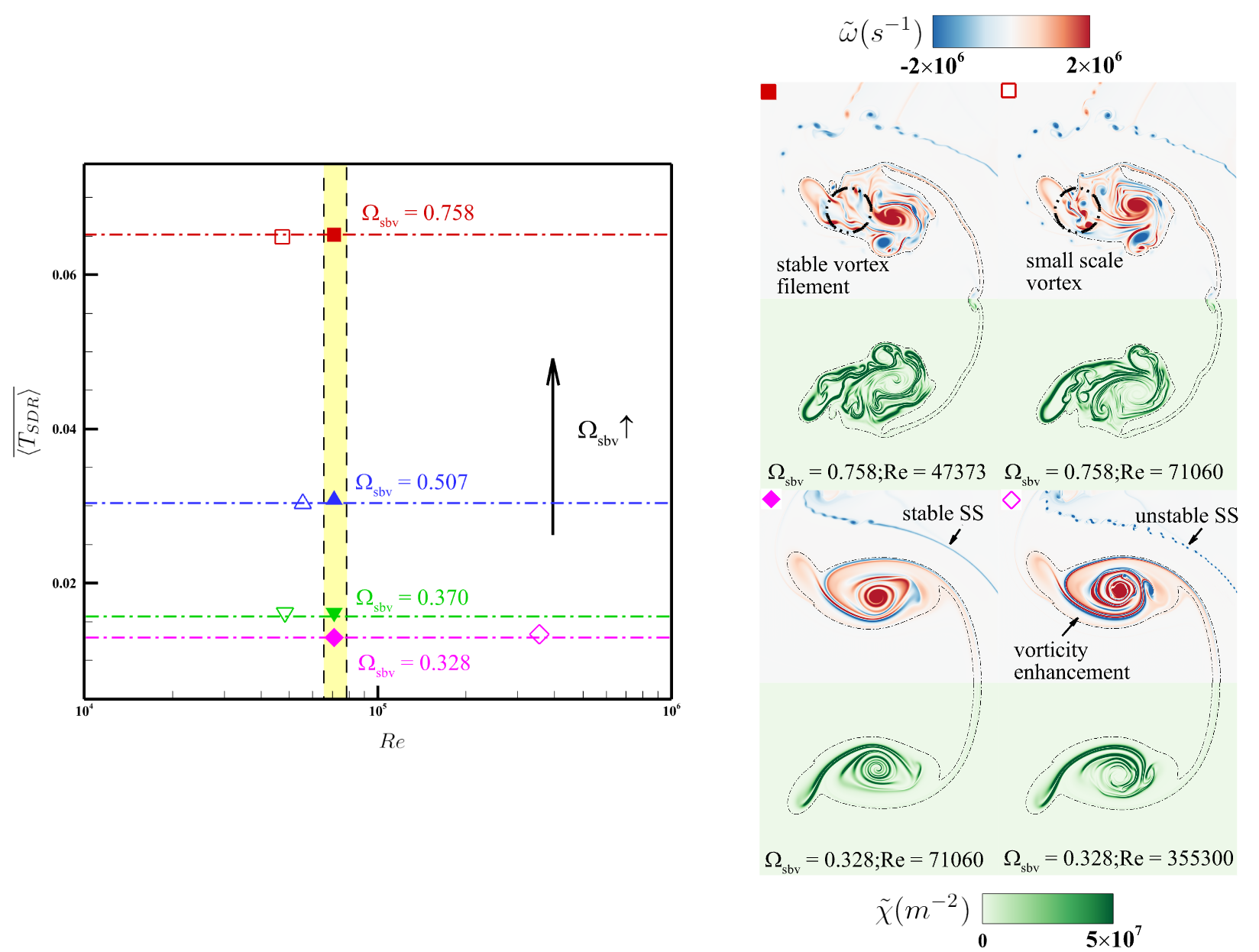}\\
  \caption{The scaling of the time averaged SDR term in the form of $\overline{\left\langle T_{SDR} \right\rangle}$ on $Re$ numbers and $\Omega_{sbv}$ numbers.}
  \label{Re scaling mixing rate}
\end{figure*}

The increment of modified mixedness $\left\langle f^{*}\right\rangle_{\Delta}$ and SDR term $\left\langle T_{SDR} \right\rangle $ are presented in Fig.\;\ref{Re mixedness and SDR}. Based on the results depicted in the figure, it is evident that there is no significant alteration in $\left\langle f^{*}\right\rangle_{\Delta}$ and $\left\langle T_{SDR} \right\rangle $ as a result of variations in Reynolds number $Re$. The time-averaged mixing rate $\overline{\left\langle T_{SDR} \right\rangle}$ is calculated according to the SDR term evolution in Fig.\;\ref{Re mixedness and SDR} (b), and the scaling behavior of $\overline{\left\langle T_{SDR} \right\rangle}$ on Reynolds number $Re$ is summarized in Fig.\;\ref{Re scaling mixing rate}.

The left-hand side of Fig.\;\ref{Re scaling mixing rate} demonstrates the distribution of $\left\langle T_{SDR} \right\rangle$ with respect to the Reynolds number $Re$ and SBV number $\Omega_{sbv}$. Solid symbols correspond to the original cases, while empty symbols correspond to the additional cases. It is apparent that when the SBV number $\Omega_{sbv}$ is held constant, the values of $\left\langle T_{SDR} \right\rangle$ remain nearly unchanged among cases with different $Re$.

The right-hand side of the figure presents the vorticity and SDR contour for cases with $\Omega_{sbv}=0.758$ and 0.328. As the Reynolds number $Re$ increases, the stable vortex filament breaks into small scale vortex in the cases with $\Omega_{sbv}=0.758$, and the stable slipstream (SS) transform into unstable SS, and the vorticity is observed to be enhanced in the cases with $\Omega_{sbv}=0.328$. These observations demonstrate that an increase in $Re$ can lead to an unstable flow field. However, as shown in the left-hand side of the figure, this instability has minimal impact on enhancing mixing, which differs from the effects of SBHI as represented by $\Omega_{sbv}$.

\subsubsection{The distinct mechanisms of two scaling behavior}

As observed in the preceding subsection \ref{scaling on Re}, the scaling behavior of the mixing rate with respect to $\Omega_{sbv}$ and $Re$ exhibits distinct characteristics. In the current subsection, we revisit these scaling behaviors to investigate their underlying mechanisms. The mixing rate is represented as  $\left\langle T_{SDR} \right\rangle = \left\langle 8 Pe \chi \right\rangle$, and therefore, the evolution of SDR $\chi$ is examined here through a budget analysis of the SDR transport equation, and the normalized form is:

\begin{equation}
    \begin{aligned}
        \frac{D\chi}{Dt}& = -2\frac{\partial Y}{\partial x_i}S_{ij}\frac{\partial Y}{\partial x_j} + 2Pe\frac{\partial Y}{\partial x_i}\frac{\partial}{\partial x_i}\left(\frac{1}{\rho}\frac{\partial \rho}{\partial x_j}\frac{\partial Y}{\partial x_j}\right) \\
        &+ 2Pe\frac{\partial Y}{\partial x_i}\frac{\partial^2}{\partial {x_j}^2}\left(\frac{\partial Y}{\partial x_i}\right) = T_{stretch} + T_{den} + T_{diff}.
    \end{aligned}
    \label{SDR equation 1}
\end{equation}
In this equation, $T_{stretch} = -2\frac{\partial Y}{\partial x_i}S_{ij}\frac{\partial Y}{\partial x_j}$ is the stretch term, referring to the increase/decrease of SDR by stretching the scalar gradient $\frac{\partial Y}{\partial x_i}$. $T_{den} = 2Pe\frac{\partial Y}{\partial x_i}\frac{\partial}{\partial x_i}\left(\frac{1}{\rho}\frac{\partial \rho}{\partial x_j}\frac{\partial Y}{\partial x_j}\right)$ is induced by the density gradient $\frac{\partial \rho}{\partial x_i}$, thus is labeled as the density term.
$T_{diff} = 2Pe\frac{\partial Y}{\partial x_i}\frac{\partial^2}{\partial {x_j}^2}\left(\frac{\partial Y}{\partial x_i}\right)$ is diffusion term, which changes SDR by the molecular diffusion.

By applying the spatial integration $\left\langle \chi \right\rangle = \iint \chi dV$, Eq.\;\ref{SDR equation 1} can be reformulated in the integrated form as follows:
\begin{equation}
    \begin{aligned}
        \frac{D\left\langle \chi \right\rangle}{Dt} &=\left\langle\frac{D\chi}{Dt}\right\rangle + \left\langle \chi \left(\frac{\partial u_k}{\partial x_k}\right)\right\rangle \\
        &= \left\langle T_{stretch}\right\rangle + \left\langle T_{den}\right\rangle + \left\langle T_{diff}\right\rangle + \left\langle T_{divU}\right\rangle.
    \end{aligned}
    \label{SDR equation 2}
\end{equation}
In this equation, $\left\langle T_{divU}\right\rangle = \left\langle \chi \left(\frac{\partial u_k}{\partial x_k}\right)\right\rangle$ arises from the velocity divergence and is therefore referred to the velocity divergence term. 

\begin{figure*}[htbp]
  \centering
  \includegraphics[clip=true,width=.8\textwidth]{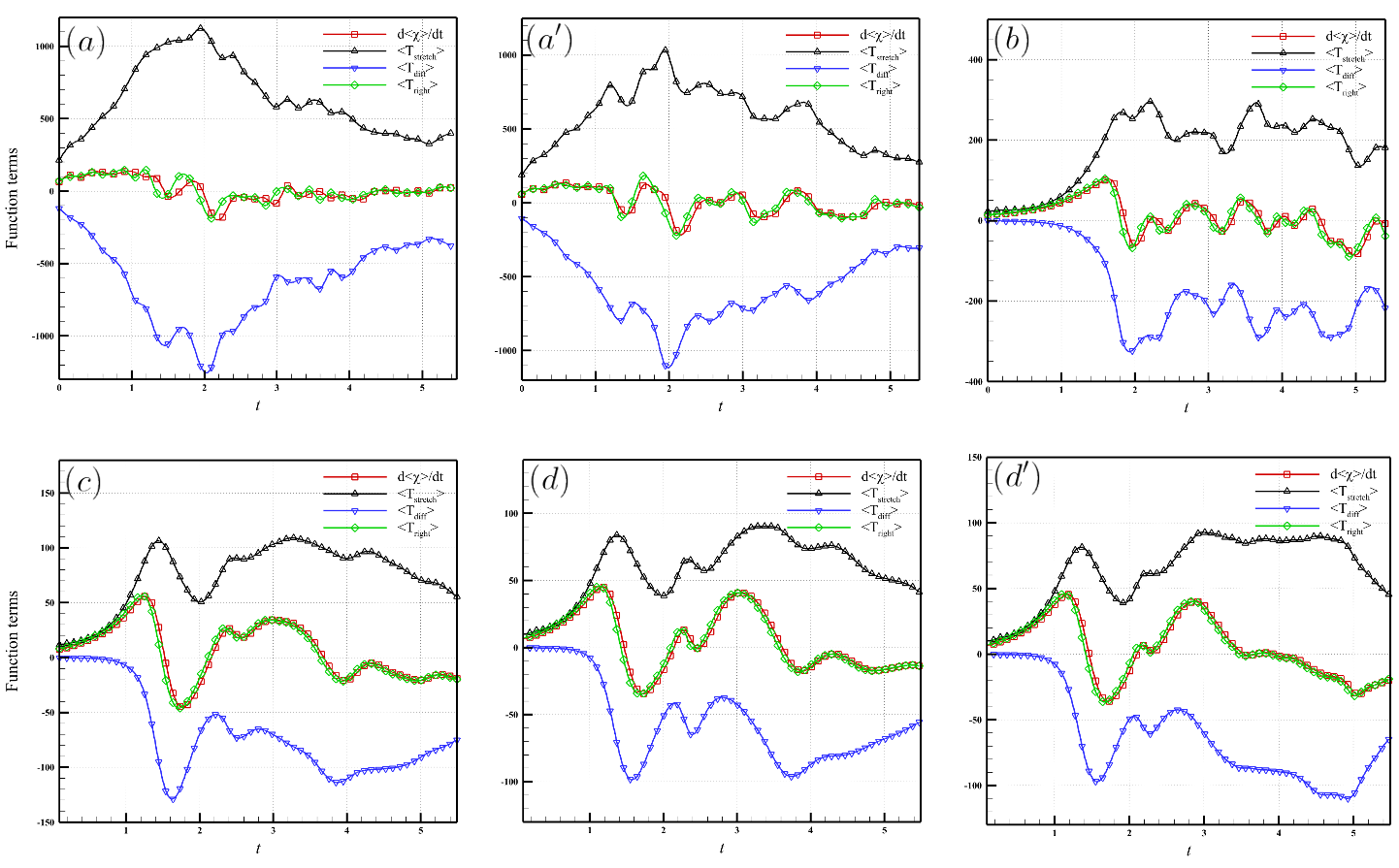}\\
  \caption{Time derivative of scalar dissipation rate $\chi$ and the corresponding source terms in Eq.\;\ref{SDR equation 2} for different SBI cases: (a) $\Omega_{sbv}=0.758$,$Re=71060$, (a') $\Omega_{sbv}=0.758$,$Re=47373$, (b) $\Omega_{sbv}=0.507$,$Re=71060$, (c) $\Omega_{sbv}=0.370$,$Re=71060$, (d) $\Omega_{sbv}=0.328$,$Re=71060$, (d') $\Omega_{sbv}=0.328$,$Re=355300$.}
  \label{time derivative of SDR}
\end{figure*}

Fig.\;\ref{time derivative of SDR} displays the time derivative of SDR and the corresponding source terms in Eq.\;\ref{SDR equation 2}.
The time derivative of SDR $\frac{D\left\langle\chi\right\rangle}{Dt}$, indicated by the red line, closely corresponds with the green line representing $\left\langle T_{right}\right\rangle$, which is the sum of the terms on the right-hand side of Eq.\;\ref{SDR equation 2}. This observation validates the reliability of the numerical results concerning the resolution of SDR dynamics within the grid, as discussed in Appendix \ref{grid resolution}. The primary source of SDR growth is the black line representing the stretch term $\left\langle T_{stretch}\right\rangle$, whereas the diffusion term $\left\langle T_{diff}\right\rangle$ dipicted by blue line primarily diminishes SDR. The density term $\left\langle T_{den} \right\rangle$ and the velocity divergence term $\left\langle T_{divU} \right\rangle$ are nearly negligible compared to the other terms and are therefore not presented in this figure.

Based on the budget analysis of SDR depicted in Fig.\;\ref{time derivative of SDR}, the primary source of SDR growth, $T_{stretch}$, is further investigated to unravel the underlying mechanisms behind the two scaling behaviors mentioned above. Consistent with previous studies \cite{tian2017numerical,tian2019density,gao2020parametric,danish2016influence,buttay2016analysis}, we decompose the stretch term using the eigenvectors of the strain rate tensor as follows:
\begin{equation}
    \begin{aligned}
        & S_{ij} = \Lambda_{i}\boldsymbol{v}_i\boldsymbol{v}_j^{T},\\
        & \Lambda_{ij} = \boldsymbol{diag}\left\{{s_i}\right\} = s_i \delta_{ij}.\\
    \end{aligned}
\end{equation}
Here, $s_i$ represents the $i$th eigenvalue of the strain rate tensor $S_{ij}$, labelled as the principal strain rate. The eigenvalues are ordered as $s_1>s_2>s_3$, with $\boldsymbol{v}_i$ denotating the corresponding eigenvector. Moreover, the scalar gradient $\frac{\partial Y}{\partial x_i}$ is expressed as:
\begin{equation}
     \frac{\partial Y}{\partial x_i}= |\nabla Y|\boldsymbol{e}_i = \sqrt{\chi}e_i.
\end{equation}
In this equation, $\boldsymbol{e}_i$ is the unit vector aligned with the scalar gradient $\frac{\partial Y}{\partial x_i}$. Consequently, the stretch term can further be transformed as follows:
\begin{equation}
    \begin{aligned}
        T_{stretch} &= -2\frac{\partial Y}{\partial x_i}S_{ij}\frac{\partial Y}{\partial x_j} = -2(\sqrt{\chi}\boldsymbol{e_i})^{T} \Lambda_{ij}\boldsymbol{v}_i\boldsymbol{v}_j^{T}\sqrt{\chi}\boldsymbol{e}_j \\
        & = -2\chi \Lambda_{ij} (\boldsymbol{v}_i^{T}\boldsymbol{e}_i)^{T}\boldsymbol{v}_j^{T}\boldsymbol{e}_j = -2\chi s_i\delta_{ij}\lambda_i\lambda_j \\
        & = -2\chi s_i {\lambda_i}^2.
    \end{aligned}
\end{equation}
Here, $\lambda_i = \boldsymbol{v}_i^{T}\boldsymbol{e}_i$ is dot product of the unit vector $\boldsymbol{e}_i$ and the eigenvector $\boldsymbol{v}_i$, reflecting the alignment of the scalar gradient $\nabla Y$ with the eigenvector $\boldsymbol{v}_i$ corresponding to the $i$th principal strain rate $s_i$.

In two-dimensional flow, $s_3 = 0$ and $\lambda_1^2 + \lambda_2^2 = 1$. Owing to the eigenvalues $s_i$ are ordered by $s_1 > s_2$, $s_1$ is the principal strain in the stretch direction, while $s_2$ is the principal strain in the compress direction. The stretch term can be expressed as:
\begin{equation}
    \begin{aligned}
        T_{stretch} &= -2\chi s_i {\lambda_i}^2 = -2\chi\left(s_1\lambda_1^2 + s_2\lambda_2^2\right) \\
        & = -2\chi\left[s_2 + (s_1-s_2)\lambda_1^2\right].
    \end{aligned}   
\end{equation}

Additionally, it is observed from the budget analysis in Fig.\;\ref{time derivative of SDR} that the velocity divergence term $T_{divU} = \chi \left(\frac{\partial u_k}{\partial x_k}\right)$ is nearly negligible. Thus, the velocity divergence $\frac{\partial u_k}{\partial x_k} = s_1 + s_2 \approx 0$, yielding the final form of the stretch term $T_{stretch}$ is:
\begin{equation}
    \begin{aligned}
        T_{stretch} &= -2\chi\left[s_2 + (s_1-s_2)\lambda_1^2\right]\\
        & = \chi\left[2s_1\left(1-2\lambda_1^2\right)\right].
    \end{aligned} 
    \label{stretch equation}
\end{equation}
By eliminating SDR $\chi$ from Eq.\;\ref{stretch equation}, the modified stretch term reflecting the stretch effect of flow on scalar gradient is given by:
\begin{equation}
    T_{stretch}^* = 2s_1\left(1-2\lambda_1^2\right),
    \label{stretch equation2}
\end{equation}
from this expression, it's evident that the stretch rate $T_{stretch}^*$ increases with a large principal strain rate $s_1$ and a small $\lambda_1$. A small value of $\lambda_1$ implies that the scalar gradient $\nabla Y$ is distant from the eigenvector $\boldsymbol{v}_1$ corresponding to principal strain $s_1$. Additionally, we employ two statistical methods, spatial integration $\left\langle\cdot\right\rangle = \iint \cdot dV$ and the probability density function $P(\cdot)$, to investigate this modified stretch term $T_{stretch}^*$. 

\begin{figure}[H]
  \centering
  \subfigure[]{\includegraphics[clip=true,width=.45\textwidth]{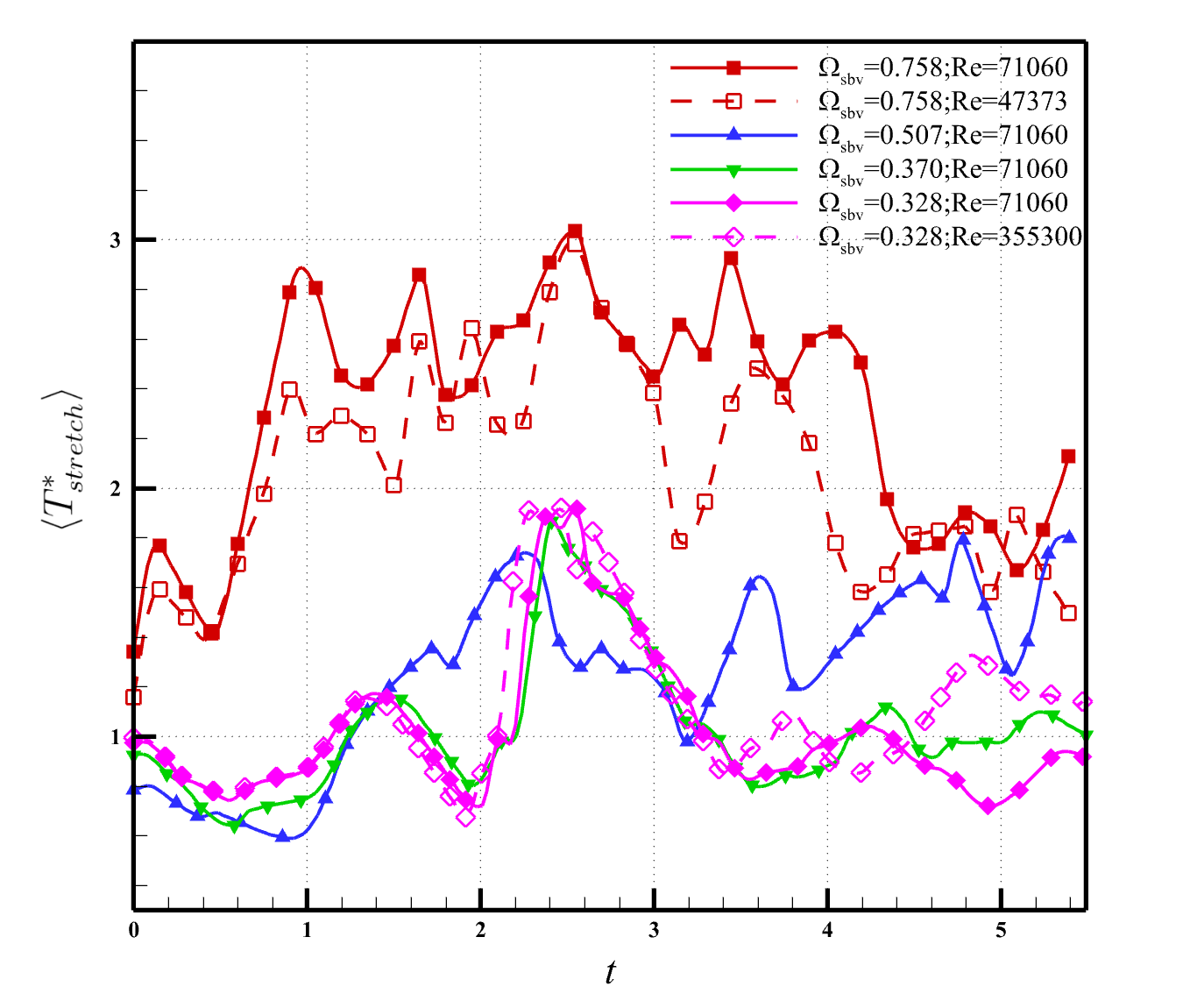}}\\
  \subfigure[]{\includegraphics[clip=true,width=.455\textwidth]{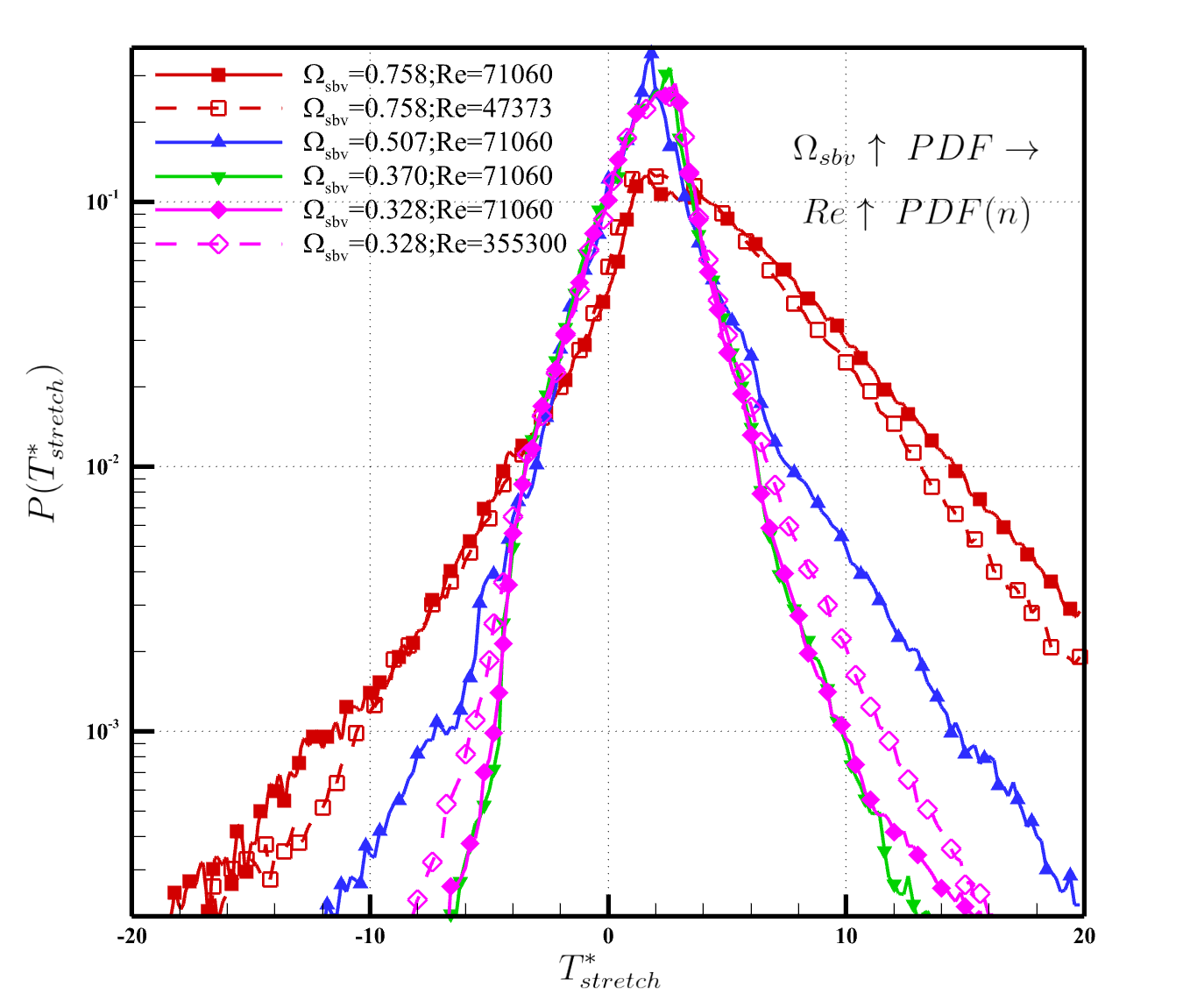}}
  \caption{Comparison of (a) the time evolution of the space averaged modified stretch term in the form of ${\left\langle T_{stretch}^*\right\rangle}$ and (b) the probability density function (PDF) of the modified stretch term $ P\left(T_{stretch}^*\right)$, $PDF(n)$ denotes a neglectable variation of the PDF curve.}
    \label{modified stretch rate}
\end{figure}

Fig.\;\ref{modified stretch rate} (a) illustrates the temporal evolution of the spatial average stretch term $\left\langle T_{stretch}^{*} \right\rangle$. The solid lines, corresponding to $\left\langle T_{stretch}^{*} \right\rangle$ for cases with the same Reynolds number $Re$, are ordered from the red solid line to the pink solid line, indicating a notable increase in the spatial average stretch term $\left\langle T_{stretch}^{*} \right\rangle$ for cases with larger $\Omega_{sbv}$. However, there is no significant variation between the dashed lines and the solid lines of the same color. This observation suggests that, for the same $\Omega_{sbv}$ value, an increase in the Reynolds number $Re$ does not result in a significant increase in the spatial average stretch term $\left\langle T_{stretch}^{*} \right\rangle$. The PDF depicted in Fig.\;\ref{modified stretch rate} (b) further supports this observation. An increase in $\Omega_{sbv}$ shifts the PDF curves to the right, indicating a greater distribution of $T_{stretch}^{*}$ in the region of large positive values. However, comparison of the solid and dashed curves shows no significant changes as $Re$ increases. These findings regarding the modified stretch term are consistent with the previously mentioned scaling behaviors of the mixing rate $\left\langle T_{SDR} \right\rangle$.

\begin{figure}[H]
  \centering
  \subfigure[]{\includegraphics[clip=true,width=.45\textwidth]{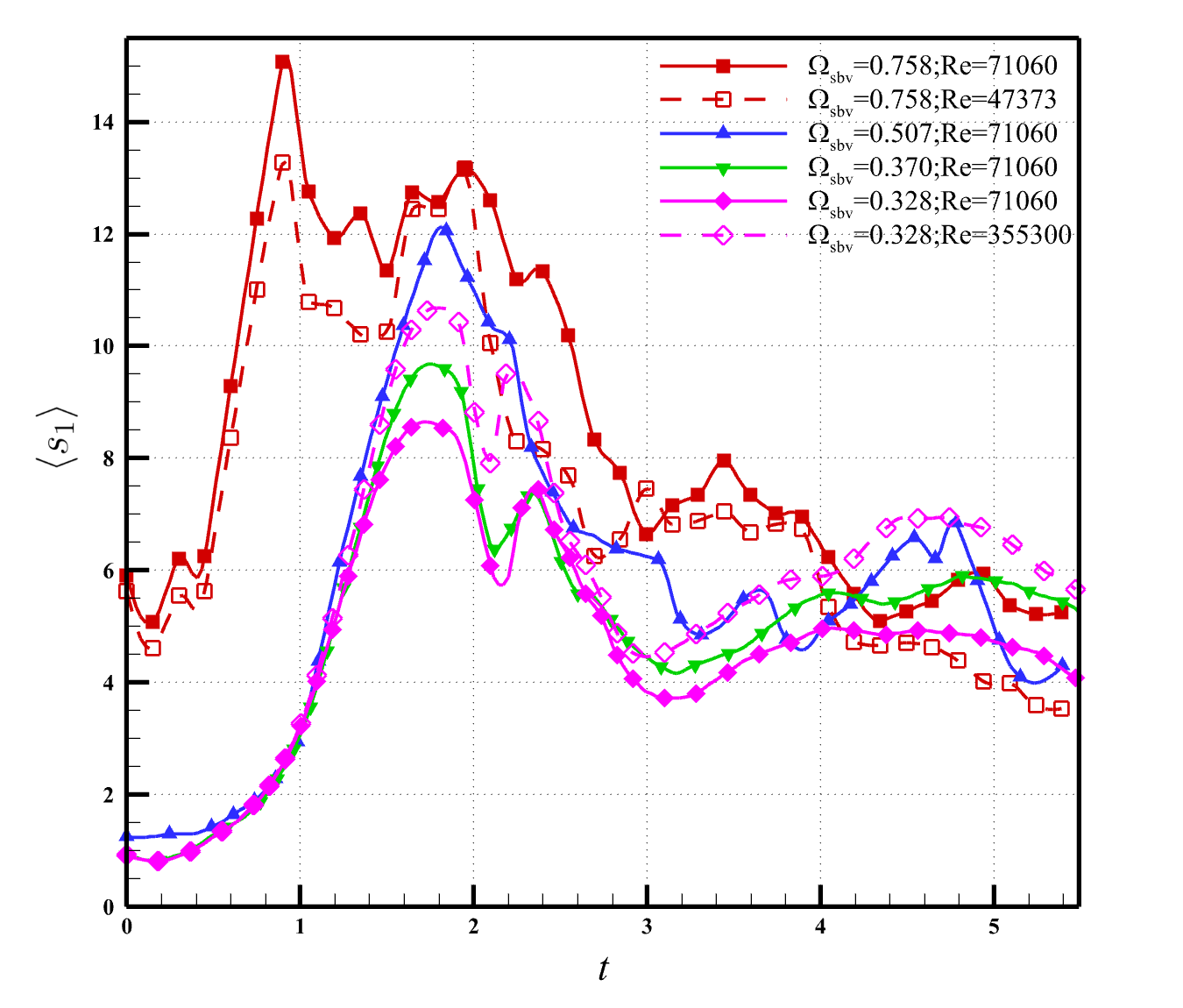}}\\
  \subfigure[]{\includegraphics[clip=true,width=.455\textwidth]{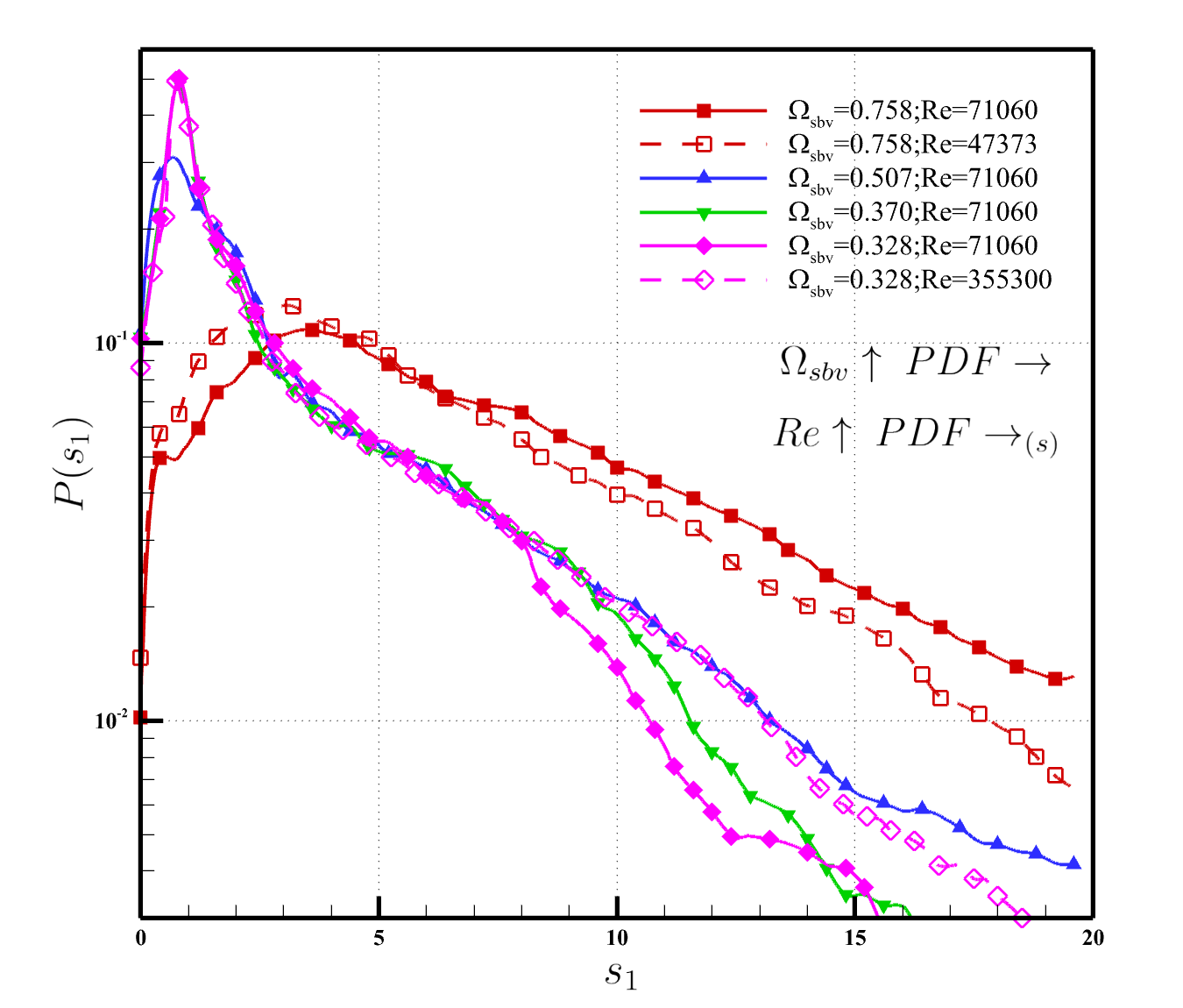}}
  \caption{Comparison of (a) the time evolution of the spatial averaged principle strain in the stretch direction in the form of $\left\langle  s_1 \right\rangle$ and (b) the probability density function (PDF) of the principle strain in the stretch direction $P\left( s_1\right)$. $\rightarrow_{(s)}$ indicates the PDF curves shift to the right slightly.}
  \label{principal strain rate}
\end{figure}

Expanding on the observation of the stretch term $T_{stretch}^{*}$ and referring to Eq.\;\ref{stretch equation}, we further analyze the principal strain rate $s_1$ and alignment $\lambda_1$. The temporal evolution of the spatial average principal strain rate $\left\langle s_1 \right\rangle$ is presented in Fig.\;\ref{principal strain rate} (a). Here, the solid lines, representing $\left\langle s_1 \right\rangle$ for cases with the same Reynolds number $Re$, are arranged from the red solid line to the pink solid line, demonstrating that the spatial average strain rate $\left\langle s_1 \right\rangle$ is greater for cases with larger $\Omega_{sbv}$ values. However, there is only a slight variation between the dashed and solid lines of the same color. This finding suggests that, for the same $\Omega_{sbv}$ value, an increase in the Reynolds number $Re$ results in a finite increase in the spatial average principal strain $\left\langle s_1 \right\rangle$. This observation is further supported by the PDF curves shown in Fig.\;\ref{principal strain rate} (b). It is apparent that the solid lines with larger $\Omega_{sbv}$ values shift to the right, indicating a greater distribution of $s_1$ in the region of large positive values. Additionally, the dashed lines closely resemble the solid lines of the same color, suggesting that flows with the same $\Omega_{sbv}$ but different $Re$ have similar distributions of $s_1$. In summary, the SBHI, as reflected by $\Omega_{sbv}$, amplifies the principal strain $s_1$, while the unsteady flow caused by large $Re$ sightly alters the principal strain $s_1$.

\begin{figure}[H]
  \centering
  \subfigure[]{\includegraphics[clip=true,width=.435\textwidth]{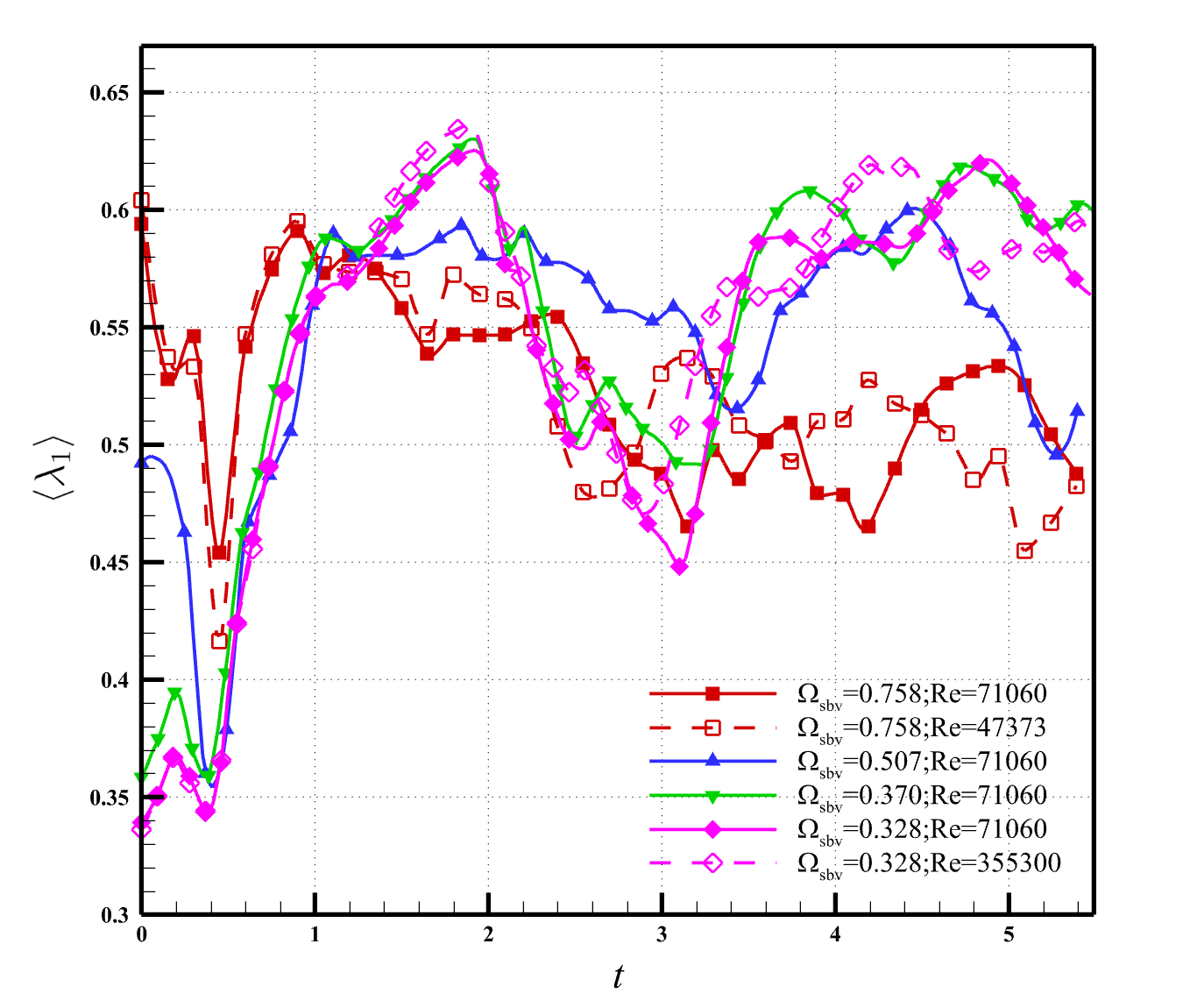}}\\
  \subfigure[]{\includegraphics[clip=true,width=.4\textwidth]{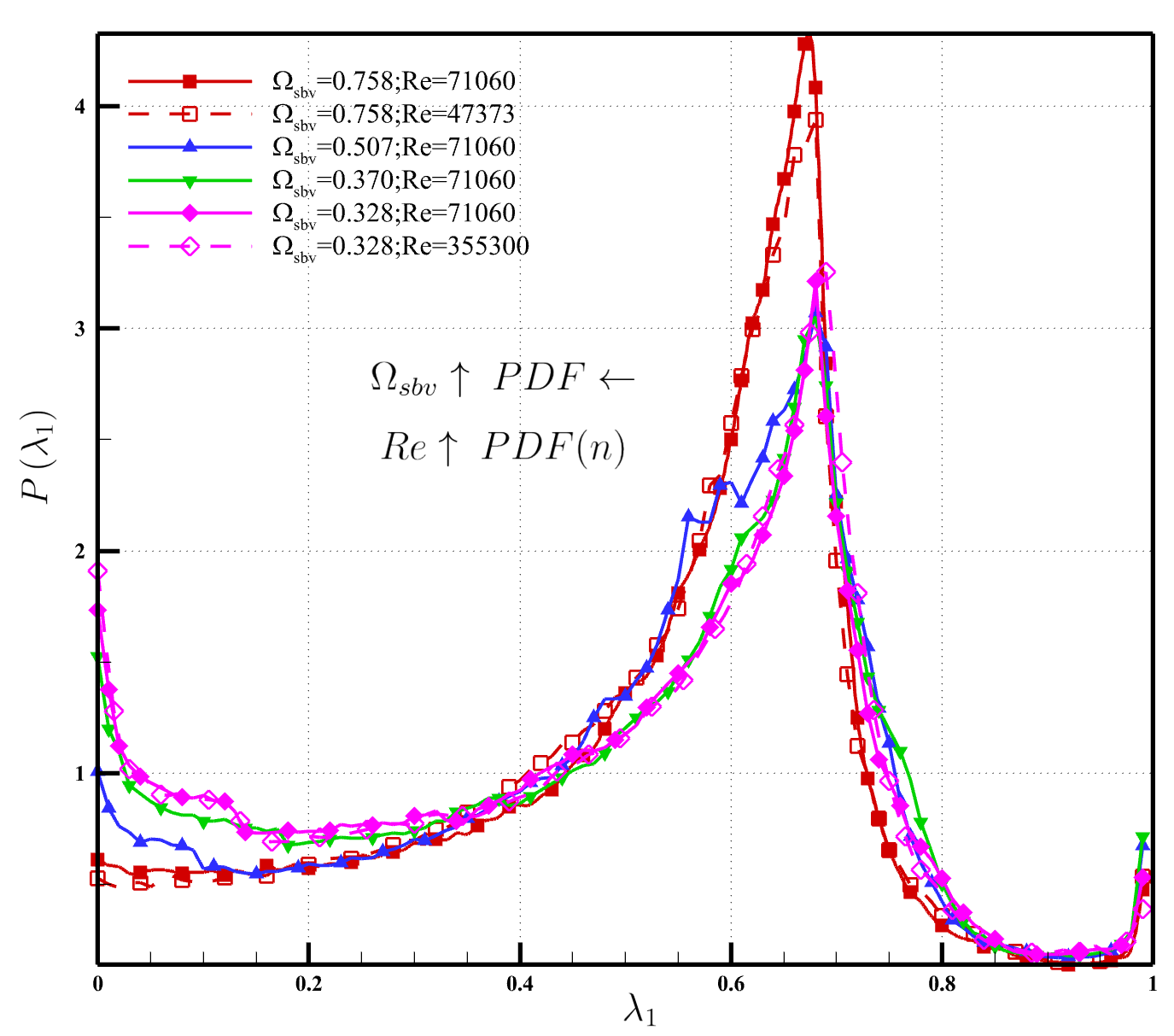}}
  \caption{Comparison of (a) the time evolution of the space averaged alignment of the scalar gradient $\nabla Y$ with the principle strain in the stretch direction $\left\langle \lambda_{1}\right\rangle$ and (b) the probability density function (PDF) of the alignment of the scalar gradient $\nabla Y$ with the principle strain in the stretch direction $P\left( \lambda_1\right)$, $PDF(n)$ denotes a neglectable variation of the PDF curve.}
  \label{alignment}
\end{figure}

As for the alignment $\lambda_1$, Fig.\;\ref{alignment} (a) displays the temporal evolution of the 
spatial average alignment $\left\langle\lambda_1\right\rangle$, differing from the observation in the principal strain $\left\langle\ s_1\right\rangle$, the solid lines, representing $\left\langle\lambda_1\right\rangle$ for cases with the same Reynolds number $Re$, are ordered from the pink solid line to the red solid line, indicating that the spatial average alignment $\left\langle\lambda_1\right\rangle$ decreases with larger $\Omega_{sbv}$ values. Furthermore, the variations between the dashed lines and solid lines of the same color are nearly neglectable, suggesting that an increase in the Reynolds number $Re$ does not alter the spatial average alignment $\left\langle\lambda_1\right\rangle$. The PDF of $\lambda_1$ is further plotted in Fig.\;\ref{alignment} (b). The peak of the solid curves shifts to the left with the increase of $\Omega_{sbv}$, implying a greater distribution of $\lambda_1$ in the region with smaller value. The dashed lines are coincide with the solid lines of the same colour, which indicates the Reynolds number $Re$ hardly alter the distribution of $\lambda_1$.

Based on the research findings outlined above, we can summarize the distinct mechanisms of two scaling behaviors as follows:

1) The strength of SBHI is measured by the SBV number $\Omega_{sbv}$. The increase in $\Omega_{sbv}$ results in greater $s_1$ and smaller $\lambda_1$, leading to the amplification of the stretch term $T_{stretch}^*$ in accordance with Eq.\;\ref{stretch equation2}. The amplification of stretch rate leads to a higher growth rate of SDR $\chi$, signifying an increased mixing rate. Consequently, the increment of mixedness $\left\langle f^{*}\right\rangle_{\Delta}$ can attain a larger value, which means the enhancement in the mixing between the helium in the bubble and the ambient air. These logical relationships can be summarized as:
\begin{equation}
    \Omega_{sbv} \uparrow \stackrel{}{\Rightarrow} \left\{
    \begin{aligned}
        & s_1 \uparrow \\
        & \lambda_1 \downarrow
    \end{aligned}
    \right.
    \stackrel{\rm{Eq.\;\ref{stretch equation2}}}{\Rightarrow} T_{stretch}^* \uparrow \stackrel{\rm{Eq.\;\ref{SDR equation 2}}}{\Rightarrow} \chi \uparrow  \stackrel{\rm{Eq.\;\ref{mixedness equation 5}}}{\Rightarrow} \left\langle f^{*}\right\rangle_{\Delta} \uparrow.
    \label{mechanism 1}
\end{equation}

2) In the case of the unstable flow caused by high Reynolds number $Re$, differing from SBHI, a large increase of $Re$ results in a finite alteration of principal strain $s_1$ that is consist with previous research \cite{pope2000turbulent}, while the alignment $\lambda_1$ undergoes no significant change. As a consequence, there is finite variation in the stretch term $T_{stretch}^{*}$. Consequently, the mixing rate SDR $\chi$ and the increment of mixedness $\left\langle f^{*}\right\rangle_{\Delta}$ are neglectable, which means the mixing between the helium in the bubble and the ambient air can hardly be enhanced effectively by increasing the Reynolds number $Re$. These logical relationships can be summarized as:
\begin{equation}
    Re \uparrow \Rightarrow \left\{
        \begin{aligned}
        & s_1 \uparrow\! (s) \\
        & \lambda_1 (n)     
        \end{aligned} \right.
    \stackrel{\rm{Eq.\;\ref{stretch equation2}}}{\Rightarrow} T_{stretch}^{*}\uparrow\! (s)  \stackrel{\rm{Eq.\;\ref{SDR equation 2}}}{\Rightarrow} \chi(n)
    \stackrel{\rm{Eq.\;\ref{mixedness equation 5}}}{\Rightarrow} \left\langle f^{*}\right\rangle_{\Delta}(n),
    \label{mechanism 2}
\end{equation}
in this equation, $\uparrow\!(s)$ indicates the increment of the parameter is slight, and $(n)$ denotes a neglectable variation of the parameter.

\section{Conclusions and future work}\label{sec:4}

In this paper, we investigate the significant effect of the hydrodynamic instability resulting from the initial diffusion on variable density mixing in shock cylindrical bubble interaction (SBI). To ditinguish each factor within this instability, we design a series of cases with varying degrees of the hydrodynamic instability, while ensuring that the total circulation $\Gamma$ is controlled to maintain constant values for the Reynolds number $Re = \Gamma/\nu$ and the P\'eclect number $Pe = \Gamma/\mathscr{D}$ by the circulation control method.

Firstly, we investigate the hydrodynamic characteristics of the hydrodynamic instability by analysing the temporal evolution of the morphology of the bubble and vorticity dynamics. Through the examination of positive circulation $\Gamma^+$ and the associated budget analysis, we ascertain that the hydrodynamic instability induced by initial diffusion arises from the baroclinic torque $\left\langle B^+\right\rangle$. Therefore, this instability is categorized as SBHI, and the magnitude of which can be measured by the difference of the peak value of positive circulation and the total circulation: $\Gamma_{sbv}^+ = \left(\Gamma^+\right)|_{peak}-\Gamma$.

Secondly, we study the effect of SBHI on mixing by checking the mixedness $f$ and SDR $\chi$. Through the temporal evolution of the increment of global modified mixedness $\left\langle f^{*}_{\Delta} \right\rangle$ and the corresponding budget analysis, we discern that the mixing rate, as reflected in the SDR term $\left\langle T_{SDR}\right\rangle$, is enhanced by SBHI. To further describe this mixing enhancement behavior, a new dimensionless parameter, the SBV number $\Omega_{sbv}$ is put forward to qualify the strength of SBHI. By examining the temporal-averaged mixing rate $T_{SDR}$ in cases with different $\Omega_{sbv}$, we observe a positive correlation between the mixing rate $\overline{\left\langle T_{SDR}\right\rangle}$ and $\Omega_{sbv}$, yielding $\overline{\left\langle T_{SDR}\right\rangle} \sim \Omega_{sbv}^{2}$.

Finally, our findings delineate the mechanisms underlying the enhancement effect of SBHI on mixing. Previous research has suggested that the unstable flow resulting from a high Reynolds number $Re$ is less effective in altering mixing. The underlying mechanisms of these two scaling behaviors are elucidated by analyzing the stretch term $\left\langle T_{stretch}\right\rangle$ in Eq.\;\ref{SDR equation 2}, which governs the growth of the mixing rate $\left\langle\chi\right\rangle$. This term is further decomposed into the principal strain $s_1$ and the alignment $\lambda_1$. Through an evaluation of the stretch term $T_{stretch}^*$ using two statistical methods, spatial integration and probability density function, we unravel the mechanisms underlying these two scaling behaviors: the SBHI, characterized by $\Omega_{sbv}$, can enhance mixing by increasing the principal strain $s_1$ and decreasing the alignment $\lambda_1$, while the unstable flow resulting from high $Re$ cannot effectively enhance mixing due to the finite alteration of the principal strain $s_1$ and the absence of significant change in alignment $\lambda_1$. These mechanisms are summerized as Eq.\;\ref{mechanism 1} and Eq.\;\ref{mechanism 2}.

The current study demonstrates the significant enhancement effect of SBHI on mixing. Consequently, adjusting the initial diffusion is an effective method for controlling mixing, making it a potential approach for enhancing mixing in practical engineering flow scenarios, such as oblique shock jet interactions. The reason why SBHI can increase the principal strain $s_1$ and decrease the alignment $\lambda_1$ deserves further investigation in the future.

\Acknowledgements{This work was supported by the Postdoctoral Fellowship Program of China Postdoctoral Science Foundation (GZC20231566), Sichuan Science and Technology Program (2025ZNSFSC0834) and Natural Science Foundation of Shanghai (25ZR1402273). We would like to express our gratitude to Prof. Pan Shucheng for raising the question at the Conference on the instability of the fluid interface and multi-species turbulence, which inspired this paper.}

\InterestConflict{The authors declare that they have no conflict of interest.}



\bibliography{mybibfile}{}
\bibliographystyle{IEEEtran}

\begin{appendix}




\renewcommand{\thesection}{Appendix}

\section{}

\subsection{\label{grid resolution} Grid resolution study}

Appendix \ref{grid resolution} introduces the grid resolution study of the simulation, and the case with $\Omega_{sbv}=29.92$ and $Re = 71060$ is listed as an example. Since the second order differential of SDR is sensitive to mesh resolution, it's crucial to select the grid size to ensure the numerical results are resolved. Consistent with our previous works \cite{yu2020scaling,yu2022effects}, in order to guarantee the grid size can capture the small-scale structures in the SBI configuration of this paper, the mesh Reynolds number $Re_{\triangle} = u^{*}\triangle/\nu$ and the mesh P\'eclect number $Pe_{\triangle} = u^{*}\triangle/\mathscr{D}$, are introduced, where $u^{*}$ is the characteristic length defined in Eq.\;\ref{characteristic time and length}, and $\triangle$ signifies the grid size. In previous research \cite{yu2020scaling,yu2022effects}, the grid resolution criteria dictate that: $Re_{\triangle} \leq 140$ and $Pe_{\triangle} \leq 23$. For this study, the grid size of $\triangle = 3.0\,\rm{\mu m}$, with the total grid number of $[N_x,N_y] = [10000,1832]$ is employed, resulting in the corresponding mesh Reynolds number and mesh P\'eclect number which match this criteria: $Re_{\triangle} = 81$ and $Pe_{\triangle} = 21$.

\begin{figure}[H]
  \centering
  \includegraphics[clip=true,width=.47\textwidth]{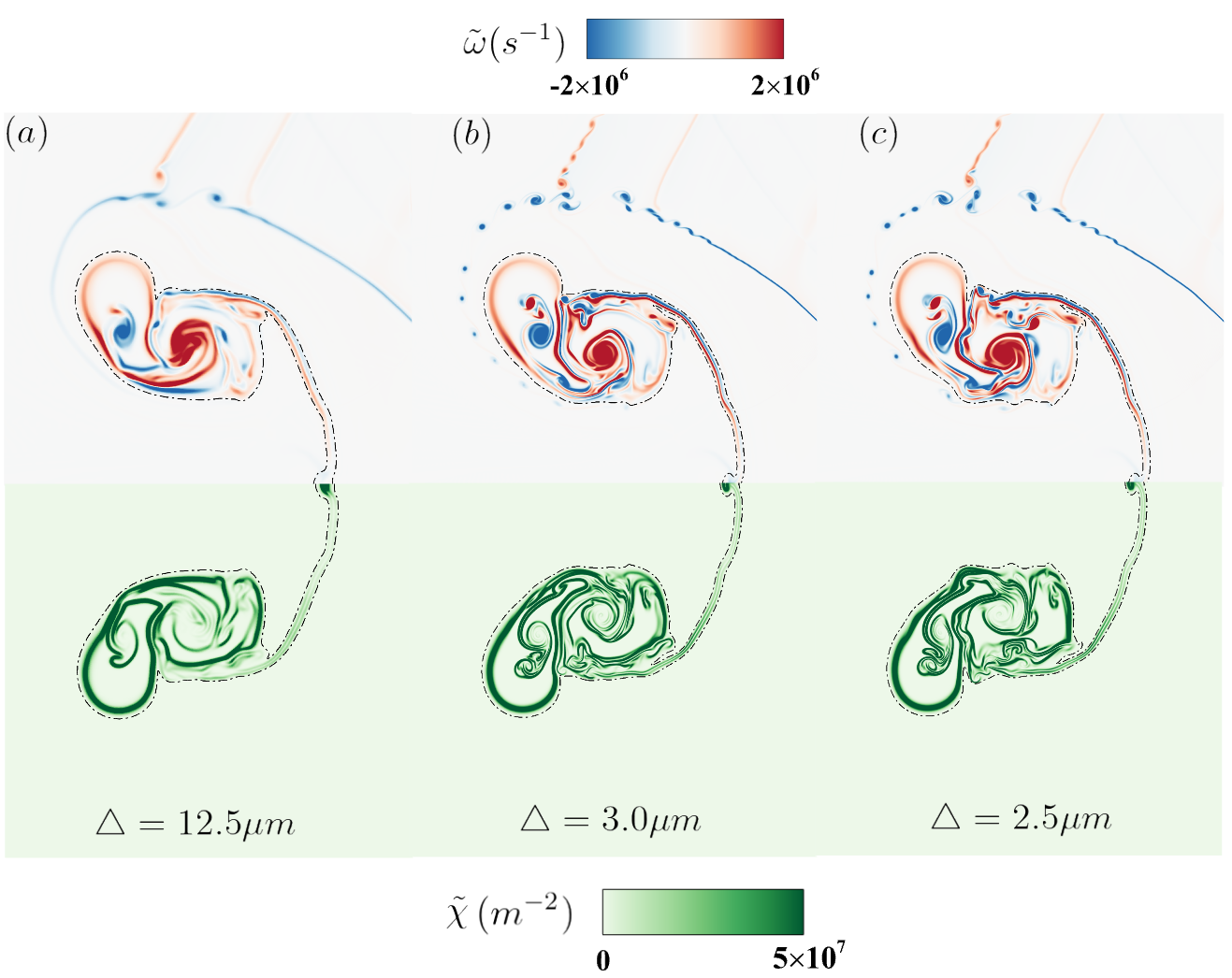}\\
  \caption{Vorticity contour (top) and SDR contour (bottom) with black dashed isoline of $Y = 0.05$ of three different mesh resolutions. (a) $\triangle = 12.5 \rm{\mu m}$; (b) $\triangle = 3.0 \rm{\mu m}$; (c) $\triangle = 2.5 \rm{\mu m}$}
  \label{MeshResolution}
\end{figure}

In order to verify if this grid size is enough to obtain a resolved numerical result, other two meshes with greater and smaller grid size are designed to compare the mesh resolution, the grid settings are displayed in Table \ref{MeshResolution}:

\begin{table}[H]
    \centering
    \begin{tabular}{c c c c c}
        \hline
         $\triangle\,(\rm{\mu m})$ & $[N_x,N_y]$ & $Re_{\triangle}$ & $Pe_{\triangle}$\\
        \hline
         12.5 & $[3200,520]$ & 338 & 88 \\
         3.0 & $[10000,1832]$ & 81 & 21 \\
         2.5 & $[12000,2199]$ & 68 & 18 \\
         \hline
    \end{tabular}
    \caption{The mesh settings applied to validate the grid resolution}
    \label{mesh setting}
\end{table}

Qualitative comparisons among the three mesh resolutions are presented in Fig.\;\ref{MeshResolution}. By comparing the vorticity and SDR contour of the three different resolutions, it is evident that the small-scale structures are smeared by numerical viscosity in the coarse mesh with a grid size of $\triangle = 12.5\,\rm{\mu m}$, whereas these structures are clearly captured in the fine meshes with grid sizes of $\triangle = 3.0\,\rm{\mu m}$ and $2.5\,\rm{\mu m}$. Additionally, a general agreement is observed between the vorticity and SDR in these two fine meshes.

\begin{figure*}[htbp]
  \centering
  \includegraphics[clip=true,width=.8\textwidth]{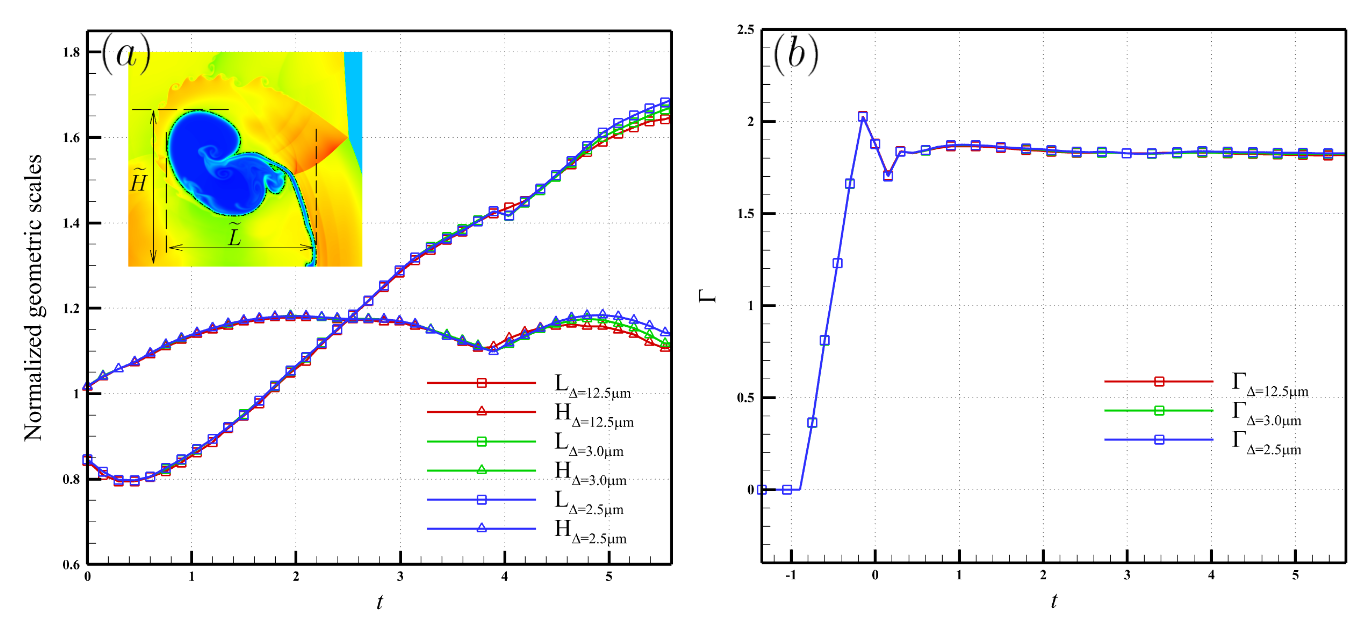}\\
  \caption{Comparison of the (a) normalized geometric scales and (b) circulation $\Gamma$ obtained from three different meshes. The black dashed line in Fig (a) is the isoline of $Y=0.05$.}
  \label{MeshResolution_largescale}
\end{figure*}

We then conducted a quantitative analysis to check the mesh resolutions. The quantities related to the large scale structures are firstly examined. In Fig. \ref{MeshResolution_largescale} (a), the temporal evolution of the normalized geometric scales $L = \widetilde{L}/x^{*}, H = \widetilde{H}/x^{*}$ is presented, where $ \widetilde{L}$ represents the length of the bubble, and  $\widetilde{H}$ represents the height. Additionally, Fig.\;\ref{MeshResolution_largescale} (b) depicts the circulation $\Gamma$. These figures illustrate that the geometric scales and circulation are independent of grid size, indicating that the qualities related to large scale structures can be accurately captured using the present fine mesh settings.

The present research focuses on the investigation of positive circulation $\Gamma^+$, mixedness $f$, and SDR $\chi$. These quantities are influenced by small-scale structures, hence, ensuring mesh independence of large-scale structures alone is insufficient, the grid resolution of small-scale structures also needs to be carefully considered. However, as indicated by the observation of the vorticity contour in Fig.\;\ref{vorticity contour}, the presence of small-scale structures signifies a strong instability within the bubble. Consequently, achieving mesh independence of small-scale structures in this type of simulation is relatively challenging. Thus, budget analysis is applied to evaluate the resolution, which has been previously utilized in research on turbulent channel flow \cite{tang2022statistical}.

\begin{figure*}[htbp]
  \centering
  \includegraphics[clip=true,width=.9\textwidth]{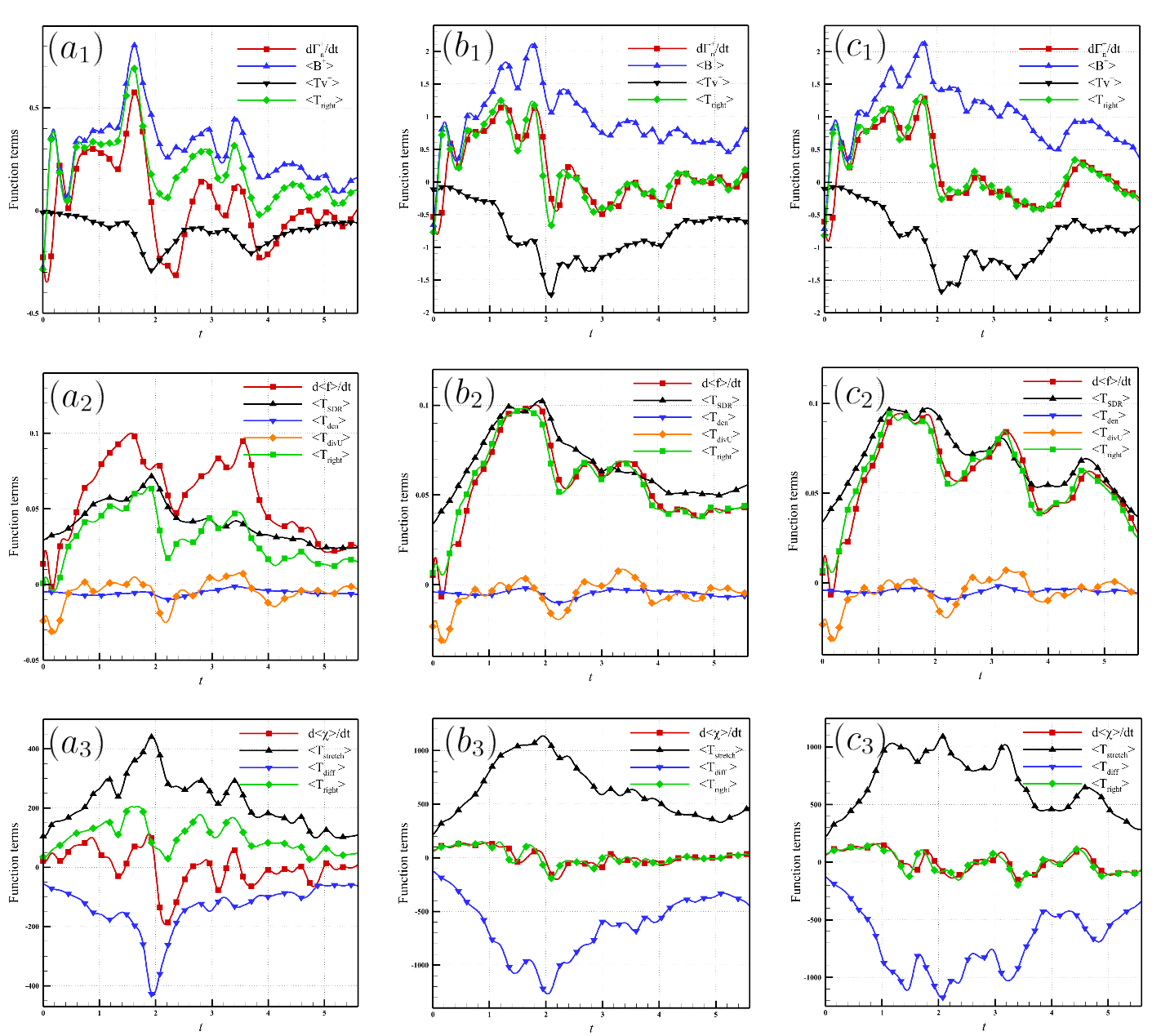}\\
  \caption{The budget analysis of the transport equations of ($a_1,b_1,c_1$) the positive circulation $\Gamma^+$ given by Eq.\;\ref{final form of circulation equation}; ($a_2,b_2,c_2$) the mixedness $f$ given by Eq.\;\ref{mixedness equation 3}; ($a_3,b_3,c_3$) the SDR $\chi$ given by Eq.\;\ref{SDR equation 2}. $a_i$ is the results obtained form the coarse mesh with grid size of $\triangle = 12.5 \rm{\mu m}$; $b_i$ corresponds to the fine mesh with grid size of $\triangle = 3.0 \rm{\mu m}$; and $c_i$ corresponds to the fine mesh with grid size of $\triangle = 2.5 \rm{\mu m}$}
  \label{MeshResolution_budget}
\end{figure*}

The transport equations for the positive circulation $\Gamma^+$, mixedness $f$, and SDR $\chi$ are given by Eq.\;\ref{final form of circulation equation}, Eq.\;\ref{mixedness equation 4}, and Eq.\;\ref{SDR equation 2}, respectively. Fig.\;\ref{MeshResolution_budget} presents the budget analysis of these equations. Taking the positive circulation budget in the first row of Fig.\;\ref{MeshResolution_budget} as an example, the temporal derivative of positive circulation $\frac{D \Gamma_n^+}{Dt}$ and the corresponding terms in the coarse mesh with a grid size of $\triangle = 12.5\,\rm{\mu m}$ are illustrated in Fig.\;\ref{MeshResolution_budget} $(a_1)$. The deviation between the red line representing $\frac{D \Gamma_n^+}{Dt}$ and the green line depicting the sum of the terms on the right-hand side of Eq.\;\ref{final form of circulation equation}, indicates that the numerical viscosity introduced by the coarse mesh is relatively large compared to the kinematic viscosity $\nu$. Consequently, the evolution of the positive circulation $\Gamma^+$ is significantly influenced by the numerical viscosity, therefore the coarse mesh lacks the capacity to effectively resolve the vorticity dynamics.

Similar analyses for the fine meshes with a grid size of $\triangle = 3.0\,\rm{\mu m}$ and $2.5\,\rm{\mu m}$ are depicted in Fig.\;\ref{MeshResolution_budget} $(b_1)$ and $(c_1)$. In contrast to the results obtained from the coarse mesh, the red line from the fine meshes aligns well with the green line, which means that the numerical viscosity introduced by the fine meshes can be neglected. Consequently, the vorticity dynamics can be considered resolved by these two fine meshes. Additionally, the transport equations of mixedness and SDR, Eq.\;\ref{mixedness equation 4} and Eq.\;\ref{SDR equation 2}, are examined in the medium and bottom rows of Fig.\;\ref{MeshResolution_budget}. With the results being similar to the positive circulation equation, we can make the conclusion that these two fine meshes have the ability to resolve the mixedness and SDR dynamics, whereas the coarse mesh does not.

Considering both the computational cost and simulation accuracy, we choose the mesh with the grid size of $\triangle = 3.0\,\rm{\mu m}$ in the present study. The budget analysis in Fig.\;\ref{time derivative of positive circulation}, Fig.\;\ref{mixedness budget 1} and Fig.\;\ref{time derivative of SDR} confirms the numerical results obtained from this mesh are resolved in the other cases in this paper. 

\subsection{\label{circulation control} Circulation control method}
The details of the selection of geometric parameters are displayed here. The cases are intricately designed to explore the impact of hydrodynamic instability induced by inital diffusion on mixing in two-dimensional SBI. As detailed in section \ref{sec:2}, the dimensionless number $Re = \Gamma/\nu$ must remain consistent throughout the cases, therefore the total circulation $\Gamma$ also needs to remain constant. The ratio $\xi = W/R_{core}$ has been set at values of 0.1, 1.0, 5.0 and 10.0 to manage the hydrodynamic instability, thus the radius of the core region of the bubble $R_{core}$ are designed here to control the total circulation $\Gamma$.

According to previous research \cite{picone1988vorticity,yang1994model,samtaney1993shock,liu2020contribution} on the circulation model:
\begin{equation}
    \left\{
    \begin{aligned}
        &\Gamma_{PB} = 2u_1^{'}R\left(1-\frac{u_1^{'}}{2W_i}ln\left(\frac{\rho_1}{\rho_2}\right) \right),\\
        &\Gamma_{YKZ} = \frac{4R}{W_i}\frac{P_1^{'}-P_1}{\rho_1^{'}}\frac{\rho_2-\rho_1}{\rho_2+\rho_1},\\
        &\Gamma_{SZ} \propto \frac{2\gamma^{1/2}}{1+\gamma}(1-\sigma^{-1/2})(1+\frac{1}{Ma}+\frac{1}{Ma^2})(Ma-1)\cdot R,
    \end{aligned}
    \right.
\end{equation}
where $W_i$ is the velocity of the incident shock, $\gamma$ is the specific heat ratio, and $Ma$ is the Mach number, the total circulation is directly proportional to the radius of the bubble: $\Gamma \propto R$. However, due to varying mass fraction distributions caused by different ratios of $\xi$, $\Gamma$ cannot remain constant while maintaining the same radius $R$. Therefore, based on the configuration illustrated in Fig.\;\ref{Initial conditions} and the mass fraction distribution in Eq.\;\ref{Mass fraction distribution}, the average radius $R_{\rho avg}$ is defined by the following equations:

\begin{equation}
    \left\{
\begin{aligned}
& Y_{avg}\frac{\pi R^2}{2} = \iint Y \,dV, \\
& \rho_{avg} = \rho_2 {Y_{avg}} +\rho_1 {(1 - Y_{avg})}, \\
& \rho_{avg}\frac{\pi R_{\rho avg}^2}{2} = \iint \rho \,dV.
\end{aligned} 
  \right.
\end{equation}

\begin{figure}[H]
  \centering
    \includegraphics[clip=true,width=.45\textwidth]{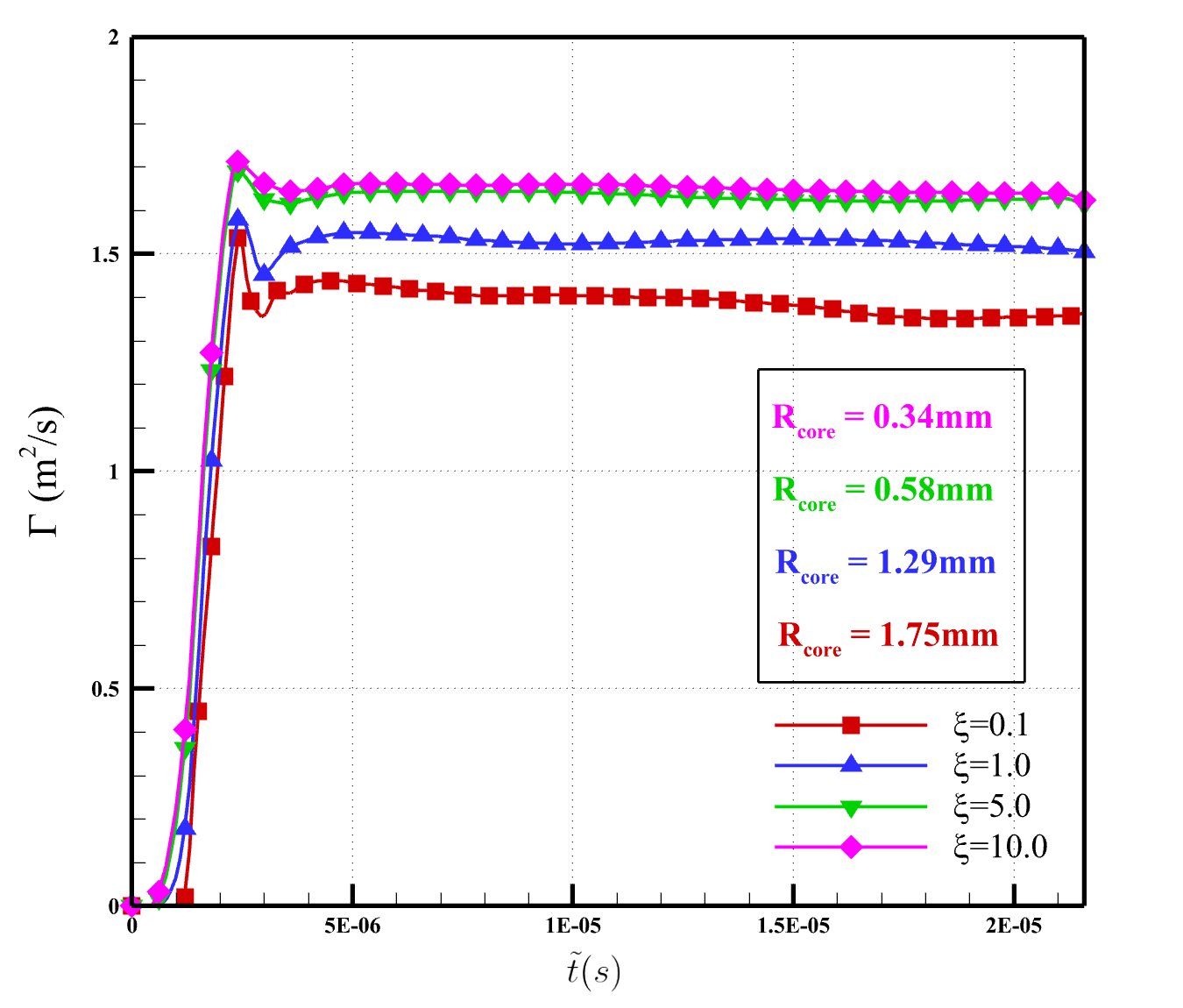}\\
    \caption{Comparison of the time variation of the total circulation $\Gamma$ of two-dimensional SBI cases in the first round simulation. }
    \label{first round circulation}
\end{figure}

The first round simulation maintains the same $R_{\rho avg}$, and the geometric parameters along with the corresponding total circulation $\Gamma$ (as shown in Fig.\;\ref{first round circulation}) are compiled in Table\;\ref{first round simulation}.

\begin{table}[H]
\caption{Geometric parameters for the two-dimensional SBI cases in the first round simulation.}
 \centering
  \begin{tabular}{c c c c c}
    \hline
    $\xi$ & $R_{core}\,\rm{(mm)}$ & $Y_{avg}$ & $R_{\rho avg}\,\rm{(mm)}$ & $\Gamma\, \rm{(m^{2}/s)}$\\ 
    \hline
    0.1 & 1.750 & 0.900 & 1.820 & 1.430\\
    1.0 & 1.290 & 0.497 & 1.820 & 1.560 \\
    5.0 & 0.580 & 0.276 & 1.820 & 1.620 \\
    10.0 & 0.238 & 0.340 & 1.820 & 1.660 \\
    \hline
  \end{tabular}
  \label{first round simulation}
\end{table}

\begin{figure}[H]
  \centering
    \includegraphics[clip=true,width=.45\textwidth]{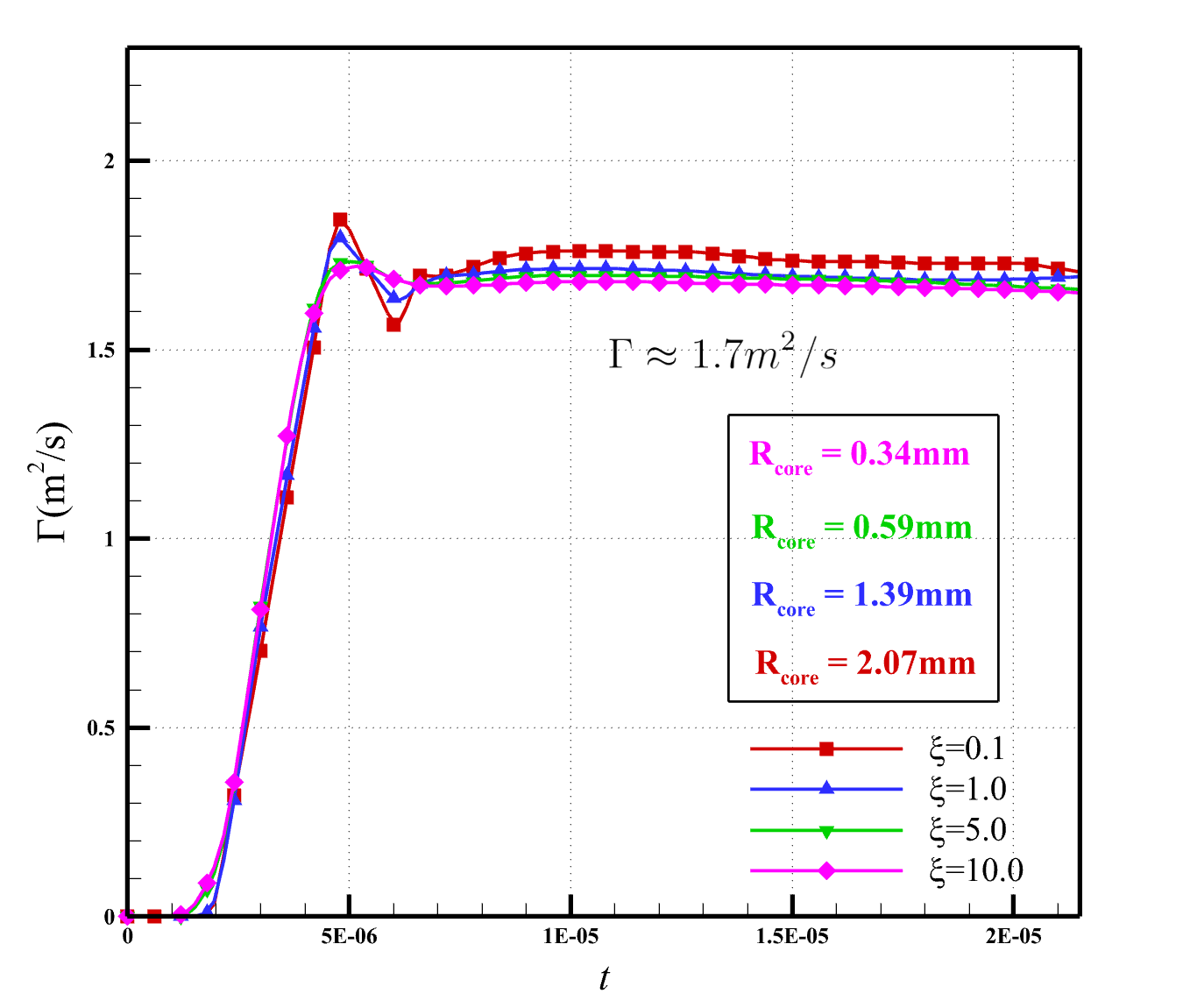}\\
    \caption{Comparison of the time variation of the total circulation $\Gamma$ of two-dimensional SBI cases in the second round simulation. }
    \label{second round circulation}
\end{figure}

In Table \ref{first round simulation} and Fig.\;\ref{first round circulation}, the circulation of SBI with a constant $R_{\rho avg}$ is relatively consistent across the cases, but still varies. Utilizing the relationship $\Gamma \propto R$, $R_{core}$ is adjusted based on the circulation in the first round simulation, and the corresponding geometric parameters in the second round simulation are listed in Table \ref{second round simulation}. As demonstrated in Fig.\;\ref{second round circulation}, the total circulation $\Gamma$ resulting from the provided parameters in Table \ref{second round simulation} approximates 1.7$\,\rm{m^2/s}$. Thus, the geometric parameters in the second round simulation are selected in this study to ensure the maintenance of the same Reynolds number $Re$ across the cases.

\begin{table}[H]
\caption{Geometric parameters for the two-dimensional SBI cases in the second round simulation.}
 \centering
  \begin{tabular}{c c c c}
    \hline
    $\xi$ & $R_{core}\,\rm{(mm)}$ & $R\,\rm{(mm)}$ & $\Gamma \,\rm{(m^{2}/s)}$\\ 
    \hline
    0.1 & 2.070 & 2.277 & 1.740\\
    1.0 & 1.390 & 2.780 & 1.680 \\
    5.0 & 0.590 & 3.540 & 1.660 \\
    10.0 & 0.238 & 3.740 & 1.660 \\
    \hline
  \end{tabular}
  \label{second round simulation}
\end{table}

\subsection{\label{Dimensionless time} Characteristic time and length}
The details regarding the selection of the characteristic time and length to normalize the flow field is introduced in this subsection.

\begin{figure}[H]
  \centering
  \includegraphics[clip=true,width=.45\textwidth]{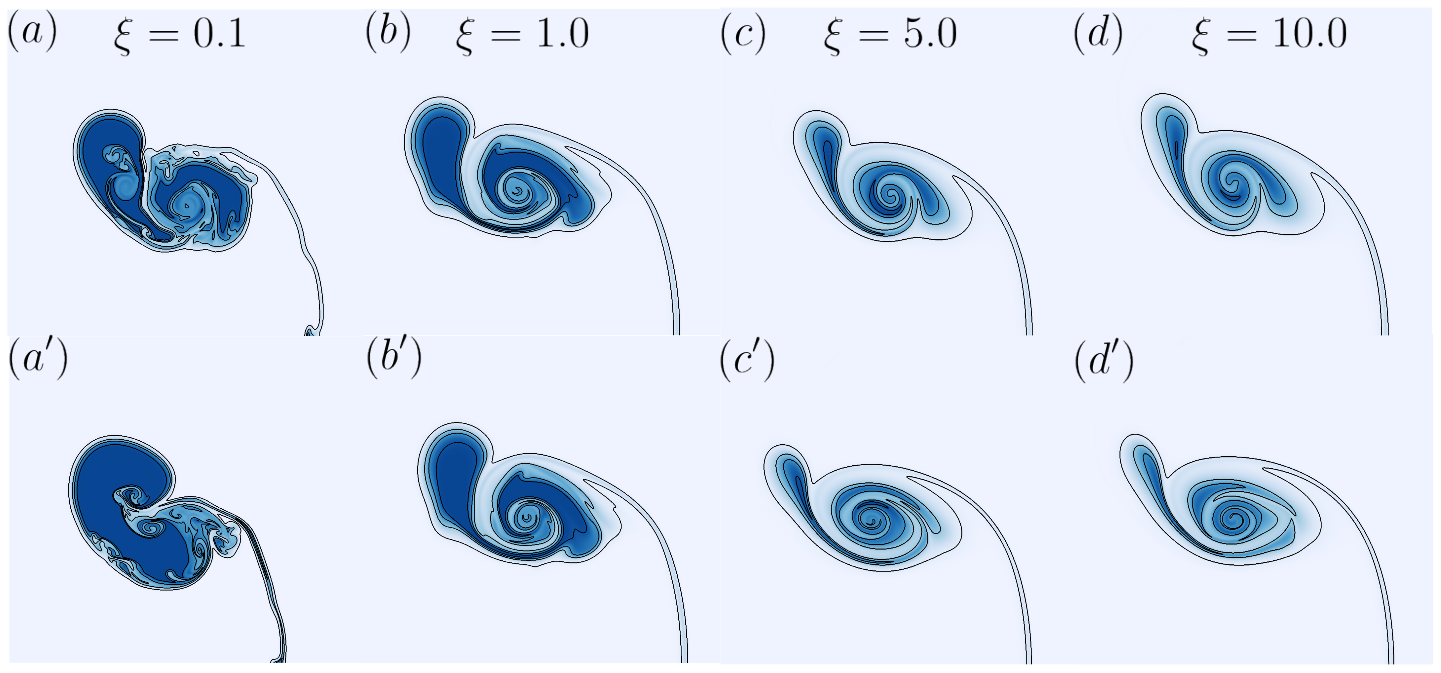}\\
  \caption{The mass fraction contour of the four cases at dimensionless time $t = 3.0$. $(a), (b), (c), (d)$: time is scaled by the post-shock air velocity: $t^{*} = \frac{R}{u_1^{'}}$. $(a'), (b'), (c'), (d')$: time is scaled by the total circulation: $t^{*} = \frac{R^2}{\Gamma}$}
  \label{time scaling flow field}
\end{figure}

As discussed in the previous research \cite{wang2018scaling,yu2022effects}, the characteristic time can be determined by the post-shock air velocity:
\begin{equation}
    t^{*} = \frac{R}{u_{1}^{'}},
    \label{time definition1}
\end{equation}
or the total circulation: 
\begin{equation}
    t^{*} = \frac{R^2}{\Gamma}.
    \label{time definition2}
\end{equation} 
Defining the initial moment $t_{0}$ is the physical time at which the shock wave passing the right edge of the bubble shown in Fig.\;\ref{density contour}, the dimensionless time is 
\begin{equation}
    t = (\tilde{t} - t_0)/t^{*}
    \label{time scaling}
\end{equation} 
The mass fraction contour of the four cases in this paper at $t = 3.0$ scaled by two characteristic time definition in Eq.\;\ref{time scaling} is shown in Fig.\;\ref{time scaling flow field}.

In Fig.\;\ref{time scaling flow field}, the characteristic time defined in Eq.\;\ref{time definition1} is suitable for normalizing the flow field, whereas the definition in Eq.\;\ref{time definition2} is less appropriate. The invalidation of the second definition can be attributed to the effect of the width of the diffusive layer on the characteristic length scale. By defining the effective radius $R_{eff} = \Gamma/u_1^{'}$, the appropriate characteristic time $t^{*} = \frac{R}{u_1^{'}}$ can be expressed as $t^{*} = \frac{R_{eff} R}{\Gamma}$, which essentially replaces the length $R$ in Eq.\;\ref{time definition2} with $\sqrt{R R_{eff}}$. Therefore, the suitable characteristic length and time considering the diffusive layer are defined as in Eq.\;\ref{characteristic time and length}.

\end{appendix}

\end{multicols}
\end{document}